arXiv:1701.06726v3 [cs.CR] 11 Feb 2017

# Instantaneous Decentralized Poker

Iddo Bentov
Cornell University

Ranjit Kumaresan
Microsoft Research

Andrew Miller
UIUC

**Abstract**

We present efficient protocols for *amortized* secure multiparty computation with penalties and secure cash distribution, of which poker is a prime example. Our protocols have an initial phase where the parties interact with a cryptocurrency network, that then enables them to interact only among themselves over the course of playing many poker games in which money changes hands.

The high efficiency of our protocols is achieved by harnessing the power of stateful contracts. Compared to the limited expressive power of Bitcoin scripts, stateful contracts enable richer forms of interaction between standard secure computation and a cryptocurrency.

We formalize the stateful contract model and the security notions that our protocols accomplish, and provide proofs in the simulation paradigm. Moreover, we provide a reference implementation in Ethereum/Solidity for the stateful contracts that our protocols are based on.

We also adopt our off-chain cash distribution protocols to the special case of stateful duplex micropayment channels, which are of independent interest. In comparison to Bitcoin based payment channels, our duplex channel implementation is more efficient and has additional features.

## 1 Introduction

As demonstrated by Cleve [12], fair multiparty computation without an honest majority is impossible in the standard model of communication. Hence, there have been numerous attempts to circumvent this theoretical impossibility result, in particular by relying on techniques such as gradual release (cf. [43] for a survey) and optimistic fair exchange [4]. With the introduction of Bitcoin [40], the academic study of decentralized cryptocurrencies gave rise to a line of research that seeks to impose fairness in secure multiparty computation (MPC) by means of monetary penalties [6]. In this model, the participating parties make security deposits, and the deposits of parties who deviate from the protocol are used to compensate the honest parties.

Still, interacting with a Proof-of-Work based decentralized network entails long waiting times due to the need to be secure against reversal of the ledger history. A recent work by Kumaresan and Bentov [30] showed a Bitcoin based amortization scheme in which the parties run an initial setup phase requiring interaction with the cryptocurrency network, but thereafter they engage in many fair secure computation executions, communicating only among themselves for as long as all parties are honest.

### 1.1 Our Contributions

**Asymptotic gains in amortized protocols.** We present new protocols that rely on stateful contracts instead of Bitcoin transactions, and thereby improve upon the previous results in several ways. First, the setup phase of [30] requires the $n$ parties to execute $O(n^2)$ PoW-based rounds of interaction with the cryptocurrency network, while our stateful protocols require $O(1)$ rounds. The protocols of [30] for secure MPC with penalties also require a security deposit of $O(n^2)$ coins per party, while our protocols require $O(n)$ coins per party. We use UC-style definitions [10] to formalize the security notions that are achieved by our amortized protocols, and provide proofs using the simulation paradigm.

**Amortized SCD.** Unlike the protocols in [30], our protocols support *secure cash distribution with penalties* (SCD), rather than only fair secure MPC with penalties. The distinction between SCD and fair MPC with penalties is that in SCD the inputs and outputs of the parties are comprised of both money and data, while fair MPC with penalties has only data for inputs and outputs (but uses money to compensate honest parties who did not learn the output).

**Real poker.** A canonical example of SCD is a mental poker game, where the outcome of the computation is not intrinsically useful, but rather determines how money should change hands. This means that following an on-chain setup phase, the parties can play any number of *instantaneous* poker games, for as long as no party has run out of money. Hence, while there is a large body of work on efficient mental poker schemes, to the best of our knowledge we are the first to provide a practical poker protocol with actual money transfers from the losers to the winners. Moreover, we accompany our poker protocol with an implementation for the Ethereum cryptocurrency.

**Highly efficient payment channels.** As a special case, our off-chain cash distribution protocols can also be used for stateful duplex payment channels. This use case does not require secure computation and yet it is particularly important. The reason for this is that micropayment channels can reduce the amount of transaction data that the decentralized cryptocurrency network maintains, and thus the long-term scalability pressures that a cryptocurrency faces can be relieved by well-functioning off-chain payment channels (see, e.g., [17] for further discussion). We compare our stateful duplex channel to Bitcoin based off-chain payment channels, and show that that the stateful approach yields better efficiency and extra features. Since micropayment channels are of independent interest, we provide a self-contained protocol and implementation of our stateful duplex off-chain channel.

## 1.2 Related Works

The first secure computation protocols that utilize Bitcoin to guarantee fairness are by Andrychowicz et al. [3, 2] and by Bentov and Kumaresan [6]. Bitcoin based protocols for reactive cash distribution and poker were given by Kumaresan, Moran, and Bentov [31]. The technique for amortized secure computation with penalties in the Bitcoin model was introduced by Kumaresan and Bentov [30]. Our protocols subsume and improve on these, providing both the amortization benefit of [30] with the cash distribution functionality of [31], and furthermore reduce the on-chain costs and the necessary amount of collateral. Several other works analyze fair (non-amortized) protocols in other cryptocurrency models, in particular the analysis of Kiayias, Zhou, and Zikas [28] and Kosba et al. [29].

The cash distribution contract we present (see Section 2.3) is closely related to an ongoing proposal in the cryptocurrency community for "state channels" (cf. Coleman [13]), wherein a group of parties agree on a sequence of "off-chain" state transitions, and resort to an on-chain reconciliation process only in the case that the off-chain communications break down. To our knowledge, no security definition has yet been provided for such applications. Furthermore, our application is much more expressive, since we can implement state transitions that depend on parties' private information, while still guaranteeing fairness.

The original mental poker protocol by Shamir, Rivest, and Adelman [46] relies on commutative encryption. However, their protocol was only for two parties and was found to have security vulnerabilities [33, 14]. Following that, many different protocols for mental poker were proposed. For example, Crépeau presented secure poker protocols that are based on probabilistic encryption [15] and zero-knowledge proofs [16], but his constructions are rather inefficient. In 2003, a breakthrough by Barnett and Smart [48] gave a far more efficient poker protocol that utilizes homomorphic encryption. Castellà-Roca et al. [11] presented a poker protocol that is similar to [48], but has a more efficient shuffle procedure.

The poker protocol that we integrate into our SCD implementation is by Wei and Wang [27], with a full version by Wei [26]. This protocol replaces the homomorphic encryption component of [11] with a faster proof of knowledge scheme, and provides a security proof using the simulation paradigm.

## 2 Overview

In Bitcoin, the full nodes maintain a data structure that is known as the "unspent transaction outputs set" (UTXO set), which represents the current state of the ledger. Each unspent output in the UTXO set incorporates a circuit (a.k.a. script or predicate), such that any party who can provide an input (a.k.a. witness) that satisfies the circuit can spend the coins amount of this output into a new unspent output. Therefore, if there is only one party who knows the witness for the circuit, then this party is in effect the holder of the coins.

Standard Bitcoin transactions use a signature as the witness. The signature is applied on data that also references the new unspent output, thereby binding the transaction to the specific receiver of the coins and thus prevents a man-in-the-middle attack by the nodes in the decentralized Bitcoin network.

However, Bitcoin allows the use of more complex circuits as well. Such circuits allow us to support quite elaborate protocols in which money changes hands, as opposed to to using Bitcoin only for simple money transfers between parties.

Specifically, protocols for fair secure computation and fair lottery can be implemented with a blackbox use of an $\mathcal{F}^{\star}_{\text{CR}}$ functionality [6, 31, 32]. Essentially, $\mathcal{F}^{\star}_{\text{CR}}$ specifies that a "sender" $P_1$ locks her coins in accordance with some circuit $\phi$, such that a "receiver" $P_2$ can gain possession of these coins if she supplies a witness $w$ that satisfies $\phi(w) = 1$ before some predefined timeout, otherwise $P_1$ can reclaim her coins. As shown in [6, 30], the $\mathcal{F}^{\star}_{\text{CR}}$ functionality can be realized in Bitcoin, as long as the circuit $\phi$ can be expressed in the Bitcoin scripting language. In the aforementioned secure computation and lottery protocols [6, 31, 32], the particular circuit that is needed verifies a signature (just as in standard transactions) and a decommitment according to some arbitrary hardcoded value. Such a circuit can be realized by using a hash function for the commitment scheme (Bitcoin supports `SHA1`,`SHA256`,`RIPEMD160`). Since signature verification is an order of magnitude more complex than hash invocation, the complexity of an $\mathcal{F}^{\star}_{\text{CR}}$ transaction is only marginally higher than that of standard Bitcoin transactions.

Thus, the $\mathcal{F}^{\star}_{\text{CR}}$ model can be regarded as a restricted version of the Bitcoin model, which is expressive enough for realizing multiparty functionalities that are impossible in the standard model.

One may ask whether it is possible to design better protocols in a model that is more expressive than the Bitcoin model. In this work we will answer the question in the affirmative.

A possible extension to the Bitcoin transaction structure is covenants [38, 41], where each unspent output specifies not only the conditions on who can spend the coins (i.e., the circuit $\phi$), but also conditions on who is allowed to receive the coins. Indeed, as shown in [38], covenants can be used to implement certain tasks that the current Bitcoin specifications do not support (e.g., vaults that protect against coin theft).

Generalizing further, each unspent output can maintain a *state*. That is, an unspent output will be comprised of a circuit $\phi$ and state variables, and parties can update the state variables by carrying out transactions that satisfy $\phi$ in accord with the current values of the state variables. Additionally, parties can deposit coins into the unspent output, and a party can withdraw some partial amount of the held coins by satisfying $\phi$ with respect to the state variables. This approach is used in Ethereum [52, 9], where an unspent output is referred to as a "contract".

With a slight abuse of terminology, the transaction format of the Bitcoin model can thus be described as "stateless". By this we mean that the coins of an unspent Bitcoin output are controlled by a hardcoded predicate that represents their current state, and anyone who can supply a witness that satisfies this predicate is able to spend these coins into an arbitrary new state.

Let us mention that the Bitcoin transaction format can still enable "smart contracts", in the sense of having coins that can be spent only if some other transaction took place (i.e., without relying on a third party). The technique for achieving this would generally involve multiple signed transactions that are prepared in advance and kept offline. Then, depending on the activity that occurs on the blockchain, some of the prepared transaction will become usable. However, in certain instances the amount of offline transactions may grow exponentially, as in the case of zero-collateral lotteries [37].

The protocols that we present in this work will be in a model that has stateful contracts. As described, this refers to unspent outputs that are controlled according to state variables. It should be emphasized that our protocols do not rely on a Turing-complete scripting language, as all the loops in the contracts that we design (in particular our poker contract) have a fixed number of iterations.

To justify our modeling choice, let us review the advantages of stateful contracts over stateless transactions. As a warmup, we begin by examining simple protocols for 2-party fair exchange.

## 2.1 Fair Exchange with Penalties between Two Parties

Suppose that $P_1, P_2$ execute an unfair secure computation that generates secret shares of the output $x = x_1 \oplus x_2$ and commitments $T_1 = h(x_1), T_2 = h(x_2)$, and delivers $(x_i, T_1, T_2)$ to party $P_i$. Consider the naive protocol for fair exchange of the shares $x_1, x_2$ via $\mathcal{F}^{\star}_{\mathrm{CR}}$ transactions:

$$P_1 \xrightarrow[q,\tau]{T_2} P_2 \tag{1}$$

$$P_2 \xrightarrow[q,\tau]{T_1} P_1 \tag{2}$$

An arrow denotes an $\mathcal{F}^{\star}_{\mathrm{CR}}$ transaction that lets $P_i$ collect $q$ coins from $P_{3-i}$ before time $\tau$, by revealing a decommitment $y$ such that $h(y) = T_i$.

The above protocol is susceptible to an "abort" attack by a malicious $P_2$ that waits for $P_1$ to make the deposit transaction (i.e., Step 1), after which $P_2$ simply does not execute Step 2 to make a deposit transaction. Instead, $P_2$ claims the first transaction, and obtains $q$ coins while $P_1$ obtains $x_2$. Now, $P_2$ simply aborts the protocol. In effect, this means that an honest $P_1$ paid $q$ coins to learn $P_2$'s share. Fairness with penalties requires that an honest party never loses any money, hence this naive approach does not work (cf. [6] for precise details).

The above vulnerability can be remedied via the following protocol:

$$P_1 \xrightarrow[q,\tau_2]{T_1 \wedge T_2} P_2 \tag{1}$$

$$P_2 \xrightarrow[q,\tau_1]{T_1} P_1 \tag{2}$$

For this improved protocol to be secure, two *sequential* PoW-based waiting periods are necessary. Otherwise, a corrupt $P_2$ may be able to reverse transaction (2) after $P_1$ claims it, so that $P_1$ would reveal her share and not be compensated.

By contrast, consider the 2-party fair exchange protocol that is based on a stateful contract, as illustrated in Figure 1. Here, both parties should deposit $q$ coins each, concurrently. If the $2q$ coins were not deposited into the contract before the timeout is reached, then an honest party who deposited into the contract can claim her coins back. In the case that the $2q$ coins were deposited, it triggers to contract to switch to a new state, where each party $P_i$ can claim her deposit by revealing the hardcoded decommitment $T_i$. In contrast to the $\mathcal{F}^{\star}_{\mathrm{CR}}$ protocol, the stateful contracts requires only one PoW-based waiting period before the honest parties may reveal their shares.

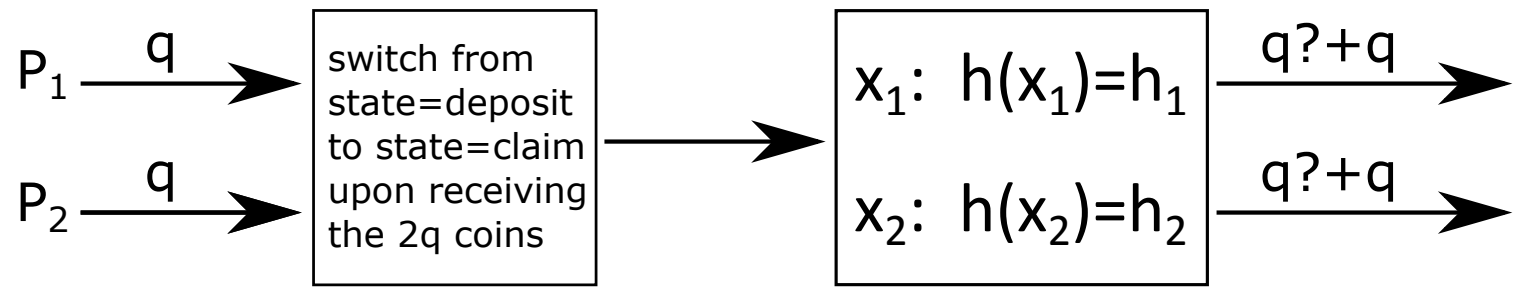

Figure 1: Stateful contract for fair secure 2-party computation with penalties.

While the quantitative difference between stateful and stateless contracts in the above discussion may appear to be unimpressive, the distinction becomes more pronounced in the case of multiparty fair exchange (a.k.a. fair reconstruction [6]), and even more so in amortized protocols. Let us demonstrate the amortized multiparty case in the next section.

## 2.2 Amortized Multiparty Fair SFE with Penalties

We illustrate in Figure 2 the stateful contract for $n$ parties who wish to engage in amortized fair secure function evaluation (SFE) with penalties. The lifespan of this contract can be thought of as having three phases:

- *Deposit Phase:* All parties should deposit $q(n-1)$ coins each. If $nq(n-1)$ coins were deposited before the initial timeout is reached, then the contrast switches into an "active" state. Otherwise, each honest party who deposited will claim her $q(n-1)$ coins back.

- *Execution Phase:* While the state is "active", the $n$ parties will not interact with the contract at all. Instead, they will engage in multiple executions of SFE. In the $i^{\text{th}}$ execution, the secure computation prepares secret shares $\{x_{i,j}\}_{j=1}^{n}$ of the output, as well as commitments $\{h_{i,j} = h(x_{i,j})\}_{j=1}^{n}$, and delivers $(x_{i,j}; h_{i,1}, h_{i,2}, \ldots, h_{i,n})$ to party $P_j$. Each party will then use her secret key (for which the corresponding public key is hardcoded in the contract) to create a signature $s_{i,j}$ for the tuple $(h_{i,1}, h_{i,2}, \ldots, h_{i,n})$, and send the signature $s_{i,j}$ to the other parties. Upon receiving all the signatures $\{s_{i,j}\}_{j=1}^{n}$, each honest party $P_j$ will send her secret share $x_{i,j}$ in the clear to the other parties.

- *Claim Phase:* In the case that a corrupt party $P_c$ did not reveal her share $x_{i,c}$ during the execution phase, each honest party $P_j$ will send $m_{i,j} = (x_{i,j}; s_{i,1}, s_{i,2}, \ldots, s_{i,n})$ to the contract, and thereby transition the contract into a "payout" state. The message $m_{i,j}$ also registers that $P_j$ deserves to receive a compensation of $q$ coins, in addition to her initial $q(n-1)$ coins deposit. Until a timeout, any party $P_\ell$ can avoid being penalized by sending $m_{i',\ell} = (x_{i',\ell}; s_{i',1}, s_{i',2}, \ldots, s_{i',n})$ with $i' \geq i$ to the contract. In case $i' > i$, this would invalidate the $q$ coins compensation that was requested via $m_{i,j}$, and instead register that $P_\ell$ is owed $q$ coins in compensation.

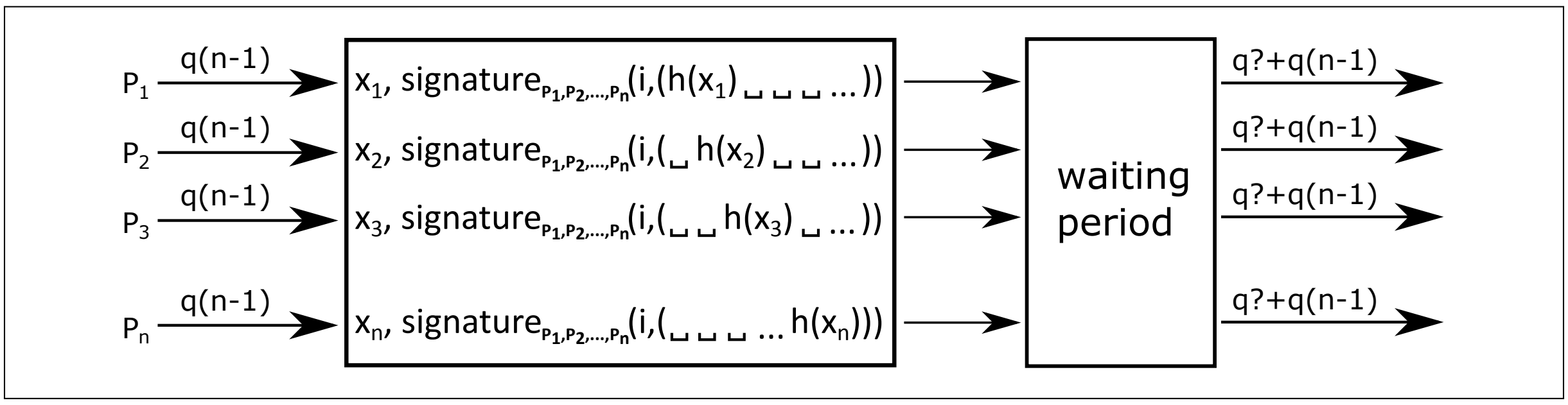

Figure 2: Stateful contract for amortized multiparty fair SFE with penalties.

As can be observed, the $n$ parties can engage in an unlimited amount of off-chain SFE executions (where the executions can compute different functions), and no interaction with the blockchain will take place as long as all parties are honest. When a corrupt party $P_c$ deviates from this protocol, each honest party will receive $q$ coins compensation, that is taken from $P_c$'s initial security deposits of $q(n-1)$ coins. The actual protocol handles more technical issues, cf. Section 4.

By contrast, achieving the same guarantees in the $\mathcal{F}^{\star}_{\text{CR}}$ model is known to possible only via an intricate "see-saw" construction, that requires $O(n^2)$ PoW-based rounds and a collateral of $O(qn^2)$ coins from each party [30]. Moreover, the stateless nature of Bitcoin transaction entail a global timeout after which the entire see-saw construction expires. Setting the global timeout to a high value enables many off-chain SFE executions, but also implies that a DoS attack by a corrupt party (who would abort before signing any secret shares of the output of the first execution) will cause each honest party to wait for a long time before being able to regain possession of her $O(qn^2)$ coins deposit. Due to the time value of money, this is obviously undesirable. The stateful approach does not require a global timeout that is measured in *absolute* terms. Instead, the contract remains operational for as long as all the parties wish to engage in the off-chain protocol, and transitioning the contract into the "payout" state will trigger an event whose expiration is *relative* to the time at which the transition occurred.

We provide an Ethereum implementation of this contract in Figure 18.

Can stateful contracts provide even more benefits? As the next section shows, the answer is "yes".

### 2.3 Stateful Off-Chain Cash Distribution Protocols

Suppose that the parties $P_1, P_2$ wish to play a multiple-round lottery game, such that either of them is allowed to quit after each round. Thus, $P_1$ enters the lottery with $m$ coins, $P_2$ enters with $w$ coins, and in the first round $P_1$ picks a random secret $x_1$ and commits to $\mathsf{com}(x_1)$, $P_2$ picks $x_2$ and commits to $\mathsf{com}(x_2)$, and then they decommit $x_1, x_2$ so that the least significant bit of $x_1 \oplus x_2$ decides whether $P_1$'s balance is incremented to $m+1$ and $P_2$'s balance is decremented to $w-1$, or $P_1$'s balance is decremented to $m-1$ and $P_2$'s balance is incremented to $w+1$. If both of them wish to continue, then they will proceed to the next round and repeat this protocol.

Obviously, the parties must not be allowed to quit in the middle of a round without repercussions. That is, if $P_1$ reveals her decommitment $x_1$ and $P_2$ aborts, then $P_1$ should be compensated. Moreover, as in Section 2.2, it is better that the parties play each round without any on-chain interaction.

Therefore, in each round $i$, $P_1$ and $P_2$ will sign the current balance $m_i, w_i$ together with the round index $i$ and the commitments $\mathsf{com}(x_{i,1}), \mathsf{com}(x_{i,2})$. After the parties exchange these signed messages, they can safely send their decommitments $x_{i,1}, x_{i,2}$ in the clear. The logic of the stateful contract allows each party to send her decommitment along with the signed message, and thus finalize the game according to the current balances. If the other party does not reveal her decommitment during a waiting period, then the contract increments the balance of the honest party. If both parties reveal, then the contract computes $x_1 \oplus x_2$ to decide who won the last round, so that the balance of the winner is incremented and the balance of the loser is decremented. During the waiting period, an honest party can send a signed message with an index $i' > i$ and thereby invalidate the message that a corrupt party sent to the contract.

It should also be noted that an honest party should not continue to play after the balance of the other party reaches 0, since the contract cannot reward the winner more money than what was originally deposited.

We illustrate the contract in Figure 3, and provide an Ethereum implementation in Figures 18 and 20. The multiparty version of our lotery code is available to run at [1].

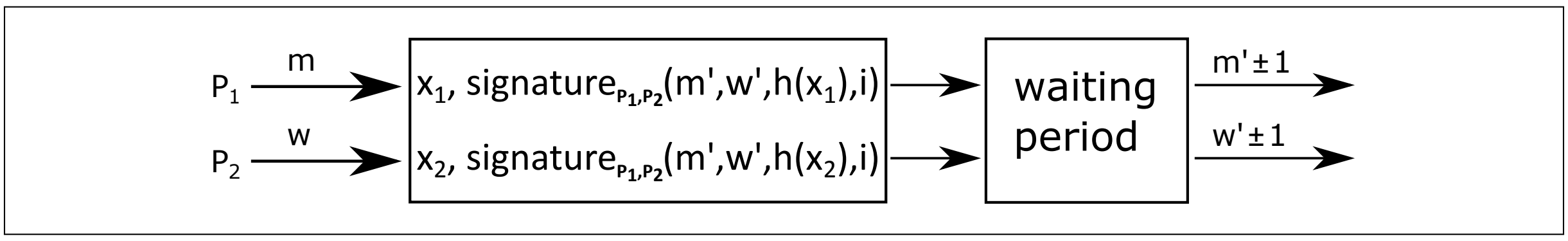

Figure 3: Off-chain 2-party lottery.

Such a 2-party lottery is a very simple example of a *secure cash distribution with penalties* (SCD) functionality [31]. Another special case of SCD is a multiparty poker game in which money (i.e., coins of the cryptocurrency system that the parties hold) is transferred from losers to winners. In Section 3 and onwards we formulate the ideal functionalities and protocol for fair MPC and SCD, and in Section 7 we provide an efficient off-chain poker protocol with implementation.

As noted in Section 1, SCD can also be realized in the $\mathcal{F}^{\star}_{\mathrm{CR}}$ model via a non-amortized (i.e., on-chain) protocol, though the construction requires a setup phase with $O(n^2)$ PoW-based rounds. To the best of our knowledge, there is no amortized SCD realization in the $\mathcal{F}^{\star}_{\mathrm{CR}}$ model.

## 2.4 Stateful Duplex Off-Chain Micropayment Channels

By removing the commit/decommit logic, the 2-party contract of Section 2.3 would function as a duplex off-chain payment channel.

Suppose that Alice may wish to buy some items that Bob sells, and vice versa. Hence, Alice will put $m$ coins into the contract, and Bob will put $w$ coins into the contract. After these two initial transactions are confirmed via PoW, Alice and Bob will be able to perform bi-directional off-chain payments between them. That is, if Bob wishes to buy an item that costs $d$ coins from Alice, then he will create the signature $S_{b,1} = \mathsf{Sign}_{sk_B}(m+d, w-d, 1)$ and send $S_{b,1}$ to Alice. Since Alice can create $S_{a,1} = \mathsf{Sign}_{sk_A}(m+d, w-d, 1)$ on her own, she effectively received a payment of $d$ coins from Bob. Later on, Alice may wish to buy an item from Bob that costs $d'$ coins, hence she will create the signature $S_{a,2} = \mathsf{Sign}_{sk_A}(m+d-d', w-d+d', 2)$ and send it to Bob.

If Alice is malicious and tries to finalize the payments by sending $S_{a,1}, S_{b,1}$ to the contract, then Bob will be able to invalidate Alice's message by sending $S_{a,2}, S_{b,2} = \mathsf{Sign}_{sk_B}(m+d-d', w-d+d', 2)$ during the waiting period of the contract. Since $2 > 1$, i.e., the message $(m+d-d', w-d+d', 2)$ has a higher index than that of the message $(m+d, w-d, 1)$, the rules of the contract will allow Bob to withdraw $w-d+d'$ coins, as it should be.

Let us note that the duplex channel cannot end up in a deadlock. Let $(m, w, i)$ be the last signed message that was sent off-chain. Suppose that Alice now signs and sends $(m-1, w+1, i+1)$ with the intention of buying an item that costs 1 coin from Bob, and at the same time Bob signs and sends $(m+1, w-1, i+1)$ with the intention of buying an item that costs 1 coin from Alice. In this case, both Alice and Bob will refuse to send the item in exchange for the payment. Instead, one of them (for example Alice) will send a signed message $(m-1, w+1, i+2)$ that specifies the higher index $i+2$, and then Bob can securely send his item to Alice.

As in Section 2.2, a duplex payment channel that is based on a stateful contract does not have global timeout: the contract remains operational as long as Alice and Bob wish to keep sending off-chain payments between them, and transitioning the contract into the waiting period will trigger an event whose timeout is *relative*. This stands in contrast to the original Bitcoin-based stateless payment channel (cf. [39, 49, 24, 50, 35]), that relies on the `CHECKLOCKTIMEVERIFY` script instruction (or on refund transactions) and therefore has an *absolute* timeout. Setting the absolute timeout to a high value enables off-chain payments over an extended period of time, but also enables DoS by a corrupt party who would abort so that the locked money of the honest party will be rendered useless during this extended time period.

Our stateful contract also allows a party who is running out of funds to deposit more coins into the contract via an on-chain transaction. As the alternative is to terminate the channel and starting a new one, this approach is more efficient and requires less fees. Furthermore, our stateful approach allows the parties to invoke on-chain incremental withdrawals, i.e., to move some of their deposited amount out of the contract while retaining the rest of the amount for future off-chain payments.

The straightforward Bitcoin-based payment channel [49, 24, 50] is uni-directional (e.g., monotonically increasing balance from Alice to Bob). Poon and Dryja [44] developed an improved payment channel with duplex functionality, that operates by generating ephemeral private keys. Independently, Decker and Wattenhofer [19] developed an alternative technique for duplex channels, where the direction of payments can be reversed (up to a bounded number of times) by having both parties sign a new "recovery" transaction with an earlier timelock value. The Poon-Dryja duplex channel supports a relative timeout (following a Bitcoin protocol fork [8]), while the Decker-Wattenhofer channel requires an absolute timeout. Both the Poon-Dryja and Decker-Wattenhofer constructions are significantly more complex than our duplex channel (due to the stateless nature of Bitcoin transactions), and neither of them supports incremental deposits/withdrawals. It should be noted that the Poon-Dryja and Decker-Wattenhofer channels also support "hashlock contracts," which effectively allow multiple channels to be chained, forming a payment network. This feature can easily be incorporated into our setting as well, but we omit it in the interest of simplicity (as it beyond the scope of our work).

Overall, the stateful contract enables a duplex off-chain channel where each (micro-)payment requires transmission of only one simple signature from the payer to the payee, which is the bare minimum that any off-chain transaction entails. See Figure 4 for an illustration of the stateful duplex off-chain channel.

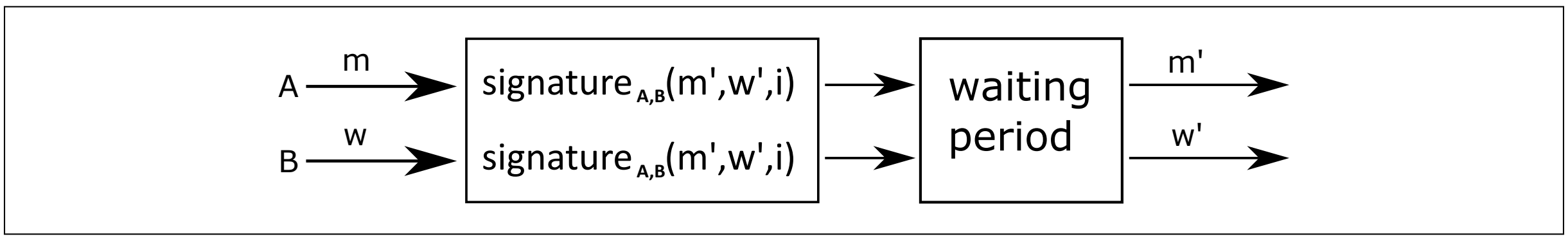

Figure 4: Duplex Off-chain Payment Channel.

As the comparison to Section 2.3 alludes to, the stateful contact for off-chain payments is a special case of our SmartAmortize contract (given in Figures 18 and 20) for off-chain secure cash distribution. However, since off-chain micropayment channels are of particular interest, we provide the self-contained implementation of a duplex off-chain channel that supports incremental deposits and withdrawals (see Section 6.1 and Figure 21).

Let us note that the universal payment channels proposal of Tremback and Hess [51] is similar to the duplex payment contract that we implement. See also Peterson [42] for an implementation of a slightly different duplex channel design. In contrast to our duplex channel, [51] and [42] do not support incremental deposits and withdrawals.

# 3 Preliminaries

We say that a function $\mu(\cdot)$ is negligible in $\lambda$ if for every polynomial $p(\cdot)$ and all sufficiently large $\lambda$'s it holds that $\mu(\lambda) < 1/|p(\lambda)|$. A probability ensemble $X = \{X(t, \lambda)\}_{t\in\{0,1\}^*, n\in\mathbb{N}}$ is an infinite sequence of random variables indexed by $a$ and $\lambda \in \mathbb{N}$. Two distribution ensembles $X = \{X(t, \lambda)\}_{\lambda\in\mathbb{N}}$ and $Y = \{Y(t, \lambda)\}_{\lambda\in\mathbb{N}}$ are said to be computationally indistinguishable, denoted $X \overset{c}{\equiv} Y$ if for every non-uniform polynomial-time algorithm $D$ there exists a negligible function $\mu(\cdot)$ such that for every $t \in \{0, 1\}^*$,

$$|\Pr[D(X(t, \lambda)) = 1] - \Pr[D(Y(t, \lambda)) = 1]| \leq \mu(\lambda).$$

All parties are assumed to run in time polynomial in the security parameter $\lambda$. We prove security in the "secure computation with coins" (SCC) model proposed in [6]. Note that the main difference from standard definitions of secure computation [22] is that now the view of $\mathcal{Z}$ contains the distribution of coins. Let $\text{IDEAL}_{f,\mathcal{S},\mathcal{Z}}(\lambda, z)$ denote the output of environment $\mathcal{Z}$ initialized with input $z$ after interacting in the ideal process with ideal process adversary $\mathcal{S}$ and (standard or special) ideal functionality $\mathcal{G}_f$ on security parameter $\lambda$. Recall that our protocols will be run in a hybrid model where parties will have access to a (standard or special) ideal functionality $\mathcal{G}_g$. We denote the output of $\mathcal{Z}$ after interacting in an execution of $\pi$ in such a model with $\mathcal{A}$ by $\text{HYBRID}^g_{\pi,\mathcal{A},\mathcal{Z}}(\lambda, z)$, where $z$ denotes $\mathcal{Z}$'s input. We are now ready to define what it means for a protocol to SCC realize a functionality.

**Definition 1.** *Let $n \in \mathbb{N}$. Let $\pi$ be a probabilistic polynomial-time $n$-party protocol and let $\mathcal{G}_f$ be a probabilistic polynomial-time $n$-party (standard or special) ideal functionality. We say that $\pi$* SCC realizes *$\mathcal{G}_f$ with abort in the $\mathcal{G}_g$-hybrid model (where $\mathcal{G}_g$ is a standard or a special ideal functionality) if for every non-uniform probabilistic polynomial-time adversary $\mathcal{A}$ attacking $\pi$ there exists a non-uniform probabilistic polynomial-time adversary $\mathcal{S}$ for the ideal model such that for every non-uniform probabilistic polynomial-time adversary $\mathcal{Z}$,*

$$\{\text{IDEAL}_{f,\mathcal{S},\mathcal{Z}}(\lambda, z)\}_{\lambda\in\mathbb{N}, z\in\{0,1\}^*} \overset{c}{\equiv} \{\text{HYBRID}^g_{\pi,\mathcal{A},\mathcal{Z}}(\lambda, z)\}_{\lambda\in\mathbb{N}, z\in\{0,1\}^*}.$$

**Definition 2.** *Let $\pi$ be a protocol and $f$ be a multiparty functionality. We say that $\pi$* securely computes *$f$* with penalties *if $\pi$ SCC-realizes the functionality $\mathcal{F}^\star_f$ according to Definition 1.*

Notation: session identifier $sid$, parties $P_1, \ldots, P_n$, adversary $\mathcal{S}$ that corrupts $\{P_s\}_{s \in C}$, *safety deposit* $d$, *penalty amount* $q$. Set $H = [n] \setminus C$ and $h = |H|$.

DEPOSIT PHASE: Initialize $\mathsf{flg} = \perp$. Wait to get message $(\mathsf{setup}, sid, ssid, j, \mathsf{coins}(d))$ from $P_j$ for all $j \in H$. Wait to get message $(\mathsf{setup}, sid, ssid, \mathsf{coins}(hq))$ from $\mathcal{S}$.

EXECUTION PHASE: Set $\mathsf{flg} = 0$. For $\mathsf{id} = 1, 2, \cdots$, sequentially do:

- If a message $(\mathsf{exit}, sid)$ is received from $P_j$, then send $(\mathsf{exit}, sid, j)$ to all parties, and go to the claim phase.
- Wait to receive a message $(\mathsf{input}, sid, ssid \| \mathsf{id}, j, y_j^{(\mathsf{id})}, g^{(\mathsf{id})})$ from $P_j$ for all $j \in H$. Then send $(\mathsf{function}, sid, ssid \| \mathsf{id}, g^{(\mathsf{id})})$ to all parties.
- Wait to receive a message $(\mathsf{input}, sid, ssid \| \mathsf{id}, \{y_s^{(\mathsf{id})}\}_{s \in C}, g^{(\mathsf{id})})$ from $\mathcal{S}$. If no such message was received, then go to the claim phase.
- Compute $z^{(\mathsf{id})} \leftarrow g^{(\mathsf{id})}(y_1^{(\mathsf{id})}, \ldots, y_n^{(\mathsf{id})})$ and send $(\mathsf{output}, sid, ssid \| \mathsf{id}, z^{(\mathsf{id})})$ to $\mathcal{S}$.
- If $\mathcal{S}$ returns $(\mathsf{continue}, sid, ssid \| \mathsf{id})$, send $(\mathsf{output}, sid, ssid \| \mathsf{id}, z^{(\mathsf{id})})$ to each $P_j$.
- Else if $\mathcal{S}$ returns $(\mathsf{abort}, sid, ssid)$, update $\mathsf{flg} = 1$, and go to the claim phase.

CLAIM PHASE:

- If $\mathsf{flg} = 0$ or $\perp$, send $(\mathsf{return}, sid, ssid, \mathsf{coins}(d))$ to $P_j$ for $j \in H$. If $\mathsf{flg} = 0$, send $(\mathsf{return}, sid, ssid, \mathsf{coins}(hq))$ to $\mathcal{S}$.
- Else if $\mathsf{flg} = 1$, send $(\mathsf{penalty}, sid, ssid, \mathsf{coins}(d + q + q_i))$ to $P_i$ for all $i \in H$ where $q_i = 0$ unless $\mathcal{S}$ sent a message $(\mathsf{extra}, sid, ssid, \{(i, q_i)\}_{i \in H}, \mathsf{coins}(\sum_{i \in H} q_i))$.

Figure 5: Special ideal functionality $\mathcal{F}^*_{\mathrm{MSFE}}$ for multiple sequential SFE with penalties.

Throughout this paper, we deal only with static adversaries and impose no restrictions on the number of parties that can be corrupted. Our schemes also make use of a digital signature scheme which we denote as $(\mathsf{SigKeyGen}, \mathsf{SigSign}, \mathsf{SigVerify})$. Please see [6, 31, 32, 30] for additional details on the model including the necessary modifications to UC. Please also see [28] which extensively treats these modifications, proposes alternative models, and uses protocol compilers different from GMW.

## 3.1 Ideal Functionalities

**Secure computation with penalties—multiple executions.** We now present the functionality $\mathcal{F}^*_{\mathrm{MSFE}}$ which we wish to realize. Recall that secure computation with penalties guarantees the following.

- An honest party never has to pay any penalty.
- If a party aborts after learning the output and does not deliver output to honest parties, then *every* honest party is compensated.

See Figure 5 for a formal description. Note that $\mathcal{F}^*_{\mathrm{MSFE}}$ directly realizes multiple invocations of *non-reactive* secure computation with penalties. In the first phase referred to as the *deposit phase*, the functionality $\mathcal{F}^*_{\mathrm{MSFE}}$ accepts safety deposits $\mathsf{coins}(d)$ from each honest party and penalty deposit $\mathsf{coins}(hq)$ from the adversary. Note that the penalty deposit suffices to compensate each honest party in the event of an abort. Once the deposits are made, parties enter the next phase referred to as the *execution phase* where parties can engage in unbounded number of secure function evaluations. In each execution, parties submit inputs and wait to receive outputs. As usual, the ideal adversary $\mathcal{S}$ gets to learn the output first and then decide whether to deliver the output to all parties. If $\mathcal{S}$ decides to abort, then no further executions are carried out, parties enter the *claim phase*, and honest parties get $\mathsf{coins}(d+q)$, i.e., their safety deposit plus the penalty amount. Now if $\mathcal{S}$ never aborts during a local execution, then the safety deposits are returned back to the honest parties, and $\mathcal{S}$ gets back its penalty deposit. Note that we explicitly allow an $\mathsf{exit}$ command that enables parties to exit the contract immediately. Prior works [31, 30] required parties to wait for a pre-specified time out parameter before parties can reclaim their deposits.

*Supporting cash distribution via* $\mathcal{F}^*_{\mathrm{MSCD}}$. See Figure 6 for a formal definition for $\mathcal{F}^*_{\mathrm{MSCD}}$. The definition for Notation: session identifier $sid$, parties $P_1, \ldots, P_n$, adversary $\mathcal{S}$ that corrupts $\{P_s\}_{s \in C}$, *safety deposit* $d$, *penalty amount* $q$, set $H = [n] \setminus C$.

Deposit phase: Initialize $\mathsf{flg} = \perp$.

- Wait to receive a message $(\mathsf{setup}, sid, ssid, j, \mathsf{coins}(d))$ from $P_j$ for all $j \in H$.
- Wait to receive a message $(\mathsf{setup}, sid, ssid, \mathsf{coins}(hq))$ from $\mathcal{S}$ where $h = |H|$.

Execution phase: Initialize $\mathsf{flg} = 0$ and $\boldsymbol{b} \leftarrow \mathbf{0}$. For $\mathsf{id} = 1, 2, \ldots$, sequentially do:

- If a message $(\mathsf{exit}, sid, ssid)$ is received from $P_j$, then send $(\mathsf{exit}, sid, ssid, j)$ to all parties, and go to the claim phase.
- If a message $(\mathsf{addmoney}, sid, ssid\|\mathsf{id}, b_j, \mathsf{coins}(b_j))$ is received from some $P_j$, and a message $(\mathsf{addmoney}, sid, ssid\|\mathsf{id}, b_j)$ was received from every $P_k$ with $k \neq j$, then send $(\mathsf{addmoney}, sid, ssid\|\mathsf{id}, b_j)$ to all parties. Update $\boldsymbol{b} \leftarrow \boldsymbol{b} + (\cdots, 0, b_j, 0, \cdots)$.
- Initialize $\mathsf{state} = \perp$. Wait to receive message $(\mathsf{function}, sid, ssid\|\mathsf{id}, g^{(\mathsf{id})})$ from $P_j$ for all $j \in H$. Then send $(\mathsf{function}, sid, ssid\|\mathsf{id}, g^{(\mathsf{id})})$ to all parties.
- Parse $g^{(\mathsf{id})} = \{g_k^{(\mathsf{id})}\}_{k \in [\rho]}$. For $k = 1, \ldots, \rho$, sequentially do:
  - Wait to receive message $(\mathsf{input}, sid, ssid\|\mathsf{id}\|k, j, y'_j)$ from $P_j$ for all $j \in H$.
  - Wait to receive a message $(\mathsf{input}, sid, ssid\|\mathsf{id}\|k, \{y'_s\}_{s \in C})$ from $\mathcal{S}$. If no such message was received, update $\mathsf{flg} = 1$ and go to the claim phase.
  - Compute $(z, \boldsymbol{b}', \mathsf{state}) \leftarrow g_k^{(\mathsf{id})}(y'_1, \ldots, y'_n; \mathsf{state}, \boldsymbol{b})$.
  - Send message $(\mathsf{output}, sid, ssid\|\mathsf{id}\|k, z, \boldsymbol{b}')$ to $\mathcal{S}$.
  - If $\mathcal{S}$ sends $(\mathsf{continue}, sid, ssid\|\mathsf{id}\|k)$, send $(\mathsf{out}, sid, ssid\|\mathsf{id}\|k, z, \boldsymbol{b}')$ to all $P_i$.
  - If $\mathcal{S}$ returns $(\mathsf{abort}, sid, ssid\|\mathsf{id}\|k)$, set $\mathsf{flg} = 1$, and go to the claim phase.
  - Update $\boldsymbol{b} \leftarrow \boldsymbol{b}'$.

Claim phase:

- If $\mathsf{flg} = 0$ or $\perp$, send $(\mathsf{return}, sid, ssid, \mathsf{coins}(d + \boldsymbol{b}_r))$ to all $P_r$ for $r \in H$. If $\mathsf{flg} = 0$, send $(\mathsf{return}, sid, ssid, \mathsf{coins}(hq + \sum_{s \in C} \boldsymbol{b}_s))$ to $\mathcal{S}$.
- Else if $\mathsf{flg} = 1$, send $(\mathsf{penalty}, sid, ssid, \mathsf{coins}(d + q + \boldsymbol{b}_i + q_i))$ to $P_i$ for all $i \in H$ where $q_i = 0$ unless $\mathcal{S}$ sent a message $(\mathsf{extra}, sid, ssid, \{q_i\}_{i \in H}, \mathsf{coins}(\sum_{i \in H} q_i))$. Send $(\mathsf{remaining}, sid, ssid, \mathsf{coins}(\sum_{s \in C} \boldsymbol{b}_s))$ to $\mathcal{S}$.

Figure 6: Special ideal functionality $\mathcal{F}^*_{\mathrm{MSCD}}$ for multiple sequential MPC with penalties.

Notation: session identifier *sid*, parties $P_1, \ldots, P_n$, *initial deposit vector* $(d_1, \ldots, d_n)$, program $\mathsf{Prog}$ and an initial state $\mathsf{st}$. We assume that the initial deposit vector is specified in $\mathsf{Prog}$.

<u>Initialization phase:</u> Wait to get message $(\mathsf{init}, sid, ssid, \mathsf{Prog}, \mathsf{st}, \mathsf{coins}(d_j))$ from $P_j$ for all $j \in [n]$. Initialize $Q \leftarrow \sum_{j\in[n]} d_j$.

<u>Execution phase:</u> Repeat until termination:

- Wait to receive a message $(\mathsf{trigger}, sid, ssid, w)$ from some party $P_j$ at time $t$ such that $\mathsf{Prog}(j, w, t; \mathsf{st}) \neq \bot$. Then, let $(\mathsf{st}, e) \leftarrow \mathsf{Prog}(j, w, t; \mathsf{st})$. Send $(\mathsf{output}, sid, ssid, j, w, t, \mathsf{st}, e)$ to each $P_k$. In addition, send $(\mathsf{coins}, sid, ssid, \mathsf{coins}(e))$ to $P_j$. Update $Q \leftarrow Q - e$. If $Q = 0$, then terminate.

Figure 7: Special ideal functionality $\mathcal{F}^*_{\mathrm{StCon}}$ for stateful contracts.

amortized secure cash distribution with penalties in the reactive setting $\mathcal{F}^*_{\mathrm{MSCD}}$ is identical to $\mathcal{F}^*_{\mathrm{MSFE}}$ except that the we repeatedly evaluate reactive functions which is composed of multiple stage functions. Now, $\mathcal{S}$ can abort between different stages of a reactive function evaluation or within a single stage. In either case, the honest parties will be compensated via the penalty deposit $\mathsf{coins}(hq)$ submitted by $\mathcal{S}$ in the deposit phase. Furthermore, $\mathcal{F}^*_{\mathrm{MSCD}}$ also supports cash distribution which makes it useful for many applications. In particular, we allow parties to dynamically add deposits. The reactive functions that are evaluated take into account the current balance which is maintained in the variable $\boldsymbol{b}$, and the output of the evaluations update $\boldsymbol{b}$ to reflect how the cash is redistributed. That is, the amount of coins are specified as input to the reactive functions, and the output will influence how the coins are redistributed. Finally we note that unlike prior formulations [31, 30], we do not require an apriori upper bound on the *number of stages* that is common to the reactive functions supported by $\mathcal{F}^*_{\mathrm{MSCD}}$. Like most prior works (with the exception of [29]), we do not attempt to hide the (updated) balance vectors or the amount of coins.

## 3.2 Stateful contract — Ideal Functionality

We now present the functionality $\mathcal{F}^*_{\mathrm{StCon}}$ which we use to abstract the smart contracts functionality provided by cryptocurrencies. At a high level, $\mathcal{F}^*_{\mathrm{StCon}}$ lets parties run a time-dependent stateful computation. In other words, the contract encodes a finite state computation where each transition could potentially be dependent on the *time of the transition*. Time-dependent transitions in our stateful contract functionality allow us to design protocols that can support early termination of contracts which could be a potentially critical feature in certain settings. Such features are not supported by claim-or-refund $\mathcal{F}^\star_{\mathrm{CR}}$.

As is typical in the penalty model, we let parties make an initial deposit to the $\mathcal{F}^*_{\mathrm{StCon}}$ functionality. Following this, parties together specify a finite state computation denoted $\mathsf{Prog}$ along with an initial state $\mathsf{st}$. The functionality then simply waits for a state transition to be triggered by any of the parties. Upon a valid trigger $w$ (i.e., for which $\mathsf{Prog}$ produces non-$\bot$ output), the contract runs $\mathsf{Prog}$ on the tuple $(j, w, t; \mathsf{st})$ where $j$ is the index of party $P_j$ who supplies the trigger $w$ at time $t$, and where $\mathsf{st}$ is the current state of $\mathsf{Prog}$. $\mathsf{Prog}$ then outputs the current state which will be stored in the variable $\mathsf{st}$ and the amount of money $e$ that $P_j$ is supposed to obtain. Both $\mathsf{st}$ and $e$ are revealed to all parties (i.e., $\mathcal{F}^*_{\mathrm{StCon}}$ has public state), and $\mathcal{F}^*_{\mathrm{StCon}}$ will distribute $\mathsf{coins}(e)$ to party $P_j$. The formal description of $\mathcal{F}^*_{\mathrm{StCon}}$ is given in Figure 7. The functionality repeatedly accepts state transitions until it has distributed all the coins that were deposited to it in the initialization phase.

Analogous to the script complexity definition for $\mathcal{F}^\star_{\mathrm{CR}}$-based protocols, we can define the *script complexity* (also referred to as "validation complexity") of a protocol in the $\mathcal{F}^*_{\mathrm{StCon}}$-hybrid model.

**Definition 3** ($\mathcal{F}^*_{\mathrm{StCon}}$ Script Complexity)**.** *Let $\Pi$ be a protocol among $P_1, \ldots, P_n$ in the $\mathcal{F}^*_{\mathrm{StCon}}$-hybrid model where $\mathcal{F}^*_{\mathrm{StCon}}$ is initialized with program* $\mathsf{Prog}$ *and initial state* $\mathsf{st}$. *We say a trigger $T = (j, w, t)$ is valid iff* $\mathsf{Prog}(j, w, t; \mathsf{st}) \neq \bot$ *where* $\mathsf{st}$ *is the current state of $\mathcal{F}^*_{\mathrm{StCon}}$. We say a state* $\mathsf{st}$ *is valid iff there exists a valid sequence of triggers starting from some initial state* $\mathsf{st}_0$ *that result in* $\mathsf{st}$ *becoming the current state of $\mathcal{F}^*_{\mathrm{StCon}}$. For a trigger $T$ acting on input state* $\mathsf{st}$*, we let $C(\mathsf{Prog}, T, \mathsf{st})$ denote the sum of the size of $T$, the*

Notation: session identifier $sid$, set of parties $\{P_1, \ldots, P_n\}$, *initial deposit vector* $d_1 = \cdots d_n = (n-1)q$, where $q$ is the penalty amount, a program $\mathsf{Prog}$, an initial state $\mathsf{st}$, and a validation function for updates $\mathsf{Update}$. During the execution phase, parties will be able to add coins via $\mathsf{Update}$.

<u>Initialization phase:</u> Wait to get $(\mathsf{init}, sid, ssid, \mathsf{Prog}, \mathsf{Update}, \mathsf{st}, d_j, \mathsf{coins}(d_j))$ from each $P_j$. Initialize $Q \leftarrow \sum_{j\in[n]} d_j$.

<u>Execution phase:</u> Repeat until termination:

- If a message $(\mathsf{update}, sid, ssid, u, b, \mathsf{coins}(b))$ is received from $P_j$ at time $t$ such that $\mathsf{Update}(j, w, t; \mathsf{st}) \neq \perp$, then set $(\mathsf{Prog}, \mathsf{Update}, \mathsf{st}) \leftarrow \mathsf{Update}(j, u, t; \mathsf{Prog}, \mathsf{st})$, accept $\mathsf{coins}(b)$, update $Q \leftarrow Q + b_j$, and send $(\mathsf{update}, sid, ssid, , j, u, t, \mathsf{st}, \mathsf{Prog}, \mathsf{Update})$ to all parties.
- If a message $(\mathsf{trigger}, sid, ssid, w)$ is received from $P_j$ at time $t$ such that $\mathsf{Prog}(j, w, t; \mathsf{st}) \neq \perp$. Then, let $(\mathsf{st}, e) \leftarrow \mathsf{Prog}(j, w, t; \mathsf{st})$. Send $(\mathsf{output}, sid, ssid, j, w, t, \mathsf{st}, e)$ to each $P_k$ and send $(\mathsf{coins}, sid, ssid, \mathsf{coins}(e))$ to $P_j$. Update $Q \leftarrow Q - e$. If $Q = 0$, then terminate.

Figure 8: Special ideal functionality $\mathcal{F}^*_{\text{StCon+}}$ for stateful contracts.

*size of the input states and the output states and the running time of* $\mathsf{Prog}$ *on input* $(T; \mathsf{st})$*. We define the* $\mathcal{F}^*_{\text{StCon}}$-validation complexity *of* $\Pi$*, or in short* transition validation complexity *of* $\Pi$ *as the maximum value of* $C(\mathsf{Prog}, T, \mathsf{st})$ *maximized over all possible choices of a valid trigger* $T$ *and a valid state* $\mathsf{st}$*.* ◇

*Remarks.* The above definition is different from the analogous definition in the $\mathcal{F}^\star_{\text{CR}}$-hybrid model. In the $\mathcal{F}^\star_{\text{CR}}$ variant, there is only one transition per $\mathcal{F}^\star_{\text{CR}}$ (and in particular $\mathcal{F}^\star_{\text{CR}}$ does not keep state), therefore it is easy to bound the total amount of work done by the miners in an information-theoretic sense. However with $\mathcal{F}^*_{\text{StCon}}$, information-theoretically speaking, there are an unbounded number of transitions. This is true for our protocol since the adversary can forge honest signatures in an information-theoretic setting. Our definition therefore focuses only on the complexity of a *single* state transition, and maximizes this quantity over all possible state transitions and all possible valid states.

*Fair scheduling.* We make a natural assumption on the scheduling of triggers to the contract. The motivation is to prevent standard DoS attacks on the contract. Thus, we assume that the scheduling of triggers is done in a fair fashion and that each party's trigger would be attempted one at a time and in a round-robin fashion. In other words, parties have fair access to $\mathcal{F}^*_{\text{StCon}}$.

**Functionality $\mathcal{F}^*_{\text{StCon+}}$.** We also present another variant of $\mathcal{F}^*_{\text{StCon}}$, which we call $\mathcal{F}^*_{\text{StCon+}}$. The main difference is that $\mathcal{F}^*_{\text{StCon+}}$ accepts coin deposits, via an $\mathsf{update}$ command even during the execution phase. We note that dynamic updates to $\mathsf{Prog}$ is supported by our definition (although we do not rely on this feature in our protocols). We note that $\mathcal{F}^*_{\text{StCon+}}$ is also supported by Ethereum.

### 3.3 Multisignature Aggregation

To reduce the on-chain complexity, the $n$ parties can utilize multisignature aggregation. This way, the verification of each transaction – which is done by all the miners in the cryptocurrency network – does not need to depend on $n$.

In the context of a fixed set of parties $P_1, P_2, \ldots, P_n$, the definition of a multisignature scheme $(\mathsf{multiGen}, \mathsf{multiSign}, \mathsf{multiVrfy})$ is as follows (cf. [25, 36]).

- The parties $P_1, P_2, \ldots, P_n$ execute the key generation algorithm $\mathsf{multiGen}$ in $\text{poly}(\lambda, n)$ time, which delivers a signing share $s_i$ to each party $P_i$, as well as an aggregated public key $pk_{\mathsf{master}}$ to all parties. The size of $pk_{\mathsf{master}}$ does not depend on $n$ but rather only on the security parameter $\lambda$.
- To sign a plaintext message $m$, the parties will use their shares $\{s_i\}_{i=1}^n$ to execute the signing algorithm $\mathsf{multiSign}$ in $\text{poly}(\lambda, n, |m|)$ time. The output of $\mathsf{multiSign}$ is an aggregated signature $\sigma$ whose size is $\text{poly}(\lambda, |m|)$.

- The verifier will execute the verification algorithm $\mathsf{multiVrfy}(pk_{\mathsf{master}}, m, \sigma)$ in poly$(\lambda, |m|)$ time, and accept if and only if the output of multiVrfy is 1.

Since the set of parties that interacts with $\mathcal{F}^*_{\mathrm{MSFE}}$ is fixed, they can deploy the Schnorr [45] based multisignature scheme of Micali, Ohta and Reyzin [36]. The parties $P_1, P_2, \ldots, P_n$ will initially execute the setup algorithm multiGen, and thereby produce their individual shares $\{s_i\}_{i=1}^n$ and the aggregated public key $pk_{\mathsf{master}}$. The key $pk_{\mathsf{master}}$ will be sent to $\mathcal{F}^*_{\mathrm{MSFE}}$, and in the implementation it will be placed inside the on-chain contract (during the invocation of the `SmartAmortize` constructor in Figure 18). To sign each off-chain message $m_j$, the parties will execute $\mathsf{multiSign}(m_j, s_1, s_2, \ldots, s_n)$ to create a signature $\sigma_j$. The time complexity of multiSign is in fact similar to that of non-aggregate signatures exchange, though it requires one additional round of interaction. Hence, in the case that a corrupt party $P_c$ deviates from the protocol in iteration $j+1$, any honest party can send the signed message $(m_j, \sigma_j)$ to $\mathcal{F}^*_{\mathrm{MSFE}}$ so that $P_c$ will either send the correct witness to $\mathcal{F}^*_{\mathrm{MSFE}}$ or be penalized. The miners in the decentralized cryptocurrency network will invoke the verification algorithm multiVrfy to check the signature $\sigma_j$ against the public key $pk_{\mathsf{master}}$. This means that the on-chain cost of validating that the $n$ parties signed $m_j$ is dominated by poly$(\lambda)$, since the time complexity of multiVrfy depends on $|m_j|$ only for the purpose of hashing $m_j$ into a digest of $O(\lambda)$ size.

We also design protocols that let parties to join and leave, via the $\mathcal{F}^*_{\mathrm{StCon+}}$functionality that allows the set of parties $A = \{P_1, P_2, \ldots, P_n\}$ to change. In the case that the parties agree to change the set from $A$ to $A'$, the parties in $A'$ will need to run multiGen and send a new aggregated key $pk'_{\mathsf{master}}$ to $\mathcal{F}^*_{\mathrm{MSFE}}$.

Let us note that the scripting language of Ethereum does not yet support Schnorr signature verification, but a more complex key generation algorithm can be used for a multisignature ECDSA scheme, see [21].

There are also multisignature schemes that do not require an intricate key generation phase whenever the set of parties changes, though such schemes are based on bilinear pairings‘[7, 34]. An efficient Schnorr-based multisignature scheme without multiple key generations is given by Bellare and Neven [5], but their scheme relies on a certificate authority.

# 4 Realizing $\mathcal{F}^*_{\mathrm{MSFE}}$ from $\mathcal{F}^*_{\mathrm{StCon}}$

In this section, we describe the protocols for amortized secure computation with penalties in the $\mathcal{F}^*_{\mathrm{StCon}}$-hybrid model. Due to lack of space we only give a brief overview. See Appendix A for full details.

The protocol for implementing $\mathcal{F}^*_{\mathrm{MSFE}}$ has three phases. In the first phase, parties interact with the on-chain stateful contract, i.e., the ideal functionality $\mathcal{F}^*_{\mathrm{StCon}}$. In particular, parties agree on setting contract parameters that fix the number of parties, the allowed state transitions of the contract, the time-out, and the compensation amounts to parties in case of an abort (cf. Figure 10). Then in the second phase (cf. Figure 9), parties perform the actual computation. This is done off-chain via local MPC executions. In addition to performing the computation, these MPC executions also provide hooks to the on-chain contract (to handle aborts). We describe the local executions first because they will introduce new variables which will serve as hooks to the contract via the contract parameters. In the next phase, we describe the process which honest parties use in case an off-chain local execution was aborted. In particular, in this phase, parties will go on to the on-chain contract to either continue the aborted local execution or claim their compensation. This phase occurs immediately following an abort in the local executions phase. Figure 11 describes how parties handle notifications received from $\mathcal{F}^*_{\mathrm{StCon}}$.

**Local executions.** The formal description of the local executions is in Figure 9. In this section, we describe this phase in more detail. Suppose the parties are interested in computing a function $g^{(\mathsf{id})}$. At a high level, parties begin by running a standard secure computation protocol (that does not guarantee fairness) which computes $z = g^{(\mathsf{id})}(y_1, \ldots, y_n)$ where $y_j$ represents the input of $P_j$. In addition this secure computation protocol also additively secret shares $z$ into $z_1, \ldots, z_n$ and computes commitments $h_j$ on each $z_j$ using uniform randomness $\omega_j$. Finally, the secure computation protocol outputs $x_j^{(\mathsf{id})} = (z_j; \omega_j)$ (i.e., the decommitment to $h_j$) and the value $\boldsymbol{h}^{(\mathsf{id})} = h_1 \| \cdots \| h_n$ to each party $P_j$. For the simulation to work, we need to use an honest-binding commitment scheme (cf. Appendix D).

1. (MPC step) Parties run a standard MPC protocol that
   - obtains inputs $y_1, \ldots, y_n$ from the parties; and
   - computes $z = g^{(\mathsf{id})}(y_1, \ldots, y_n)$; and
   - secret shares $z$ into $z_1, \ldots, z_n$; and
   - computes $h_j = \mathsf{com}(x_j^{(\mathsf{id})} = (z_j; \omega_j))$; where each $\omega_j$ is chosen uniformly at random; and
   - outputs $(x_j^{(\mathsf{id})}, \boldsymbol{h}^{(\mathsf{id})} = h_1 \| \cdots \| h_n)$ to each $P_j$.
2. (Signature broadcast) Each $P_j$ computes a signature $\sigma_j$ on message $(\mathsf{id}, \boldsymbol{h}^{(\mathsf{id})})$ under the signing key $sk_j$ and then broadcasts it. Let $S_j$ denote the set of parties whose signature on message $(\mathsf{id}, \boldsymbol{h}^{(\mathsf{id})})$ was received by $P_j$. If $S_j = [n]$, then let $\boldsymbol{\sigma}^{(\mathsf{id})} = \sigma_1 \| \cdots \| \sigma_n$ and update $\mathsf{best}_j \leftarrow (\mathsf{id}, ((j, x_j^{(\mathsf{id})}), \boldsymbol{h}^{(\mathsf{id})}, \boldsymbol{\sigma}^{(\mathsf{id})}))$. Else, parties abort the local execution and go to the on-chain contract for resolution (cf. Figure 11).
3. (Share broadcast) Each $P_j$ broadcasts $x_j^{(\mathsf{id})}$. Let $S_j^{(\mathsf{id})}$ denote the set of parties whose share (authenticated against $\boldsymbol{h}^{(\mathsf{id})}$) was received by $P_j$ and let $X_j^{(\mathsf{id})}$ denote the corresponding set of decommitments received by $P_j$. Each $P_j$ updates $\mathsf{best}_j \leftarrow (\mathsf{id}, (X_j, \boldsymbol{h}, \boldsymbol{\sigma}))$. If $|S_j^{(\mathsf{id})}| = [n]$, then $P_j$ computes the output of the $\mathsf{id}$-th local execution as $\bigoplus_{k=1}^{n} z_k$ where we parse $x_k^{(\mathsf{id})} \in X_j^{(\mathsf{id})}$ as $(z_k; \omega_k)$. Else, parties abort the local execution and go to the on-chain contract for resolution (cf. Figure 11).

Figure 9: $\mathsf{id}$-th off-chain local execution for implementing $\mathcal{F}^*_{\mathrm{MSFE}}$.

Note that there could be aborts here and in every subsequent stage of the local execution. In each step, we assume that parties stop the protocol if they do not receive valid messages (i.e., including signatures) from any other party. Importantly, there is an additional (implicit) time-interval parameter $\delta$ which is used to detect aborts. In more detail, we say that a party aborted the protocol if (1) it is its turn to send a message, and (2) if the party does not send a valid protocol message within time-interval $\delta$ of the previous event. (We assume that all honest parties act immediately, i.e., within time-interval $\delta$.) In the event of an abort (as defined above), parties go up to the on-chain contract for resolution (cf. Figure 11). Once the local secure computation protocol ends, we ask each party $P_j$ to compute a signature $\sigma_j$ on the message $(\mathsf{id}, \boldsymbol{h}^{(\mathsf{id})})$ under its secret signing key $sk_j$ and then broadcast it to all parties. In the next step, each party $P_j$ broadcasts the decommitment $x_j^{(\mathsf{id})}$ to the value $h_j$ contained in $\boldsymbol{h}^{(\mathsf{id})}$ (which in particular includes the secret share $z_j$ of the output $z$). Once this step is completed, then all parties can recover the output of the $\mathsf{id}$-th computation as $\bigoplus_{k \in [n]} z_k$. Please see Appendix A for more details.

**Contract parameters.** See Figure 10 for a formal description. The state parameters are established in the following way. The variables $d_1, \ldots, d_n$ represent the amount $\mathsf{coins}((n-1)q)$ that the parties are expected to put in as the initial deposit to the contract.

*State components and initialization.* The variable $\mathsf{st}$ denotes the current state of the contract. The values $\boldsymbol{pk}$ and $\Delta$ are constant parameters to the contract and are always maintained as part of the state. The parameter $\boldsymbol{pk}$ represents the set of public keys of all the parties. The parameter $\Delta$ represents the length of the time interval within which parties need to act in order to keep the contract from defaulting. In addition, each state variable has five components: (1) $\mathsf{st.mode}$ represents the current mode in which the contract is in, and is one of {"init", "exec", "exit", "payout", "abort", "inactive"}; (2) $\mathsf{st.id}$ represents the $\mathsf{id}$ of the execution that is being continued currently on the on-chain contract; (3) $\mathsf{st}.\textsc{tt}$ represents the current transcript of the execution that is being continued on the chain; (4) $\mathsf{st}.t$ represents the time when the on-chain contract was triggered and either (a) was moved to "exit" or (b) resulted in a change of the variable $\mathsf{st}.\textsc{tt}$; (5) $\mathsf{st}.L$ is a boolean array that represents which parties have already withdrawn their deposits and compensations from $\mathcal{F}^*_{\mathrm{StCon}}$.

We represent the state variable $\mathsf{st}$ as a five tuple ($\mathsf{st.mode}$, $\mathsf{st.id}$, $\mathsf{st}.\textsc{tt}$, $\mathsf{st}.t$, $\mathsf{st}.L$). Also, $\mathsf{st}.\textsc{tt}$ is either $\perp$ (denoting the null transcript) or is a tuple of the form $(X, \boldsymbol{h}, \boldsymbol{\sigma})$. The initial state is $\mathsf{st}$ and its components

**Notation.** The variable $\mathsf{st}$ denotes the current state of the contract. We represent the state variable $\mathsf{st}$ as a five tuple ($\mathsf{st.mode}$, $\mathsf{st.id}$, $\mathsf{st}.\textsc{tt}$, $\mathsf{st}.t$, $\mathsf{st}.L$). We omit $\boldsymbol{pk}$ and $\Delta$ from the state to keep the presentation simple. The initial deposits are $d_1 = \cdots = d_n = (n-1)q$. The initial state is ("init", $-1$, $\perp$, $-1$, $\mathbf{1}$).

**Subroutine $\mathsf{pred}$.** Let $\mathsf{pred}(j, w, t; \mathsf{st}) = 1$ if

- $w$ is parsed as $(\mathsf{id}, \textsc{tt} = (X, \boldsymbol{h}, \boldsymbol{\sigma}))$ with $\boldsymbol{h} = h_1 \| \cdots \| h_n$, $\boldsymbol{\sigma} = \sigma_1 \| \cdots \| \sigma_n$; and
- for each $k \in [n]$ it holds that $\mathsf{SigVerify}((\mathsf{id}, \boldsymbol{h}), \sigma_k; pk_k) = 1$; and
- for each $k$ such that $X[k] \neq \mathsf{NULL}$ it holds that $h_k = \mathsf{com}(X[k])$; and
- either (1) $\mathsf{st.mode}$ = "init"; or (2) $\mathsf{st.mode}$ = "exec"/"exit" and $t \leq \mathsf{st}.t + \Delta$ and either (2.1) $\mathsf{id} > \mathsf{st.id}$ or (2.2) $\mathsf{id} = \mathsf{st.id}$ and $X \not\subseteq \mathsf{st}.\textsc{tt}.X$.

**State transitions.** $\mathsf{Prog}(j, w, t; \mathsf{st})$: Initialize $e \leftarrow 0$.

- If $w = (\mathsf{id}, \textsc{tt})$ and $\mathsf{pred}(j, w, t; \mathsf{st}) = 1$: If $\mathsf{st.id} = \mathsf{id}$, then update $\mathsf{st}.\textsc{tt}.X \leftarrow \mathsf{st}.\textsc{tt}.X \cup \textsc{tt}.X$, else set $\mathsf{st}.\textsc{tt} \leftarrow \textsc{tt}$. Set $\mathsf{st} \leftarrow$ ("exec", $\mathsf{id}$, $\mathsf{st}.\textsc{tt}$, $t$, $\mathsf{st}.L$).
- Else if $w = \mathsf{exit}$:
  - If (1) $\mathsf{st.mode}$ = "init" or (2) $\mathsf{st.mode}$ = "exec" and $|\mathsf{st}.\textsc{tt}.X| = n$: Set $\mathsf{st.mode} \leftarrow$ "exit" and $\mathsf{st}.t \leftarrow t$.
  - If $\mathsf{st.mode}$ = "exec" or "abort", and $t > \mathsf{st}.t + \Delta$ and $\mathsf{st}.L[j] = 1$ and $|\mathsf{st}.\textsc{tt}.X| \neq n$: Then update $\mathsf{st}.L[j] \leftarrow 0$ and $\mathsf{st.mode} \leftarrow$ "abort" and $\mathsf{st}.L[k] \leftarrow 0$ for all $k$ such that $\mathsf{st}.\textsc{tt}.X[k] = \perp$. Further, if $\mathsf{st}.\textsc{tt}.X[j] \neq \perp$, then set $e \leftarrow n(n-1)q/|\mathsf{st}.\textsc{tt}.X|$.
  - If $\mathsf{st.mode}$ = "exit" or "payout", and $t > \mathsf{st}.t + \Delta$ and $\mathsf{st}.L[j] = 1$: Set $e \leftarrow (n-1)q$ and update $\mathsf{st.mode} \leftarrow$ "payout" and $\mathsf{st}.L[j] \leftarrow 0$.

If $\mathsf{st}.L[k] = 0$, for all $k \in [n]$ then we update $\mathsf{st.mode} \leftarrow$ "inactive".

Figure 10: $\mathcal{F}^*_{\text{StCon}}$ parameters for $\mathcal{F}^*_{\text{MSFE}}$.

are initialized in the following way: (1) $\mathsf{st.mode}$ = "init"; (2) $\mathsf{st.id} = -1$; (3) $\mathsf{st}.\textsc{tt} = \mathsf{NULL}$; (4) $\mathsf{st}.t = -1$; and (5) $\mathsf{st}.L = (1, \ldots, 1)$. Recall that $\mathsf{st}$ also contains the list of all public keys of the participants $\boldsymbol{pk}$, and the global time-out parameter $\Delta$.

*Triggering state transitions.* During course of the execution, the state of the contract would either (1) remain in the initial state with $\mathsf{st.mode}$ = "init"; or (2) be in exit mode, i.e., with $\mathsf{st.mode}$ = "exit", where contract participants are trying to get their initial deposit out of the contract (and terminate the contract); or (3) be trying to continue an incomplete off-chain local execution by keeping track of the current state of the local execution computation, i.e., with $\mathsf{st.mode}$ = "exec"; or (4) be in payout mode with $\mathsf{st.mode}$ = "payout" where parties have successfully completed all executions so far and are waiting to get their initial deposits out of $\mathcal{F}^*_{\text{StCon}}$; or (5) be in "abort" mode where an execution was aborted and honest parties are waiting to get their initial deposits and compensation out of $\mathcal{F}^*_{\text{StCon}}$; or (6) be in "inactive" mode where $\mathcal{F}^*_{\text{StCon}}$ no longer accepts any further state transitions and in particular, has given out all the money that was deposited to it.

Transitions to different states will be triggered by a witness $(j, w, t)$. Here $j$ represents the party triggering the contract, i.e., party $P_j$. The value $t$ represents the time at which the contract is triggered. Note that when $\mathsf{st.mode}$ = "payout"/"abort"/"inactive", the triggering witness $w$ is simply the token value $\mathsf{exit}$. As we will see later, the transitions from these states depends only on $\mathsf{st}.L$ and the triggering time $t$ and $\mathsf{st}.t$. The more interesting case is when $\mathsf{st.mode}$ = "init"/"exit"/"exec". In this case, the triggering witness $w$ provides the most recent state of the current local execution. We will use a separate subroutine $\mathsf{pred}$ to determine the validity of a trigger $(j, w, t)$ when the witness $w$ represents a transcript of an execution.

*Subroutine* $\mathsf{pred}$. The predicate $\mathsf{pred}$ essentially decides if a trigger to the contract is a valid continuation of the computation on the chain. Now, $\mathsf{pred}$ takes a trigger $(j, w, t)$ and examines it in conjunction with the current state of the contract. First, $\mathsf{pred}$ parses the witness $w$ as $(\mathsf{id}, \textsc{tt} = (X, \boldsymbol{h}, \boldsymbol{\sigma}))$ where $\mathsf{id}$ represents the (off-chain) execution that is being attempted to be continued on the chain by party $P_j$. The value $\text{TT} = (X, \boldsymbol{h}, \boldsymbol{\sigma})$ essentially provide (along with a proof) the most recent state of a computation (typically the last incomplete off-chain computation). In particular and in the context of non-reactive functionalities, the value $X$ maintains the set of parties who have completed their step of the computation on the chain along with their broadcasted decommitments to the secret share of the final output. The values $\boldsymbol{h}$ and $\boldsymbol{\sigma}$ essentially authenticate to the contract that values in $X$ are legitimate values corresponding to the id-th off-chain computation. In more detail, $\boldsymbol{h} = h_1\| \cdots \|h_n$ is the set of commitments (that is public to all parties). The value $\boldsymbol{h}$ should be consistent with the broadcast values $X$ in the sense that $\mathsf{com}(X[k]) = h_k$ for all $k$ such that $X[k] \neq \perp$. Likewise, the commitment values $\boldsymbol{h}$ need to be accompanied with $\boldsymbol{\sigma} = \sigma_1\| \cdots \|\sigma_n$ where $\sigma_i$ is the signature of party $P_i$ attesting to the correctness of $\boldsymbol{h}$. Note that the signatures also tie the value of $\boldsymbol{h}$ to id.

Clearly, pred should output 1 if the witness is valid and if $(j, w, t)$ happens to be the very first trigger to the contract. On the other hand, if $(j, w, t)$ is not the first trigger to the contract, then we have to ensure that the trigger $(j, w, t)$ provides a valid update to the contract state. Now the contract state could be in exit mode, i.e., st.mode = "exit", and in this case the trigger $(j, w, t)$ with a valid witness $w$ could be by an honest party to continue an incomplete off-chain execution. This is to ensure that a malicious party cannot subvert the continuation of the off-chain execution on the on-chain contract by trying to exit prematurely (i.e., when st.mode = "init". Likewise a malicious party might also submit an old completed execution (even while the current off-chain execution has not yet completed). Thus, we must have pred output 1 when the new id present in $w$ is greater than st.id.

Now the contract could be in exec mode, i.e., st.mode = "exec", in which case the contract is typically waiting for the on-chain execution to be completed. There are essentially two cases: (1) the current state does not correspond to or continue the most recent off-chain execution; in this case, the id in the new trigger must satisfy id > st.id (i.e., the contract is essentially reset to the "correct last computation"), and (2) the new trigger continues the current state of the contract and for this id = st.id must hold and also we need $X$ to contain at least one value which is not in $\mathsf{st}.\text{TT}.X$, i.e., there is some $k \in [n]$ such that $X[k] \neq \perp = \mathsf{st}.\text{TT}.X$. In either case, the new trigger must appear within the time interval $\Delta$ of the previous trigger (i.e., before time $\mathsf{st}.t + \Delta$).

*State transitions.* The state transition function Prog takes as input the trigger $(j, w, t)$ and the current state st. First, we check if the witness provided corresponds to an execution transcript. In this case, we invoke the predicate pred and if pred outputs 1, then we update $\mathsf{st}.\text{TT}$ depending on whether (1) st.id = id, in which case we update $\mathsf{st}.\text{TT}.X$ to include decommitments specified in $X$; or (2) st.id < id, in which case we update $\mathsf{st}.\text{TT} \leftarrow \text{TT}$. If $w$ does not correspond to an execution transcript, then we assume that it is a token value exit. There are effectively three cases to handle:

- If st.mode equals "init" or equals "exec" with a fully completed transcript, then we change st.mode to "exit" and store the triggering time $t$ in $\mathsf{st}.t$. This transition is provided to ensure that honest parties' deposits are not "locked in" and to enable them to withdraw their deposits from $\mathcal{F}^*_{\text{StCon}}$.
- If st.mode = "exec"/"abort", then we check if $t > \mathsf{st}.t + \Delta$ and if $|\mathsf{st}.\text{TT}.X| \neq n$ to confirm that the execution has indeed been aborted. In this case, if $\mathsf{st}.L[j] = 1$, then we will allow $P_j$ to take money out of $\mathcal{F}^*_{\text{StCon}}$. We further need to check whether $P_j$ was a malicious party that did not contribute to completing the execution. We do this by checking if $\mathsf{st}.\text{TT}.X[j] \neq \perp$. If all checks pass, we let $P_j$ to withdraw its initial deposit plus compensation, i.e., a total of $n(n-1)q/|\mathsf{st}.\text{TT}.X|$ from $\mathcal{F}^*_{\text{StCon}}$.
- If st.mode = "exit"/"payout", then we check if $t > \mathsf{st}.t + \Delta$. This is to prevent situations where a malicious party tries to subvert continuing the off-chain aborted execution on the chain. (That is, honest parties get an additional time $\Delta$ to get $\mathcal{F}^*_{\text{StCon}}$ out of the exit mode.) If $t > \mathsf{st}.t + \Delta$ indeed holds, then we allow party $P_j$ to take its initial deposit $(n-1)q$ out of the contract if it was not already paid before (i.e., $\mathsf{st}.L[j] = 1$).

**Main protocol.** The formal description can be found in Figure 11. In this section we will describe the main protocol that makes use of the local execution subprotocol of Figure 9 and also how parties interact with $\mathcal{F}^*_{\text{StCon}}$ according to the parameters described in Figure 10. Parties basically start by initializing the Parties initialize the parameters as in Figure 10. Then for $\mathsf{id} = 1, 2, \cdots$, parties run the local execution prescribed in Figure 9. Recall that each party $P_j$ maintains a variable $\mathsf{best}_j$ during the local executions which is initialized as $\perp$.

1. If a party $P_j$ wants to exit the contract and reclaim its initial deposit, then it sends $w = \mathsf{exit}$ to $\mathcal{F}^*_{\mathrm{StCon}}$.
2. If there is an abort during a local off-chain execution, then each party $P_j$ triggers $\mathcal{F}^*_{\mathrm{StCon}}$ with the value $\mathsf{best}_j$.
3. Each party $P_j$ waits and responds to state changes in $\mathcal{F}^*_{\mathrm{StCon}}$ depending on the current state $\mathsf{st}$:
   (a) If $\mathsf{st.mode}$ = "payout"/"abort" and $\mathsf{st}.L[j] = 1$, send $w = \mathsf{exit}$ to $\mathcal{F}^*_{\mathrm{StCon}}$.
   (b) If (1) $\mathsf{st.id} < \mathsf{best}_j.\mathsf{id}$, or (2) $\mathsf{st.id} = \mathsf{best}_j.\mathsf{id}$ and $\mathsf{st}.\textsc{tt}.X[j] = \perp$, then submit $\mathsf{best}_j$ to $\mathcal{F}^*_{\mathrm{StCon}}$.
   (c) If $\mathsf{st.id} > \mathsf{best}_j.\mathsf{id}$, then submit $\mathsf{best}_j \leftarrow (\mathsf{st.id},\ ((j, x_j^{(\mathsf{st.id})}),\ \mathsf{st}.\textsc{tt}.\boldsymbol{h},\ \mathsf{st}.\textsc{tt}.\boldsymbol{\sigma}))$ to $\mathcal{F}^*_{\mathrm{StCon}}$.

Finally, parties also keep track of whether $\mathsf{st.mode}$ = "exit" or "exec" and the current time $t$ is such that $t > \mathsf{st}.t + \Delta$. In this case, parties send $w = \mathsf{exit}$ in order to claim their payout or compensation.

Figure 11: Main protocol for implementing $\mathcal{F}^*_{\mathrm{MSFE}}$.

$\mathcal{F}^*_{\mathrm{StCon}}$ parameters as in Figure 10. Following this, they begin off-chain local executions. Recall that each party $P_j$ maintains a variable $\mathsf{best}_j$ which denotes the transcript corresponding to the latest *active* execution (both on-chain and off-chain) *according to the local view of party* $P_j$ (see Figure 9). This value will provide the necessary hook to the on-chain contract to handle off-chain aborts. In particular, the value $\mathsf{best}_j$ will be submitted by party $P_j$ in order to recover from aborted off-chain executions.

In our main protocol, we essentially deal with three different scenarios: (1) when parties want to exit the contract and get back their deposits and compensation, (2) when parties want to continue an aborted off-chain execution on the chain, and (3) when parties are notified of state changes in $\mathcal{F}^*_{\mathrm{StCon}}$.

*Exiting the contract.* First, we deal with the situation when parties would like to terminate the protocol and retrieve their initial deposits from the contract. To do so, we simply let parties submit a token value $w = \mathsf{exit}$ to trigger and put the contract into exit mode. Note that malicious parties might revert $\mathcal{F}^*_{\mathrm{StCon}}$ to go into exec mode. In this case, $\mathcal{F}^*_{\mathrm{StCon}}$ will notify honest parties of the change. Honest parties will be able to recover from this and put the contract back into exit mode. This will be described when we discuss how parties react to notifications from $\mathcal{F}^*_{\mathrm{StCon}}$.

*Continuing an aborted off-chain execution.* This is where the value $\mathsf{best}_j$ comes in handy as it stores the most recent state of the off-chain executions. We instruct parties to trigger $\mathcal{F}^*_{\mathrm{StCon}}$ with the value $\mathsf{best}_j$ which includes $P_j$'s decommitment $x_j^{(\mathsf{id})}$ which in turn ensures (by the logic in Figure 10) that $P_j$ will not be penalized.

*Responding to notifications from* $\mathcal{F}^*_{\mathrm{StCon}}$. First, if $\mathsf{st.mode}$ = "payout"/"abort", then parties send a token value $w = \mathsf{exit}$ to get their deposits out of $\mathcal{F}^*_{\mathrm{StCon}}$. In addition, if $\mathsf{st.mode}$ = "abort", then parties would also get compensation from $\mathcal{F}^*_{\mathrm{StCon}}$. Second, if the on-chain execution does not correspond to the most recent execution, then we ask parties to submit $\mathsf{best}_j$ to the contract. (This will also handle the case when honest parties try to exit the contract but a malicious party feeds an older execution to $\mathcal{F}^*_{\mathrm{StCon}}$.) Checking if the on-chain execution corresponds to the most recent execution is handled by checking first if $\mathsf{st.id} < \mathsf{best}_j.\mathsf{id}$ and then if $\mathsf{st.id} = \mathsf{best}_j.\mathsf{id}$ but $\mathsf{st}.\textsc{tt}.X[j] = \perp$ (i.e., party $P_j$'s decommitment is not yet part of the on-chain execution transcript). Finally, we also need to handle the corner case when $\mathsf{st.id} > \mathsf{best}_j.\mathsf{id}$. This scenario is actually possible when party $P_j$ is honest but only when $\mathsf{st.id} = \mathsf{best}_j.\mathsf{id} + 1$. We now describe the sequence of events which lead to this case. Suppose in Step 2 of Figure 9 some malicious party did not broadcast its signature on $\boldsymbol{h}^{(\mathsf{id})}$, then party $P_j$ will not update $\mathsf{best}_j$. Thus $\mathsf{best}_j.\mathsf{id} = \mathsf{id} - 1$ where $\mathsf{id}$ is the execution id of the current execution. Note that each honest $P_j$ would have submitted its signature on $\boldsymbol{h}^{(\mathsf{id})}$ in Step 2. Therefore, malicious parties would possess a valid $\boldsymbol{h}^{(\mathsf{id})}$ and $\boldsymbol{\sigma}^{(\mathsf{id})}$ for execution $\mathsf{id}$. That is, a malicious party $P_k$ is able to trigger $\mathcal{F}^*_{\mathrm{StCon}}$ with a witness $w = (\mathsf{id}, \textsc{tt} = ((k, x_k^{(\mathsf{id})}), \boldsymbol{h}^{(\mathsf{id})}, \boldsymbol{\sigma}^{(\mathsf{id})})$ which will result in $\mathsf{pred}(k, w, t) = 1$. Thus, we need a mechanism to allow honest parties to continue the $\mathsf{id}$-th execution (i.e., continue $\text{TT}$) and ensure that they don't get penalized. This is indeed possible since honest parties already obtain $x_j^{(\mathsf{st.id})}$ from Step 1 of Figure 9. That is, we let honest parties submit $w = (\mathsf{st.id}, \text{TT}_j = ((j, x_j^{(\mathsf{st.id})}),$ $\boldsymbol{h}^{(\mathsf{id})} = \mathsf{st}.\text{TT}.\boldsymbol{h}, \boldsymbol{\sigma}^{(\mathsf{id})} = \mathsf{st}.\text{TT}.\boldsymbol{\sigma}))$ to $\mathcal{F}^*_{\text{StCon}}$.

Due to lack of space, we prove the following theorem in Appendix A.

**Theorem 1.** *Let $\lambda$ be a computational security parameter. Assume the existence of one-way functions. Then there exists a protocol that* SCC*-realizes (cf. Definition 1) $\mathcal{F}^*_{\text{MSFE}}$ in the ($\mathcal{F}^*_{\text{StCon}}$,$\mathcal{F}_{\text{OT}}$)-hybrid model whose script complexity (cf. Definition 3) is independent of the number of secure function evaluations and depends only on the length of outputs of the functions evaluated in $\mathcal{F}^*_{\text{MSFE}}$ and is otherwise independent of them. Furthermore, in the optimistic case when all parties are honest, there are a total of $O(n)$ state transitions each having complexity $O(n\lambda)$.*

# 5 Realizing $\mathcal{F}^*_{\text{MSCD}}$ from $\mathcal{F}^*_{\text{StCon+}}$

We now discuss how to implement $\mathcal{F}^*_{\text{MSCD}}$. Since we are now dealing with reactive functionalities, we make use of an MPC protocol that handles reactive functionalities (say GMW). Since we are dealing with cash distribution, we will let the MPC protocol take, in addition to regular inputs, values that represent the current balance of each party. Note that this balance could change at the end of each stage of the reactive function evaluation. We stress that the reactive functions take only strings as inputs/outputs (and do not handle coins), and the amount of coins and current balance are merely specified as strings. We assume that the updated balance vectors can be obtained directly from the transcript of the protocol implementing the reactive function.

The overall protocol structure closely mimics our protocol for implementing $\mathcal{F}^*_{\text{MSFE}}$. We give a high level overview of the protocol and detail the main differences from the implementation of $\mathcal{F}^*_{\text{MSFE}}$.

**Local executions.** See Figure 12 for a formal description. The local executions begin by allowing parties to add coins to their deposit (which will be redistributed depending on the output of the stage functions of the reactive function) but only between different reactive function evaluations. In order to synchronize (in order to make the simulation go through) and ensure that coins are not added while a reactive function is being evaluated, we ask parties to obtain signatures from all parties. (Then we design $\mathcal{F}^*_{\text{StCon+}}$ such that it accepts coins only when the submitting party has signatures from all participants.) Then, in the next step, we ask parties to agree on the transcript validation function for the reactive protocol $\pi$ implementing $g^{(\mathsf{id})}$ that they are going to execute. That is, in the $\mathsf{id}$-th local execution, parties agree on $\mathsf{tv}^{(\mathsf{id})}$ and each party signs this value under its signing key and broadcasts it to all parties. This is different from the previous case where while implementing $\mathcal{F}^*_{\text{MSFE}}$, we needed parties to sign on $\boldsymbol{h}^{(\mathsf{id})}$ in the $\mathsf{id}$-th execution but *after* the secure computation protocol was run. Note that we also need parties to agree on the updated balance vector $\boldsymbol{b}$ before beginning the $\mathsf{id}$-th local execution.

Once the signatures are done, parties begin evaluating each stage of the reactive computation in sequence. Parties then run a secure computation protocol for each stage sequentially until the entire reactive protocol completes. Note that a typical protocol for a reactive computation maintains state across different stages by secret sharing this value among the participants. That is, when parties are ready to begin the secure computation protocol for the next stage, they supply along with the inputs for the new stage also an authenticated secret share of the previous state. Note that the balance vectors are supplied as input to the reactive MPC (in order to calculate the updated balance). We abstract these details and assume that the authenticated secret shares of intermediate states corresponding to party $P_j$ is part of its input $y_j$ for this stage. The next message function $\mathsf{nmf}$ takes the current available transcript as input, along with the parties' input for this stage, the randomness, the current balance vector, and also the secret signing key of this party. Note that $\mathsf{nmf}$ and $\mathsf{tv}$ are such that for every partial transcript $\text{TT}'$ such that $|\text{TT}'| = (j-1) \bmod n$ and $\mathsf{tv}(\text{TT}') = 1$, we have that $(m, \sigma) \leftarrow \mathsf{nmf}(\text{TT}'; (y_j, \omega_j, \boldsymbol{b}, sk_j))$ satisfies $\mathsf{tv}(\text{TT}\|(m, \sigma)) = 1$. Observe that $\mathsf{nmf}$ produces signed messages that continue that transcript, and $\mathsf{tv}$ checks whether messages are signed appropriately, and if the newly extended transcript is valid according to the underlying reactive MPC. Such

1. (Adding new coins) If some party $P_j$ wants to add $\mathsf{coins}(b_j)$ to $\mathcal{F}^*_{\text{StCon+}}$, then it sends $\boldsymbol{b}' \leftarrow \boldsymbol{b} + (\cdots, 0, b_j, 0, \cdots)$ to all parties. Each party $P_k$ then generates a signature $\psi'_k \leftarrow \mathsf{Sign}(\mathsf{id}, \boldsymbol{b}')$ and broadcasts it. If $P_j$ receives signatures from all parties, then $P_j$ sends $u = (\mathsf{update}, \boldsymbol{b}', \boldsymbol{\psi}, \mathsf{coins}(b_j))$ to $\mathcal{F}^*_{\text{StCon+}}$. Other parties wait to receive notification from $\mathcal{F}^*_{\text{StCon+}}$ of the updated balance. If either (1) $P_j$ did not obtain signatures from other parties; or (2) the remaining parties did not receive notification from $\mathcal{F}^*_{\text{StCon+}}$, then all honest parties go to the on-chain contract with the intention of exiting the contract. See Figure 14. On the other hand, if the above step was successfully completed, then parties update $\boldsymbol{b} \leftarrow \boldsymbol{b}'$ and execute the next step.
2. (Parameter agreement) Initialize transcript $\text{TT} = \perp$. Parties agree on the reactive function $g^{(\mathsf{id})} = \{g_k^{(\mathsf{id})}\}_k$ to be executed next. Parties also agree on a specific MPC protocol $\pi$ using which they will securely compute $g^{(\mathsf{id})}$. Let the transcript verification predicate for this reactive MPC protocol be dentoed by $\mathsf{tv}^{(\mathsf{id})}$. We use $|\mathsf{tv}^{(\mathsf{id})}|$ to denote the number of messages in a valid transcript that corresponds to a *completed* execution of $g^{(\mathsf{id})}$. Once they agree on $\mathsf{tv}^{(\mathsf{id})}$, each $P_j$ computes $\sigma'_j \leftarrow \mathsf{Sign}((\mathsf{id}, \mathsf{tv}^{(\mathsf{id})}, \boldsymbol{b}); sk_j)$ and broadcasts this value to all parties. Each party sets $\boldsymbol{\sigma}^{(\mathsf{id})} = (\sigma'_1, \ldots, \sigma'_n)$. If not all signatures were obtained, then parties stop the local execution and go to the on-chain contract for resolution. Else each party $P_j$ updates $\mathsf{best}_j \leftarrow (\mathsf{transcript},\ \mathsf{id},\ \perp,\ \mathsf{tv}^{(\mathsf{id})},\ \boldsymbol{b},\ \boldsymbol{\sigma}^{(\mathsf{id})})$
3. (Reactive MPC execution) Parse $g^{(\mathsf{id})} = \{g_k^{(\mathsf{id})}\}_{k \in [\rho]}$. For $k = 1, \ldots, \rho$:
   - Let $r_k^{(\mathsf{id})}$ denote the number of rounds in a standard (unfair) MPC protocol that implements $g_k^{(\mathsf{id})}$. We denote party $P_j$'s inputs to $g_k^{(\mathsf{id})}$ by $y_j$ and the corresponding randomness by $\omega_j$. For $r = 1, \ldots, r_k^{(\mathsf{id})}$ sequentially:
     - Let $j = r \bmod n$. Party $P_j$ computes its next message $(m_r, \sigma_r) \leftarrow \mathsf{nmf}(\text{TT}; (y_j, \omega_j, \boldsymbol{b}, sk_j))$, and broadcasts $(m_r, \sigma_r)$ to all parties.
     - If no message was received, then parties abort the local execution and go to the on-chain contract for resolution. Else, each $P_j$ updates the transcript $\text{TT} \leftarrow \text{TT} \| (m_r, \sigma_r)$ and sets $\mathsf{best}_j \leftarrow (\mathsf{transcript},\ \mathsf{id},\ \text{TT},\ \mathsf{tv}^{(\mathsf{id})},\ \boldsymbol{b}',\ \boldsymbol{\sigma})$.
   - Parties compute the output of $g_k^{(\mathsf{id})}$ using the completed transcript TT. Note that this output specifies the new balance vector $\boldsymbol{b}'$. Parties update $\boldsymbol{b} \leftarrow \boldsymbol{b}'$.

Figure 12: The $\mathsf{id}$-th off-chain local executions for implementing $\mathcal{F}^*_{\text{MSCD}}$.

**Parameters.** The variable $\mathsf{st}$ denotes the current state of the contract. The values $\boldsymbol{pk}$ and $\Delta$ are constant parameters to the contract and are always maintained as part of the state. We represent the state variable $\mathsf{st}$ as a seven tuple ($\mathsf{st.mode}$, $\mathsf{st.id}$, $\mathsf{st.}\textsc{tt}$, $\mathsf{st.}t$, $\mathsf{st.}L$, $\mathsf{st.tv}$, $\mathsf{st.}\boldsymbol{b}$, $\mathsf{st.}B$). We omit $\boldsymbol{pk}$ and $\Delta$ and the total deposits so far from the state to keep the presentation simple. The initial deposits are $d_1 = \cdots = d_n = (n-1)q$. The initial state is (“init”, $-1$, $\perp$, $-1$, $\mathbf{1}$, $\perp$, $\mathbf{0}$, $\mathbf{0}$). We also use a function $\mathsf{cash}$ (specified as part of $\mathsf{st.tv}$) that takes a valid transcript $\textsc{tt}$ and $j$ as input, and outputs the amount of coins that need to be given to $P_j$.

**Subroutine $\mathsf{pred}$.** Let $\mathsf{pred}(j, w, t; \mathsf{st}) = 1$ if

- $w$ is parsed as $(\mathsf{message}, \mathsf{id}, (m, \sigma))$ with $\mathsf{id} = \mathsf{st.id}$, $\mathsf{st.tv}(\mathsf{st.}\textsc{tt} \| (m, \sigma)) = 1$; or
- $w$ is parsed as $(\mathsf{transcript}, \mathsf{id}, \textsc{tt}, \mathsf{tv}, \boldsymbol{b}, \boldsymbol{\sigma})$ with (1) $\mathsf{tv}(\textsc{tt}) = 1$; and (2) $\boldsymbol{\sigma} = (\sigma_1, \ldots, \sigma_n)$ and for each $k \in [n]$ it holds that $\mathsf{SigVerify}((\mathsf{id}, \mathsf{tv}, \boldsymbol{b}), \sigma_k; pk_k) = 1$ and either (1) $\mathsf{st.mode} =$ “init”; or (2) $\mathsf{st.mode} =$ “exec”/“exit” and $t \leq \mathsf{st.}t + \Delta$ and either (2.1) $\mathsf{id} > \mathsf{st.id}$ or (2.2) $\mathsf{id} = \mathsf{st.id}$ and $|\textsc{tt}| > |\mathsf{st.}\textsc{tt}|$.

**Adding money.** $\mathsf{Update}(j, u, t; \mathsf{Prog}, \mathsf{st})$ is defined as follows: If $u = (\boldsymbol{b}', \boldsymbol{\psi}, \mathsf{coins}(b_j))$, then parse $\boldsymbol{\psi}$ as $(\psi', \ldots, \psi'_n)$ and verify whether $b_j \neq 0$ and $\mathsf{SigVerify}((j, \boldsymbol{b}'), \psi_k; pk_k) = 1$ for all $k \in [n]$. Then check if $\sum_{k\in[n]} \boldsymbol{b}'_k = b_j + \sum_{k\in[n]} \mathsf{st.}B[k]$. If all checks pass and if $\mathsf{st.mode} \notin \{$“exit”, “abort”, “payout”$\}$, then update $\mathsf{st.}B[j] \leftarrow \mathsf{st.}B[j] + b_j$. Else output $\perp$.

**State transitions.** $\mathsf{Prog}(j, w, t; \mathsf{st})$ is defined as follows: Initialize $e \leftarrow 0$.

- If $w = (\mathsf{transcript}, \mathsf{id}, \textsc{tt}, \mathsf{tv}, \boldsymbol{b}, \boldsymbol{\sigma})$ and $\mathsf{pred}(j, w, t; \mathsf{st}) = 1$: Set $\mathsf{st} \leftarrow$ (“exec”, $\mathsf{id}$, $\textsc{tt}$, $t$, $\mathsf{st.}L$, $\mathsf{tv}$, $\boldsymbol{b}$, $\mathsf{st.}B$).
- Else if $w = (\mathsf{message}, \mathsf{id}, (m, \sigma))$ and $\mathsf{pred}(j, w, t; \mathsf{st}) = 1$: Update $\mathsf{st.}\textsc{tt} \leftarrow \mathsf{st.}\textsc{tt} \| (m, \sigma)$.
- Else if $w = \mathsf{exit}$:
  - If (1) $\mathsf{st.mode} =$ “init”, or (2) $\mathsf{st.mode} =$ “exec” and $|\mathsf{st.}\textsc{tt}| = |\mathsf{st.tv}|$: Set $\mathsf{st.mode} =$ “exit”, $\mathsf{st.}t \leftarrow t$.
  - If $\mathsf{st.mode} =$ “exec”/“abort”, $t > \mathsf{st.}t + \Delta$, $\mathsf{st.}L[j] = 1$, $|\mathsf{st.}\textsc{tt}| \neq |\mathsf{st.tv}|$ and $j \neq j_a = 1 + |\mathsf{st.}\textsc{tt}| \bmod n$: Set $e \leftarrow nq + \mathsf{st.}\boldsymbol{b}_j$, $\mathsf{st.}L[j] \leftarrow 0$, $\mathsf{st.}L[j_a] \leftarrow 0$ and $\mathsf{st.mode} \leftarrow$ “abort”.
  - If $\mathsf{st.mode} =$ “exit”/“payout”, and $t > \mathsf{st.}t + \Delta$ and $\mathsf{st.}L[j] = 1$: Set $e \leftarrow (n-1)q + \mathsf{cash}(j, \mathsf{st.}\textsc{tt})$, $\mathsf{st.mode} \leftarrow$ “payout”, and $\mathsf{st.}L[j] \leftarrow 0$.

If at the end of a transition, it holds that $\mathsf{st.}L[k] = 0$ for all $k \in [n]$, then we update $\mathsf{st.mode} \leftarrow$ “inactive”.

Figure 13: $\mathcal{F}^*_{\text{StCon+}}$ parameters for $\mathcal{F}^*_{\text{MSCD}}$.

modifications to the underlying reactive MPC protocol (namely, adding signatures in $\mathsf{nmf}$ and verifying them in $\mathsf{tv}$, and getting updated balance vectors) were also present in previous protocols that dealt with the reactive case [31, 30]. Like in the implementation of $\mathcal{F}^*_{\text{MSFE}}$, here too we ask each party $P_j$ to maintain a value $\mathsf{best}_j$ which essentially maintains the transcript corresponding to the current execution. Note that $\mathsf{best}_j$ contains both $\mathsf{tv}^{(\mathsf{id})}$ as well signatures on it from all parties denoted by $\boldsymbol{\sigma}^{(\mathsf{id})}$.

**$\mathcal{F}^*_{\text{StCon+}}$ parameters.** See Figure 13 for a formal description. The overall structure is similar to the $\mathcal{F}^*_{\text{StCon}}$ parameters for $\mathcal{F}^*_{\text{MSFE}}$, and in particular we interpret the state $\mathsf{st}$ has having multiple components which keep track of the current mode of the state, the current transcript, the current execution id, time of last exec mode transition, and which parties have already withdrawn money from $\mathcal{F}^*_{\text{StCon+}}$. The main addition is now we also explicitly keep track of the transcript validation function of the current execution as part of the state. We denote this variable by $\mathsf{st.tv}$. We also keep track of how much each party deposited in the variable $\mathsf{st.}B$ and the latest redistribution of cash (i.e., before $\mathsf{st.id}$-th execution began) in the variable $\mathsf{st.}\boldsymbol{b}$. As with $\mathcal{F}^*_{\text{StCon}}$ parameters for $\mathcal{F}^*_{\text{MSFE}}$, here too we make use of a subroutine $\mathsf{pred}$ that effectively determines if the trigger witness $w$ extends the state of the current/latest off-chain execution. For the sake of presentation, we allow parties to submit a trigger witness $w$ which extends $\mathsf{st.}\textsc{tt}$. Alternatively, and in cases where $\mathsf{st.id}$ does not correspond to the latest execution, we also let parties submit a trigger witness $w$ which provides the entire transcript of an off-chain execution. In this case, $\mathsf{pred}$ outputs 1 if the trigger was submitted at time $t \leq \mathsf{st.}t + \Delta$ and if $\mathsf{id} > \mathsf{st.id}$ or $\mathsf{id} = \mathsf{st.id}$ but $\textsc{tt}$ is a longer transcript than the (partial) transcript $\mathsf{st.}\textsc{tt}$.

The state transition function Prog will make use of pred described above. If pred outputs 1 then, the contract moves into exec mode and records the trigger time and updates the transcript with the transcript contained in the trigger. If the trigger is a token value exit, then if the current mode is either "init" or "exec" and $\mathsf{st}.\mathrm{TT}$ is a completed transcript (we check this by checking if $|\mathsf{st}.\mathrm{TT}| = |\mathsf{st.tv}|$), then we put the contract into exit mode and record the time. The rest of the contract specification is quite similar to the $\mathcal{F}^*_{\text{MSFE}}$ case. In particular, when the trigger is a token value $w = \mathsf{exit}$ and if $t > \mathsf{st}.t + \Delta$ and the triggering party has not yet withdrawn money from $\mathcal{F}^*_{\text{StCon+}}$, we move the contract into "payout" or "abort" (depending on the current mode) and refund the deposit (plus current balance plus compensation if applicable) to the triggering party as long as it had not contributed to an off-chain/on-chain aborted execution. To detect whether the triggering party $P_j$ aborted an execution, we simply check if $j = 1 + |\mathsf{st}.\mathrm{TT}| \bmod n$ and $|\mathsf{st}.\mathrm{TT}| \neq |\mathsf{st.tv}|$ holds. In this case, we penalize the party by not giving its deposit back but instead distributing it among the remaining parties. This is slightly different from the $\mathcal{F}^*_{\text{MSFE}}$ case, where we could potentially penalize multiple corrupt parties depending on whether they contributed their output secret share to the execution. Here, on the other hand, we only penalize one party $j_a = 1 + |\mathsf{st}.\mathrm{TT}| \bmod n$. Note that we also redistribute the deposits made by the parties depending on the output of the latest complete execution. If the execution was completed on-chain, then we use the function $\mathsf{cash}_j$ (which we assume is a part of tv for simplicity) applied on $\mathsf{st}.\mathrm{TT}$ to determine how much money party $P_j$ is supposed to obtain. On the other hand, if the on-chain execution was also aborted, then (in addition to paying compensation to the honest parties) we distribute the initial deposits depending on the latest balance prior to this execution (which is stored in variable $\mathsf{st}.\boldsymbol{b}_j$).

Finally, the Update function provides an interface for parties to add coins between different reactive MPCs. Upon receiving a witness $u = (\boldsymbol{b}', \boldsymbol{\psi}, \mathsf{coins}(b_j))$, it checks if the provided coins $b_j$ plus the coins already deposited with $\mathcal{F}^*_{\text{StCon+}}$ (i.e., sum of elements in $\mathsf{st}.B$) matches the amount specified in the balance vector (i.e., sum of elements in $\boldsymbol{b}'$). Note that the signatures also include the party index $j$ (to avoid situations where a different party abuses the broadcasted signatures). Also, note that the deposits $\mathsf{st}.B$ only increase which ensures that there are no replay attacks. This is because once $\mathcal{F}^*_{\text{StCon+}}$ starts giving back coins (i.e., $\mathsf{st.mode} \in \{$"payout", "abort"$\}$), then it does not accept any more coins.

**Main protocol: handling aborts and notifications from $\mathcal{F}^*_{\text{StCon+}}$.** The formal description can be found in Figure 14. As with implementing $\mathcal{F}^*_{\text{MSFE}}$, here too parties begin by initializing $\mathcal{F}^*_{\text{StCon+}}$ with the parameters as in Figure 13, then continue executing off-chain as in Figure 12. Each party $P_j$ maintains a local variable $\mathsf{best}_j$ which represents the most recent transcript of the current off-chain execution. This value will be helpful while recovering from an aborted off-chain execution (cf. Step 2 of Figure 14). While in the $\mathcal{F}^*_{\text{MSFE}}$ case, party $P_j$ triggered $\mathcal{F}^*_{\text{StCon+}}$ with $\mathsf{best}_j$ when its decommitment did not appear in the transcript in $\mathcal{F}^*_{\text{StCon+}}$, here in the $\mathcal{F}^*_{\text{MSCD}}$ case, party $P_j$ triggers $\mathcal{F}^*_{\text{StCon+}}$ with $\mathsf{best}_j$ when $\mathsf{best}_j$ contains a longer transcript than the one that is current on the contract. Like in the $\mathcal{F}^*_{\text{MSFE}}$ case, we need to handle the corner case when $\mathsf{st.id} = \mathsf{best}_j.\mathsf{id} + 1$. This happens when honest parties have completed Step 1 of the st.id-th local execution phase but did not receive signatures on $\mathsf{tv}^{(\mathsf{st.id})}$ from all corrupt parties. In this case, each party $P_j$ will choose new input and fresh randomness and continue the protocol from the transcript $\mathsf{st}.\mathrm{TT}$. Note that $P_j$ responds only when $j = 1 + |\mathsf{st}.\mathrm{TT}| \bmod n$ (i.e., it is its turn) and when the execution has not already completed (i.e., $|\mathsf{st}.\mathrm{TT}| \neq |\mathsf{st.tv}|$). Finally, there is one other case which is unique to $\mathcal{F}^*_{\text{MSCD}}$ implementation. Unlike the $\mathcal{F}^*_{\text{MSFE}}$ case, each party might have to send out multiple messages (corresponding to the reactive MPC protocol) within a single execution. In particular, once the aborted off-chain execution goes on-chain, it remains on-chain (i.e., parties have to respond within time $\Delta$ of the previous step in order to avoid paying a penalty) and needs to be completed by the parties.[1] This brings us to the final case where $\mathsf{st.id} = \mathsf{best}_j.\mathsf{id}$ where in Step 3(d) of Figure 14, honest party $P_j$ uses the next message function in order to continue the transcript $\mathsf{st}.\mathrm{TT}$.

We prove the following theorem in Appendix C.

**Theorem 2.** *Let $\lambda$ be a computational security parameter. Assume the existence of enhanced trapdoor*

[1]Alternatively, when a contract goes on-chain, it is possible to make it come back off-chain right after getting the next message from the party that aborted the off-chain execution. Our protocol does not do this but can be easily modified to behave as described above. Note that this modification does not change our theorem statements.

Parties initialize the parameters as in Figure 13. Then for $\mathsf{id} = 1, 2, \cdots$: Parties run the local execution prescribed in Figure 12. Recall that each party $P_j$ maintains a variable $\mathsf{best}_j$ which is initialized as $\perp$.

1. If a party $P_j$ wants to exit the contract, then it sends $w = \mathsf{exit}$ to $\mathcal{F}^*_{\text{StCon}+}$.
2. If there is an abort at any stage during a local off-chain execution, then parties do not continue any more local executions and instead trigger $\mathcal{F}^*_{\text{StCon}+}$ with the value $\mathsf{best}_j$ if it is non-null.
3. Each party $P_j$ waits and responds to state changes in $\mathcal{F}^*_{\text{StCon}+}$ depending on the current state $\mathsf{st}$:
   (a) If $\mathsf{st.mode}$ = "payout"/"abort" and $\mathsf{st}.L[j] = 1$, send $w = \mathsf{exit}$ to $\mathcal{F}^*_{\text{StCon}+}$.
   (b) If (1) $\mathsf{st.id} < \mathsf{best}_j.\mathsf{id}$, or (2) $\mathsf{st.id} = \mathsf{best}_j.\mathsf{id}$ and $|\mathsf{best}_j.\text{TT}| > |\mathsf{st}.\text{TT}|$, then submit $\mathsf{best}_j$ to $\mathcal{F}^*_{\text{StCon}+}$.
   (c) If $\mathsf{st.id} = \mathsf{best}_j.\mathsf{id} + 1$ and $j = 1 + |\mathsf{st}.\text{TT}| \bmod n$, then choose input $y_j$ and use fresh randomness $\omega_j$ and compute $(m, \sigma) \leftarrow \mathsf{nmf}(\mathsf{st}.\text{TT}; (y_j, \omega_j, \boldsymbol{b}, sk_j))$ and send $(\mathsf{message}, \mathsf{st.id}, (m, \sigma))$ to $\mathcal{F}^*_{\text{StCon}+}$ and update $\mathsf{best}_j \leftarrow (\mathsf{transcript}, \mathsf{st.id}, \mathsf{st}.\text{TT}\|(m, \sigma), \mathsf{st.tv}, \mathsf{st}.\boldsymbol{b}, \mathsf{st}.\boldsymbol{\sigma})$.
   (d) If $\mathsf{st.id} = \mathsf{best}_j.\mathsf{id}$ and if $j = 1 + |\mathsf{st}.\text{TT}| \bmod n$ and if $|\mathsf{st}.\text{TT}| \neq |\mathsf{st.tv}|$, then compute $(m, \sigma) \leftarrow \mathsf{nmf}(\mathsf{st}.\text{TT}; (y_j, \omega_j, \boldsymbol{b}, sk_j))$ where $y_j, \omega_j$ are inputs to the current stage. Send $\mathsf{best}_j \leftarrow (\mathsf{message}, \mathsf{st.id}, (m, \sigma))$ to $\mathcal{F}^*_{\text{StCon}+}$ and update $\mathsf{best}_j \leftarrow (\mathsf{transcript}, \mathsf{st.id}, \mathsf{st}.\text{TT}\|(m, \sigma), \mathsf{st.tv}, \mathsf{st}.\boldsymbol{b}, \mathsf{st}.\boldsymbol{\sigma})$.

Finally, parties also keep track of whether $\mathsf{st.mode}$ = "exit" or "exec" and the current time $t$ is such that $t > \mathsf{st}.t + \Delta$. In this case, parties send $w = \mathsf{exit}$ in order to claim their payout or compensation and do not participate in any further local executions.

Figure 14: Main protocol implementing $\mathcal{F}^*_{\text{MSCD}}$.

*permutations. Then there exists a protocol that* SCC-realizes *(cf. Definition 1)* $\mathcal{F}^*_{\text{MSCD}}$ *in the* $(\mathcal{F}^*_{\text{StCon}+}, \mathcal{F}_{\text{OT}})$*-hybrid model whose script complexity (cf. Definition 3) is independent of the number of secure computations. Furthermore, in the optimistic case (i.e., all parties are honest), there are a total of* $O(n)$ *state transitions (i.e., discounting updates) each of complexity* $O(n\lambda)$*.*

# 6 Implementation in Ethereum

In Figure 18, we provide our reference implementation of smart contract code that correspond to the $\mathcal{F}^*_{\text{MSFE}}$ ideal functionality, in Ethereum's high-level Solidity programming language. A Solidity contract program consists of data declarations and top-level functions. Type annotations must be provided for each variable. Standard types include 256-bit signed or unsigned integers, arrays, and associative mappings. Users interact with a smart contract by publishing a transaction containing a procedure call, including the address of the contract, the name of the function to invoke, and arguments to pass. When a user creates a transaction, they must pay a "gas" fee, which roughly corresponds to the computational cost (e.g., storage plus number of opcodes) of executing the smart contract code.

Like many Ethereum smart contracts, our SmartAmortize implementation is structured as a time-based state machine, where a deadline is set for each transition between states. For now, assume that the contract has already been initialized and received a sufficient deposits from each party; to simplify presentation, we handle this initialization step using a separate contract module as we explain shortly. The state machine begins the "claim" phase of $\mathcal{F}^*_{\text{MSFE}}$ (i.e., enters the time from `T1` to `T2`) when any party calls the `trigger` method, which causes an `EventTriggered` notification to be delivered to each party that has "subscribed" for notifications about this contract.

Within this first time interval (before `T2`), each party must respond to the contract with `submitClaim`, corresponding to the most recent execution for which all the signed commitments have been received. The contract verifies each signature, stores each commitment, and keeps track of the index `r` of the latest execution (lines 33 through 44).

Once the first interval ends (i.e., after `T2`), each party now has until time `T3` to open the commitment corresponding to the latest execution, `bestRound`, by calling `openCommitment` with the appropriate value.

After the final deadline, `T3`, any party may invoke the `finalize` method, which disburses all the appropriate payouts. In the case that every party has revealed the commitment, (i.e., when `anyCorrupt` is false), every party receives their `DEPOSIT` back but no other compensation is necessary (line 65). Since all of the commitments for this round have been opened, the output value can be reconstructed from the information published on the blockchain. However, if any party has failed to open their commitment, then every honest party receives their `DEPOSIT` back, along with an additional `COMPENSATION` (line 63).

We now explain several details about initialization. Our smart contract defines a "constructor" (lines 19-24), which must be invoked by some party by creating a special Ethereum transaction that creates the contract. As mentioned, our implementation also needs some logic for collecting deposits from each of the participating parties, and handling the case where some party aborts before deposits are collected. To accomplish this, we abstract this chunk of logic as a separate reusable contract, `PrepareDeposits`, shown in Figure 19. In short, the `PrepareDeposits` contract is parameterized by a threshold deposit amount, a deadline, and a recipient contract (the address of the `SmartAmortize` contract). Parties can deposit money into this contract at any time. If the threshold is not reached by the deadline, then every party can claim back her deposit (lines 22-27). Otherwise, the entire deposited balance is forwarded on to the `SmartAmortize` contract itself (lines 18-20). This can be done because the `SmartAmortize` contract is created first, so that the recipient contract address is already known. Note that the SmartAmortize contract does not need to implement any special logic to receive the payment from `PrepareDeposits` contract; instead, honest parties should only interact the SmartAmortize contract if and when the `PrepareDeposits` contract succeeds.

**Implementation of the secure cash distribution functionality.** Consider that each state in the reactive functionality is associated with a "balance" for each party, and thus state transitions may alter the balance (e.g., transfer a payment from one party's balance to another). This is the natural setting for a game such as poker, where parties can choose how much of their remaining chips to gamble after each hand ends.

It is easy to add a line to the SmartAmortize contract so that the `finalize` function reconstructs the final state by combining all of the opened commitments (or else pays `COMPENSATION` if these commitments cannot be opened). Recall that we do not need to give `COMPENSATION` for the largest amount that a party can win after an unbounded number of stages, but instead just the largest amount that a party's balance could increase after a single stage. Hence, our code requires each party submit the last *completed* state transition, in case the current state transition ends up incomplete. If the current state transition fails, then each party receives the payment they earned in the most recent state, along with the compensation.

Our extended smart contract is shown in Figure 20. We omit the data and function declarations that remain the same as in Figure 18, and color in red just the lines that have changed. The final *state* is reconstructed from the decommitted shares of either the `bestRound` or `secondBestRound`, depending on if the commitments for `bestRound` have not been opened (lines 40-50). In the case that at least one party is honest, one of these sets of commitments will all be opened (or else no stage of the execution has been completed, and the state defaults to a hard-coded `INITIAL_STATE` value). Each party's payment is determined as an application-specific function `applicationSpecificPayment` of the final *state* (line 51).

**Example application: repeated lotteries.** As an example application, let us describe a multiparty "repeated lottery" game that generalizes the 2-party off-chain lottery of Section 2.3. Each party deposits an initial amount of money (their starting balance), and can play a potentially unbounded number of lottery game iterations. In each lottery iteration, one party's balance is incremented by $N-1$ coins while each other party loses one coin. These iterations continue until any party runs out of money or decides to quit. Since the maximum any party can earn in an iteration is $N-1$ coins, this determines the appropriate value of `COMPENSATION`, implying that the required collateral deposit for each party is $(N-1)(N-1)$. Note that this collateral cost is independent of the starting balance and the number of iterations played. We provide a reference code for this example in [1].

Table 1: Worst-case gas costs for participating in a 4-party contract. Simple application (Figure 18) and extended application (Figure 20).

| | Figure 18 (4 parties) | | Figure 20 (4 parties) | |
|---|---|---|---|---|
| | Gas | (USD) | Gas | (USD) |
| `submitClaim` | 153349 | (3.8¢) | 394003 | (10.0¢) |
| `openCommitment` | 43819 | (1.0¢) | 248046 | (6.2¢) |
| `finalize` | 61567 | (1.5¢) | 70190 | (1.8¢) |

**Implementation and testing.** The smart contract component of our implementation is written in the Solidity language. However, our implementation also includes local Python code for the parties of the protocol to execute, which includes behavior like receiving signatures and broadcasting openings of the commitments, and responding to events in the smart contract. To test our protocol, we use the `pyethereum.tester` blockchain simulator. This simulator runs the actual Ethereum virtual machine on compiled contracts and manually-constructed transactions. By feeding different sequences of transactions to the simulator, we can test the functionality of all the code branches of the contract, and measure the amount of gas required.

To facilitate reproducibility (especially since the Ethereum tool suite is in a state of rapid development), in our final version we will include a Docker image for our experiments, including the Solidity compiler version and `pyethereum` dependency versions that we used.

Gas costs in Ethereum are determined by several factors. First, there is a fixed mapping for each instruction to "gas units," as defined in the Ethereum "yellow paper" reference document [52]. Second, miners enforce a minimum Ether price per gas unit, which today is approximately 2.5E-8 Ether per gas unit. Finally, the market price of Ethereum today is approximately $10 per Ether. Putting these together allows us to express the actual cost of our smart contract in dollars. In Table 1, we show the worst-case costs (per party) to run each phase of our protocol (for an execution involving 4 parties), including both the basic protocol in Figure 18 and the extended application in Figure 20 (assuming the state is represented by 32 bytes). In total, no party would need to pay more than 20 cents in transaction fees to participate in the contract.

## 6.1 Duplex Payment Channel Implementation

In our framework, a duplex micropayment channel can be expressed as a special case of $\mathcal{F}^*_{\text{MSCD}}$, in which the number of parties is two and the functionality has no secret inputs. However, our generic implementation would be overkill for this application; since there is no private state, there is no need for the separate "commit/reveal" phases or for collateral deposits in case one party aborts. Therefore, we give an implementation of a simpler contract for the duplex micropayment channel in Figure 21.

We provide here the detail of our implementation. The parties maintain an off-chain state, which consists of a current balance `net` (positive indicates that Alice has paid Bob, negative vice versa), a round number `r`, and an incremental `withdrawal` approval. Each party is able to do an incremental withdrawal by mutual agreement, i.e., by signatures from both parties. In the case that a malicious party refuses to go along, the honest party can invoke the `trigger` function to finalize the channel and withdraw the total amount that she is owed.

Hence, each party ensures that the withdrawals and round numbers increase monotonically, refusing to sign any state to the contrary. The contract maintains a view of some prior agreed-upon state, which can be updated through the `update` function with an increasing round number. The contract additionally maintains monotonic `deposits` and `withdrawn` fields that keep track of how much total amount of money each party has deposited or withdrawn.

It can be seen that the `withdraw()` function in line 49 has two cases. The first case is for an incremental withdrawal (line 54), that can be reached only in the case that both parties agreed to allow the withdrawal by signing an off-chain message of the form $m$ = (`r, net, withdrawA, withdrawB`). After any party invokes the `update` function with an input $m$ that is signed by both parties, she will be able to perform the incremental withdrawal. The second case is for a final withdrawal (line 59), that can be reached after any party calls the `trigger` function and thereby sets in motion the destruction of the channel.

Our duplex channel reference implementation is available at [1].

# 7 Efficient Poker Protocol

A tailor-made protocol for a poker game in which money changes hands was presented by Kumaresan, Moran, and Bentov in [31, Section 6]. However, that protocol is not efficient enough in practice, due to two distinct reasons. The first reason is that [31, Section 6] was designed to work in the $\mathcal{F}^{\star}_{\mathrm{CR}}$ model, which incurs an expensive setup procedure and does not support off-chain amortization in its SCD variant (cf. Section 2). Furthermore, the $\mathcal{F}^{\star}_{\mathrm{CR}}$ verification circuits that [31, Section 6] uses are quite complex and not supported by the current Bitcoin scripting language. The second reason is that preprocessing shuffle protocol that [31, Section 6] employs is a generic secure MPC that delivers hash-based commitments to shares of the shuffled cards. It would be impractical to run a general-purpose secure MPC protocol (typically among 3 to 9 parties in a poker game) that performs the shuffle and the hash invocations, and indeed there are special-purpose poker protocols that perform much better.

See Section 1.2 for a survey of the various poker protocols. Our implementation uses the poker protocol of Wei and Wang [27, 26], which improves an earlier work of Castellà-Roca et al. [11] by using a zero-knowledge proof of knowledge scheme instead of homomorphic encryption.

A potential disadvantage of special-purpose poker protocols is the on-chain verification cost: the generic secure MPC approach would allow us to define an on-chain predicate that verifies that a pre-image (corresponding to a share of a shuffled card) hashes to the commitment value, and penalize a corrupt party who would not supply the correct pre-image. By contrast, the efficient poker protocols rely on constructions that are algebraic in nature, which implies that the on-chain verification predicate would be significantly more complex.

In the case that all parties are honest, on-chain verification will never occur. In the case that corrupt parties deviate, they can force an honest party to send a transaction containing a witness that satisfies the complex on-chain predicate. The on-chain fallback procedure introduces an additional cost in the form of a transaction fee the party supplying the witness pays (to the miners). While the on-chain fallback also introduces a delay that all of the parties would bear, a malicious party may still cause an honest party to pay the fee. Fortunately, the cost of transaction fees is quite minor (cf. Section 6). Still, our implementation provides an improvement by employing a technique that shares the transaction fee across all parties. In Ethereum this is not straightforward, as the initiator of the transaction must provide all of the transaction fees up front; however, our technique compensates this party by paying funds collected from all of the parties in advance.

By using the efficient scheme of Wei and Wang, the main steps of our SCD poker protocol are as follows:

1. The parties will execute a deck preparation and shuffle protocols, that output group elements (cf. [27, Protocols 1,3,4]).

2. The parties will sign an off-chain message that commits to these group elements (cf. Section 3.3). This signed message could later be sent to the on-chain contract, in case that a corrupt party deviates from the protocol.

3. The parties will run the poker game according to the predefined rules, where in each round a specific party performs a valid action (e.g., raise/call/fold).

4. After each round, all the parties will sign an off-chain message that encodes the state of the poker table.

5. When a party draws a private card from the deck, the parties will execute the card drawing protocol of [27, Protocol 6].

Table 2: Gas costs for zero-knowledge proof of knowledge of equality of discrete logarithms.

| | NIZK Verify | Scalar Multiplication | Built-in Instruction |
|---|---|---|---|
| Gas cost | 1287858 | 303401 | $\leq 3000$ |
| Dollar cost | 30¢ | 7.5¢ | $< 0.1$¢ |

Per the above discussion regarding the complexity of the on-chain predicate, it can be seen that the verification procedure for drawing a private card is dominated by a zero-knowledge proof of knowledge of equality of discrete logarithms (cf. [48, 47]). To reduce the round complexity and avoid the HVZK concern, in Step 5 the parties will use a non-interactive proof of knowledge. While there are provably secure methods to obtain the NIZK (see [18]), our efficient implementation uses the Fiat-Shamir heuristic.

Our Ethereum implementation of the NIZK verifier is based on the `Secp256k1` ellitpic curve group, the same used in Ethereum and in Bitcoin for digital signatures. The Ethereum language does not provide opcodes for working with elliptic curve points. Furthermore, although it features a dedicated opcode for `secp256k1` signature verification, this opcode is signature-specific and cannot (to our knowledge) be re-purposed for the NIZK scheme. Thus we are forced to implement our NIZK the "hard way," making use of an `Secp256k1` elliptic curve library (due to Andreas Olofsson) built from low-level Ethereum opcodes that implement the elliptic curve exponentiation (`_mul`) operation. Our Ethereum implementation, which is shown in Figure 22, bears an obvious resemblance to the high-level proof of knowledge of discrete logarithms protocol [48, 47].

Using the pyethereum simulator framework, we found that the total gas cost of the NIZK verifier is $1.3M$, which corresponds to about 30¢ the time of writing. The cost of the verifier is dominated by the cost of four scalar multiplications. In contrast, the signature verification opcode is only 3000 (a hundred times cheaper), despite performing a scalar multiplication anyway. Thus if Ethereum were modified to support more general elliptic curve arithmetic, we would anticipate a hundred-fold improvement with respect to the transaction fees. In Table 2 we summarize the likely costs of the subroutines that dominate the on-chain verification complexity.

Note that the off-chain signatures in Step 4 include only the current state and not the transcript history, because the proof of knowledge NIZKs do not branch. That is, at a specific round a party will need to provide a NIZK that depends only on public values: the intermediate result of the card drawing protocol, and the group elements that the parties committed in the first step.

The poker protocol of Wei and Wang that we deploy supports all the requirements that were suggested by Crépeau [15]. For example, complete confidentiality of strategy is supported, since the proof of knowledge verification would not reveal the cards at the end of the game. Thus, our implementation enables poker variants such as Texas hold 'em and five-card draw, where private cards are drawn after the game is already in progress. See Wei [26] for benchmarks that measure the running time of the initial shuffling (which is done off-chain).

Our poker implementation is available at [1].

# 8 Conclusion

The combination of standard secure computation and stateful contracts enables fair protocols that are highly efficient, since the interaction with the cryptocurrency network is kept to a minimum. The same is also true for secure cash distribution protocols, and in particular the efficient poker protocol that we provide. Stateful contracts can also be valuable for protocols that do not require secure computation, as demonstrated by our design of a duplex payment channel.

# Appendix A  Full Details on Implementing $\mathcal{F}^*_{\mathrm{MSFE}}$

In this section, we describe the protocols for amortized secure computation with penalties in the $\mathcal{F}^*_{\mathrm{StCon}}$-hybrid model. Note that [30] already constructed protocols for the same in the weaker $\mathcal{F}^\star_{\mathrm{CR}}$-hybrid model. Here, we are going to leverage the power of $\mathcal{F}^*_{\mathrm{StCon}}$ to construct protocols with better properties.

## A.1 General Protocol Structure

The protocol for amortization has three phases.

**Setting up contract parameters.** In the first phase, parties interact with the on-chain stateful contract, i.e., the ideal functionality $\mathcal{F}^*_{\mathrm{StCon}}$. In particular, parties agree on setting contract parameters that fix the number of parties, the allowed state transitions of the contract, the time-out, and the compensation amounts to parties in case of an abort. We describe these contract parameters in Section A.3. Please see Figure 10.

**Local executions.** Then in the second phase, parties perform the actual computation. This is done off-chain via local MPC executions. In addition to performing the computation, these MPC executions also provide hooks to the on-chain contract (to handle aborts). These local executions are described in Section A.2. Please also see Figure 9. We describe the local executions first because they will introduce new variables which will serve as hooks to the contract via the contract parameters described in Section A.3.

**Handling aborts.** In the next phase, we describe the process which honest parties use in case an off-chain local execution was aborted. In particular, in this phase, parties will go on to the on-chain contract to either continue the aborted local execution or claim their compensation. This phase occurs immediately following an abort in the local executions phase. We describe this phase in Section A.4. Please also see Figure 11 which describes how parties handle notification received from $\mathcal{F}^*_{\mathrm{StCon}}$.

## A.2 Local Executions

The formal description of the local executions is in Figure 9. In this section, we describe this phase in more detail. Suppose the parties are interested in computing a function $g^{(\mathsf{id})}$. At a high level, parties begin by running a standard secure computation protocol (that does not guarantee fairness) which computes $z = g^{(\mathsf{id})}(y_1, \ldots, y_n)$ where $y_j$ represents the input of $P_j$. In addition this secure computation protocol also additively secret shares $z$ into $z_1, \ldots, z_n$ and computes commitments $h_j$ on each $z_j$ using uniform randomness $\omega_j$. Finally, the secure computation protocol outputs $x_j^{(\mathsf{id})} = (z_j; \omega_j)$ (i.e., the decommitment to $h_j$) and the value $\boldsymbol{h}^{(\mathsf{id})} = h_1 \| \cdots \| h_n$ to each party $P_j$.

Note that there could be aborts here and in every subsequent stage of the local execution. In each step, we assume that parties stop the protocol if they do not receive valid messages (i.e., including signatures) from any other party. Importantly, there is an additional (implicit) time-interval parameter $\delta$ which is used to detect aborts. In more detail, we say that a party aborted the protocol if (1) it is its turn to send a message, and (2) if the party does not send a valid protocol message within time-interval $\delta$ of the previous event. (We assume that all honest parties act immediately, i.e., within time-interval $\delta$.) In the event of an abort (as defined above), parties go up to the on-chain contract for resolution (cf. Figure 11). To keep the presentation simple, we will discuss handling aborts in Section A.4.

Once the local secure computation protocol ends, we ask each party $P_j$ to compute a signature $\sigma_j$ on the message $(\mathsf{id}, \boldsymbol{h}^{(\mathsf{id})})$ under its secret signing key $sk_j$ and then broadcast it to all parties.

In the next step, each party $P_j$ broadcasts the decommitment $x_j^{(\mathsf{id})}$ to the value $h_j$ contained in $\boldsymbol{h}^{(\mathsf{id})}$ (which in particular includes the secret share $z_j$ of the output $z$). Once this step is completed, then all parties can recover the output of the $\mathsf{id}$-th computation as $\bigoplus_{k \in [n]} z_k$.

**Hooks to the on-chain smart contract.** Now we explicitly describe the local execution values which will serve as hooks to the contract parameters in Section A.3. Clearly, the on-chain contract needs to be independent of the actual computation $g^{(\mathsf{id})}$ and must be equipped to support any non-reactive computation. The idea to make the on-chain contract independent of $g^{(\mathsf{id})}$ is to make it operate only on the secret shares of the output of the computation. This is a standard trick used to reduce fair secure computation to fair reconstruction of an additive secret sharing scheme (see also [23, 6]). For the idea to work in the smart contract setting, we provide hooks that check for authenticity of the output shares. This is supported by the value $\boldsymbol{h}^{(\mathsf{id})}$ which is further signed by each party. Thus, the pair $\boldsymbol{h}^{(\mathsf{id})}, \boldsymbol{\sigma}^{(\mathsf{id})}$ take care of checking authenticity of the output shares. Note that in order to verify that the signatures are authentic, we need to provide the set of public keys $\boldsymbol{pk} = (pk_1, \ldots, pk_n)$ as a hook to the on-chain contract.

It is important to note that the value $\mathsf{id}$ is included in the signature $\sigma_i$ (contained in $\boldsymbol{\sigma}^{(\mathsf{id})}$). This is to ensure that the most recent off-chain computation (which is possibly aborted) is continued by the on-chain contract. Thus, the on-chain contract will also need the $\mathsf{id}$ (implicit in $\boldsymbol{\sigma}^{(\mathsf{id})}$) as a hook. To help description of the protocol, we use an additional variable $\mathsf{best}_j$ maintained locally by each party $P_j$ to denote the transcript corresponding to the latest *active* execution. An off-chain execution is said to be *active* once $P_j$ broadcasts its signature $\sigma_j$ to all parties. (The idea is that an *active* execution could possibly get initiated by a malicious party (but not by an honest party) and get completed on the on-chain contract.) This concludes the description of the hooks to the on-chain contract from the local executions. The actual contract parameters includes a time-out parameter and also includes a state variable that keeps track of the continued off-chain computation (in particular, the set of parties who have behaved honestly, and the set of parties who have already been compensated, etc.). We now move on to describe the contract parameters in the next section.

## A.3 Contract Parameters

The contract parameters are formally described in Figure 10. Recall that we want to allow parties to evaluate non-reactive functionalities an unbounded number of times and without a time-limit on the contract. The state parameters are established in the following way. The variables $d_1, \ldots, d_n$ represent the amount $\mathsf{coins}((n-1)q)$ that the parties are expected to put in as the initial deposit to the contract.

*State components and initialization.* The variable $\mathsf{st}$ denotes the current state of the contract. The values $\boldsymbol{pk}$ and $\Delta$ are constant parameters to the contract and are always maintained as part of the state. The parameter $\boldsymbol{pk}$ represents the set of public keys of all the parties. The parameter $\Delta$ represents the length of the time interval within which parties need to act in order to keep the contract from defaulting. In addition, each state variable has five components: (1) $\mathsf{st.mode}$ represents the current mode in which the contract is in, and is one of {"init", "exec", "exit", "payout", "abort", "inactive"}; (2) $\mathsf{st.id}$ represents the $\mathsf{id}$ of the execution that is being continued currently on the on-chain contract; (3) $\mathsf{st}.\textsc{tt}$ represents the current transcript of the execution that is being continued on the chain; (4) $\mathsf{st}.t$ represents the time when the on-chain contract was triggered and either (a) was moved to "exit" or (b) resulted in a change of the variable $\mathsf{st}.\textsc{tt}$; (5) $\mathsf{st}.L$ is a boolean array that represents which parties have already withdrawn their deposits and compensations from $\mathcal{F}^*_{\text{StCon}}$.

We represent the state variable $\mathsf{st}$ as a five tuple $(\mathsf{st.mode}, \mathsf{st.id}, \mathsf{st}.\textsc{tt}, \mathsf{st}.t, \mathsf{st}.L)$. Also, $\mathsf{st}.\textsc{tt}$ is either $\perp$ (denoting the null transcript) or is a tuple of the form $(X, \boldsymbol{h}, \boldsymbol{\sigma})$. The initial state is $\mathsf{st}$ and its components are initialized in the following way: (1) $\mathsf{st.mode}$ = "init"; (2) $\mathsf{st.id} = -1$; (3) $\mathsf{st}.\textsc{tt} = \mathsf{NULL}$; (4) $\mathsf{st}.t = -1$; and (5) $\mathsf{st}.L = (1, \ldots, 1)$. Recall that $\mathsf{st}$ also contains the list of all public keys of the participants $\boldsymbol{pk}$, and the global time-out parameter $\Delta$.

*Triggering state transitions.* During course of the execution, the state of the contract would either (1) remain in the initial state with $\mathsf{st.mode}$ = "init"; or (2) be in exit mode, i.e., with $\mathsf{st.mode}$ = "exit", where contract participants are trying to get their initial deposit out of the contract (and terminate the contract); or (3) be trying to continue an incomplete off-chain local execution by keeping track of the current state of the local execution computation, i.e., with $\mathsf{st.mode}$ = "exec"; or (4) be in payout mode with $\mathsf{st.mode}$ = "payout" where parties have successfully completed all executions so far and are waiting to get their initial deposits out of $\mathcal{F}^*_{\text{StCon}}$; or (5) be in "abort" where an execution was aborted and honest parties are waiting to get their initial deposits and compensation out of $\mathcal{F}^*_{\text{StCon}}$; or (6) be in "inactive" where $\mathcal{F}^*_{\text{StCon}}$ no longer accepts any further state transitions and in particular, has given out all the money that was deposited to it.

Transitions to different states will be triggered by a witness $(j, w, t)$. Here $j$ represents the party triggering the contract, i.e., party $P_j$. The value $t$ represents the time at which the contract is triggered. Note that when $\mathsf{st.mode}$ = "payout"/"abort"/"inactive", the triggering witness $w$ is simply the token value $\mathsf{exit}$. As we will see later, the transitions from these states depends only on $\mathsf{st}.L$ and the triggering time $t$ and $\mathsf{st}.t$. The more interesting case is when $\mathsf{st.mode}$ = "init"/"exit"/"exec". In this case, the triggering witness $w$ provides the most recent state of the current local execution. We will use a separate subroutine $\mathsf{pred}$ to determine the validity of a trigger $(j, w, t)$ when the witness $w$ represents a transcript of an execution.

*Subroutine* $\mathsf{pred}$. The predicate $\mathsf{pred}$ essentially decides if a trigger to the contract is a valid continuation of the computation on the chain. In more detail, let $(j, w, t)$ denote the trigger encountered by $\mathcal{F}^*_{\text{StCon}}$. Now, $\mathsf{pred}$ takes this trigger and examines it in conjunction with the current state of the contract. First, $\mathsf{pred}$ parses the witness $w$ as $(\mathsf{id}, \textsc{tt} = (X, \boldsymbol{h}, \boldsymbol{\sigma}))$ where $\mathsf{id}$ represents the (off-chain) execution that is being attempted to be continued on the chain by party $P_j$. The value $\textsc{tt} = (X, \boldsymbol{h}, \boldsymbol{\sigma})$ essentially provide (along with a proof) the most recent state of a computation (typically the last incomplete off-chain computation). In particular and in the context of non-reactive functionalities, the value $X$ maintains the set of parties who have completed their step of the computation on the chain along with their broadcasted decommitments to the secret share of the final output. The values $\boldsymbol{h}$ and $\boldsymbol{\sigma}$ essentially authenticate to the contract that values in $X$ are legitimate values corresponding to the $\mathsf{id}$-th off-chain computation. In more detail, $\boldsymbol{h} = h_1 \| \cdots \| h_n$ is the set of commitments (that is public to all parties). The value $\boldsymbol{h}$ should be consistent with the broadcast values $X$ in the sense that $\mathsf{com}(X[k]) = h_k$ for all $k$ such that $X[k] \neq \perp$. Likewise, the commitment values $\boldsymbol{h}$ need to be accompanied with $\boldsymbol{\sigma} = \sigma_1 \| \cdots \| \sigma_n$ where $\sigma_i$ is the signature of party $P_i$ attesting to the correctness of $\boldsymbol{h}$. Note that the signatures also tie the value of $\boldsymbol{h}$ to $\mathsf{id}$. In summary, the witness $w$ essentially provides the current state of the most recent execution that needs to be completed on the chain.

Clearly, $\mathsf{pred}$ should output 1 if the witness is valid and if $(j, w, t)$ happens to be the very first trigger to the contract. On the other hand, if $(j, w, t)$ is not the first trigger to the contract, then we have to ensure that the trigger $(j, w, t)$ provides a valid update to the contract state. Now the contract state could be in exit mode, i.e., $\mathsf{st.mode}$ = "exit", and in this case the trigger $(j, w, t)$ with a valid witness $w$ could be by an honest party to continue an incomplete off-chain execution. This is to ensure that a malicious party cannot subvert the continuation of the off-chain execution on the on-chain contract by trying to exit prematurely (i.e., when $\mathsf{st.mode}$ = "init". Likewise a malicious party might also submit an old completed execution (even while the current off-chain execution has not yet completed). Thus, we must have $\mathsf{pred}$ output 1 when the new $\mathsf{id}$ present in $w$ is greater than $\mathsf{st.id}$.

Now the contract could be in exec mode, i.e., $\mathsf{st.mode}$ = "exec", in which case the contract is typically waiting for the on-chain execution to be completed. We now describe the cases when the trigger $(j, w, t)$ is a valid continuation of the current state of the contract. There are essentially two cases: (1) the current state does not correspond to or continue the most recent off-chain execution; in this case, the $\mathsf{id}$ in the new trigger must satisfy $\mathsf{id} > \mathsf{st.id}$ (i.e., the contract is essentially reset to the "correct last computation"), and (2) the new trigger continues the current state of the contract and for this $\mathsf{id} = \mathsf{st.id}$ must hold and also we need $X$ to contain at least one value which is not in $\mathsf{st}.\textsc{tt}.X$, i.e., there is some $k \in [n]$ such that $X[k] \neq \perp = \mathsf{st}.\textsc{tt}.X$. In either case, the new trigger must appear within the time interval $\Delta$ of the previous trigger (i.e., before time $\mathsf{st}.t + \Delta$). This completes the description of $\mathsf{pred}$.

*State transitions.* We are now ready to describe the state transition function $\mathsf{Prog}$ that takes as input the trigger $(j, w, t)$ and the current state $\mathsf{st}$. First, we check if the witness provided corresponds to an execution transcript. In this case, we invoke the predicate $\mathsf{pred}$ and if $\mathsf{pred}$ outputs 1, then we update $\mathsf{st}.\textsc{tt}$ depending on whether (1) $\mathsf{st.id} = \mathsf{id}$, in which case we update $\mathsf{st}.\textsc{tt}.X$ to include decommitments specified in $X$; or (2) $\mathsf{st.id} < \mathsf{id}$, in which case we update $\mathsf{st}.\textsc{tt} \leftarrow \textsc{tt}$. If $w$ does not correspond to an execution transcript, then we assume that it is a token value $\mathsf{exit}$. There are effectively three cases to handle:

- If $\mathsf{st.mode}$ equals "init" or equals "exec" with a fully completed transcript, then we change $\mathsf{st.mode}$ to "exit" and store the triggering time $t$ in $\mathsf{st}.t$. This transition is provided to ensure that honest parties' deposits are not "locked in" and to enable them to withdraw their deposits from $\mathcal{F}^*_{\mathrm{StCon}}$.
- If $\mathsf{st.mode}$ = "exec"/"abort", then we check if $t > \mathsf{st}.t + \Delta$ and if $|\mathsf{st}.\textsc{tt}.X| \neq n$ to confirm that the execution has indeed been aborted. In this case, if $\mathsf{st}.L[j] = 1$, then we will allow $P_j$ to take money out of $\mathcal{F}^*_{\mathrm{StCon}}$. We further need to check whether $P_j$ was a malicious party that did not contribute to completing the execution. We do this by checking if $\mathsf{st}.\textsc{tt}.X[j] \neq \perp$. If all checks pass, we let $P_j$ to withdraw its initial deposit plus compensation, i.e., a total of $n(n-1)q/|\mathsf{st}.\textsc{tt}.X|$ from $\mathcal{F}^*_{\mathrm{StCon}}$.
- If $\mathsf{st.mode}$ = "exit"/"payout", then we check if $t > \mathsf{st}.t + \Delta$. This is to prevent situations where a malicious party tries to subvert continuing the off-chain aborted execution on the chain. (That is, honest parties get an additional time $\Delta$ to get $\mathcal{F}^*_{\mathrm{StCon}}$ out of the exit mode.) If $t > \mathsf{st}.t + \Delta$ indeed holds, then we allow party $P_j$ to take its initial deposit $(n-1)q$ out of the contract if it was not already paid before (i.e., $\mathsf{st}.L[j] = 1$).

Finally, once $\mathcal{F}^*_{\mathrm{StCon}}$ has given back all the coins that it initially held, we allow $\mathcal{F}^*_{\mathrm{StCon}}$ to get into an inactive mode where it stops accepting any further state transitions. This concludes the description of state transitions in $\mathcal{F}^*_{\mathrm{StCon}}$.

*Additional remarks.* Note that when $\mathsf{pred}(j, w, t; \mathsf{st})$ outputs 1, we update the last trigger time to $t$. Note that in this case, no coins will be distributed to the parties. The key point to note is that the contract remains in exec mode even when $\mathsf{st}.\textsc{tt}$ represents a fully completed transcript. This is important since a malicious adversary might try to submit an older off-chain execution transcript to the contract and prevent the latest incomplete off-chain execution from being continued on the chain. Therefore, we should wait till time interval $\Delta$ before releasing the coins. This also serves the purpose of giving the honest parties sufficient time to rectify the situation and submit the latest incomplete off-chain execution onto the chain. Additionally, keeping the contract in exec mode also enables re-use of the contract by the participants (i.e., to perhaps complete *future* aborted off-chain local executions). For situations where we do not want to enable re-use of the contract, the honest parties can simply move the contract into exit mode. One could also change the contract so that this happens automatically whenever $|S| = n$ (and $\mathsf{pred}$ outputs 1). Note that even though the contract is in exit mode, it can always be reverted back into continuing an off-chain local execution (i.e., the honest parties have to act within time $\Delta$ from the time that the contract went into exit mode). This is important since a malicious adversary may be trying to make the contract go into exit mode in order to avoid paying a penalty.

This concludes the explanation of Figure 10.

## A.4 Main Protocol

The formal description can be found in Figure 11. In this section we will describe the main protocol that makes use of the local execution subprotocol of Figure 9 and also how parties interact with $\mathcal{F}^*_{\text{StCon}}$ according to the parameters described in Figure 10. Parties basically start by initializing the $\mathcal{F}^*_{\text{StCon}}$ parameters as in Figure 10. Following this, they begin off-chain local executions. Recall that each party $P_j$ maintains a variable $\mathsf{best}_j$ which denotes the transcript corresponding to the latest *active* execution (both on-chain and off-chain) *according to the local view of party* $P_j$ (see Figure 9). This value will provide the necessary hook to the on-chain contract to handle off-chain aborts. In particular, the value $\mathsf{best}_j$ will be submitted by party $P_j$ in order to recover from aborted off-chain executions.

In our main protocol, we essentially deal with three different scenarios: (1) when parties want to exit the contract and get back their deposits and compensation, (2) when parties want to continue an aborted off-chain execution on the chain, and (3) when parties are notified of state changes in $\mathcal{F}^*_{\text{StCon}}$.

*Exiting the contract.* First, we deal with the situation when parties would like to terminate the protocol and retrieve their initial deposits from the contract. To do so, we simply let parties submit a token value $w = \mathsf{exit}$ to trigger and put the contract into exit mode. Note that malicious parties might revert $\mathcal{F}^*_{\text{StCon}}$ to go into exec mode. In this case, $\mathcal{F}^*_{\text{StCon}}$ will notify honest parties of the change. Honest parties will be able to recover from this and put the contract back into exit mode. This will be described when we discuss how parties react to notifications from $\mathcal{F}^*_{\text{StCon}}$.

*Continuing an aborted off-chain execution.* Here, we deal with the case when an off-chain local execution was aborted. This is where the value $\mathsf{best}_j$ comes in handy as it stores the most recent state of the off-chain executions. We simply instruct parties to trigger $\mathcal{F}^*_{\text{StCon}}$ with the value $\mathsf{best}_j$. Note that the value $\mathsf{best}_j$ in particular includes $P_j$'s decommitment $x_j^{(\mathsf{id})}$ which in turn ensures (by the logic in Figure 10) that $P_j$ will not be penalized.

*Responding to notifications from* $\mathcal{F}^*_{\text{StCon}}$. First, if $\mathsf{st.mode}$ = "payout"/"abort", then parties send a token value $w = \mathsf{exit}$ to get their deposits out of $\mathcal{F}^*_{\text{StCon}}$. In addition, if $\mathsf{st.mode}$ = "abort", then parties would also get compensation from $\mathcal{F}^*_{\text{StCon}}$. Second, if the on-chain execution does not correspond to the most recent execution, then we ask parties to submit $\mathsf{best}_j$ to the contract. (This will also handle the case when honest parties try to exit the contract but a malicious party feeds an older execution to $\mathcal{F}^*_{\text{StCon}}$.) Checking if the on-chain execution corresponds to the most recent execution is handled by checking first if $\mathsf{st.id} < \mathsf{best}_j.\mathsf{id}$ and then if $\mathsf{st.id} = \mathsf{best}_j.\mathsf{id}$ but $\mathsf{st}.\textsc{tt}.X[j] = \perp$ (i.e., party $P_j$'s decommitment is not yet part of the on-chain execution transcript). Finally, we also need to handle the corner case when $\mathsf{st.id} > \mathsf{best}_j.\mathsf{id}$. This scenario is actually possible when party $P_j$ is honest but only when $\mathsf{st.id} = \mathsf{best}_j.\mathsf{id} + 1$. We now describe the sequence of events which lead to this case. Suppose in Step 2 of Figure 9 some malicious party did not broadcast its signature on $\boldsymbol{h}^{(\mathsf{id})}$, then party $P_j$ will not update $\mathsf{best}_j$. Thus $\mathsf{best}_j.\mathsf{id} = \mathsf{id} - 1$ where $\mathsf{id}$ is the execution id of the current execution. Note that each honest $P_j$ would have submitted its signature on $\boldsymbol{h}^{(\mathsf{id})}$ in Step 2. Therefore, malicious parties would possess a valid $\boldsymbol{h}^{(\mathsf{id})}$ and $\boldsymbol{\sigma}^{(\mathsf{id})}$ for execution $\mathsf{id}$. That is, a malicious party $P_k$ is able to trigger $\mathcal{F}^*_{\text{StCon}}$ with a witness $w = (\mathsf{id}, \textsc{tt} = ((k, x_k^{(\mathsf{id})}), \boldsymbol{h}^{(\mathsf{id})}, \boldsymbol{\sigma}^{(\mathsf{id})})$ which will result in $\mathsf{pred}(k, w, t) = 1$. Thus, we need a mechanism to allow honest parties to continue the $\mathsf{id}$-th execution (i.e., continue $\textsc{tt}$) and ensure that they don't get penalized. This is indeed possible since honest parties already obtain $x_j^{(\mathsf{st.id})}$ from Step 1 of Figure 9. That is, we let honest parties submit $w = (\mathsf{st.id}, \textsc{tt}_j = ((j, x_j^{(\mathsf{st.id})}),$ $\boldsymbol{h}^{(\mathsf{id})} = \mathsf{st}.\textsc{tt}.\boldsymbol{h}, \boldsymbol{\sigma}^{(\mathsf{id})} = \mathsf{st}.\textsc{tt}.\boldsymbol{\sigma}))$ to $\mathcal{F}^*_{\mathrm{StCon}}$.

## A.5 Proof of Security

**Theorem 1.** Let $\lambda$ be a computational security parameter. Assume the existence of one-way functions. Then there exists a protocol that SCC-realizes (cf. Definition 1) $\mathcal{F}^*_{\mathrm{MSFE}}$ in the ($\mathcal{F}^*_{\mathrm{StCon}}$,$\mathcal{F}_{\mathrm{OT}}$)-hybrid model whose script complexity (cf. Definition 3) is independent of the number of secure function evaluations and depends only on the length of outputs of the functions evaluated in $\mathcal{F}^*_{\mathrm{MSFE}}$ and is otherwise independent of them. Furthermore, in the optimistic case when all parties are honest, there are a total of $O(n)$ state transitions each having complexity $O(n\lambda)$.

*Proof.* We briefly discuss the optimistic case when all parties are honest. In this case, there are no aborts in any of the local executions. Finally, when parties agree to exit the contract, they each send $w = \mathsf{exit}$ to $\mathcal{F}^*_{\mathrm{StCon}}$, get back their initial deposits and put $\mathcal{F}^*_{\mathrm{StCon}}$ in inactive mode. Note that we calculate the complexity of a state transition (cf. Definition 3) as the sum of the size of the input states and output states. Since the input/output states contain $\boldsymbol{pk}$, we have that the complexity of each state transition is $O(n\lambda)$. We now focus on the more interesting case where some of the parties are corrupt.

We describe the simulator $\mathcal{S}$ and then prove indistinguishability of the simulation and the real world execution. The full description of the simulator can be found in Figures 17, 15, 16. Note that in the simulation, $\mathcal{S}$ would act as $\mathcal{F}^*_{\mathrm{StCon}}$ and also as $\mathcal{F}_{\mathrm{OT}}$. Acting as $\mathcal{F}^*_{\mathrm{StCon}}$, $\mathcal{S}$ waits for each corrupt party to make the initial deposits $\mathsf{coins}((n-1)q)$. It then uses $\mathsf{coins}(hq)$ from this deposit amount to submit the initial deposit to $\mathcal{F}^*_{\mathrm{MSFE}}$. (Note that there is at least one corrupt party, and so $\mathcal{S}$ will be able to complete this step.) $\mathcal{S}$ initializes $\boldsymbol{pk}$ with public keys from corrupt parties, and supplies randomly generated public keys for honest parties (and stores the corresponding secret keys). Then for $\mathsf{id} = 1, 2, \cdots$: $\mathcal{S}$ waits to receive the function $g^{(\mathsf{id})}$ from $\mathcal{F}^*_{\mathrm{MSFE}}$. Following this, $\mathcal{S}$ notifies the corrupt parties of $g^{(\mathsf{id})}$. Then $\mathcal{S}$ invokes the SFE simulator $\mathcal{S}_{\mathsf{sfe}}$ for the MPC protocol for evaluating $g^{(\mathsf{id})}$ (which works in the $\mathcal{F}_{\mathrm{OT}}$-hybrid model). $\mathcal{S}_{\mathsf{sfe}}$ extracts the inputs of the corrupt parties $\{y_s^{(\mathsf{id})}\}_{s\in C}$ where $C$ represents the set of corrupt parties. As output of the SFE protocol for $g^{(\mathsf{id})}$, $\mathcal{S}$ instructs $\mathcal{S}_{\mathsf{sfe}}$ to output random values $x_s^{(\mathsf{id})} = (z_s; \omega_s)$ for each $s \in C$, and $\boldsymbol{h}^{(\mathsf{id})} = h_1 \| \cdots \| h_s$ where for $s \in C$, it holds that $h_s = \mathsf{com}(x_s^{(\mathsf{id})})$ and for $j \in H$, the value $h_j$ is generated using the equivocation algorithm for the honest binding commitment scheme $\mathsf{com}$. This completes the simulation of the MPC step of the $\mathsf{id}$-th local execution. Following this, $\mathcal{S}$ sends signatures on $(\mathsf{id}, \boldsymbol{h}^{(\mathsf{id})})$ under $sk_j$ for $j \in H$. Then $\mathcal{S}$ waits to receive signatures on $(\mathsf{id}, \boldsymbol{h}^{(\mathsf{id})})$ from corrupt parties. Once this is done, $\mathcal{S}$ sends $\{y_s^{(\mathsf{id})}\}_{s\in C}$ to $\mathcal{F}^*_{\mathrm{MSFE}}$ and receives the output $z^{(\mathsf{id})}$. Acting as honest parties, $\mathcal{S}$ then sends the decommitments to the output shares. To do this, $\mathcal{S}$ leverages the equivocation property of the honest binding commitment scheme to generate $\{x_j^{(\mathsf{id})} = (z_j; \omega_j)\}_{j\in H}$ such that $z^{(\mathsf{id})} = \bigoplus_{k\in[n]} z_k$ holds. This concludes the description of the simulation of the local executions as long as there is no abort or notification from the parties to exit the contract.

Now we discuss how to handle the case when parties try to exit the contract. Suppose $\mathcal{S}$ receives notification of the form $(\mathsf{exit}, sid, j)$ from $\mathcal{F}^*_{\mathrm{MSFE}}$ for $j \in H$, then acting as $\mathcal{F}^*_{\mathrm{StCon}}$, $\mathcal{S}$ sends a notification $(\mathsf{output}, sid, ssid, j, w = \mathsf{exit}, t, \mathsf{st}, e)$ to each corrupt party where $e = 0$ and $\mathsf{st}$ represents the current state of the simulated $\mathcal{F}^*_{\mathrm{StCon}}$. Note that in this case, $\mathcal{S}$ would also receive its deposit $\mathsf{coins}(hq)$ from $\mathcal{F}^*_{\mathrm{MSFE}}$ and it will be able to distribute $\mathsf{coins}((n-1)q)$ to all corrupt parties. On the other hand, suppose $\mathcal{S}$ acting as $\mathcal{F}^*_{\mathrm{StCon}}$ receives a message $w = \mathsf{exit}$ from the adversary $\mathcal{A}$, then it changes the state of $\mathcal{F}^*_{\mathrm{StCon}}$ using the function $\mathsf{Prog}(k, w, t; \mathsf{st})$ (as specified in Figure 10) where $\mathsf{st}$ represents the state of the simulated contract, and $k$ represents the index of the corrupt party on behalf of whom $\mathcal{A}$ submitted the trigger. Let $\mathsf{st}, e$ be the output of $\mathsf{Prog}$. Now $\mathcal{S}$ simulates honest parties response to the notification of $\mathsf{st}, e$ from $\mathcal{F}^*_{\mathrm{StCon}}$. That is, it simulates the honest parties' execution of Step 3 of Figure 11. Note that the value $\mathsf{best}_j$ for each honest $j$ is prepared using the decommitments generated by $\mathcal{S}$ using the equivocation simulator for the honest binding commitment scheme (that guarantees that the output is consistent with the output of $\mathcal{F}^*_{\mathrm{MSFE}}$). Now Step 3 instructs honest parties to send their messages to $\mathcal{F}^*_{\mathrm{StCon}}$. Therefore, $\mathcal{S}$ acting as $\mathcal{F}^*_{\mathrm{StCon}}$ needs to react to the messages received from the simulated honest parties. That is, $\mathcal{S}$ acting as $\mathcal{F}^*_{\text{StCon}}$ will invoke $\mathsf{Prog}$ using the inputs submitted by the simulated honest parties, and then distribute the output $\mathsf{st}, e$ from $\mathcal{F}^*_{\text{StCon}}$ to $\mathcal{A}$. This process continues until $\mathcal{F}^*_{\text{StCon}}$ is in inactive mode. Next we will discuss how to handle aborts in the off-chain execution. When $\mathcal{S}$ is simulating the off-chain executions, it sends the simulated honest parties' messages immediately, and waits for time period $\delta$ to receive messages from $\mathcal{A}$. If not all messages corresponding to corrupt parties are received within this time period, then this counts as an abort in the off-chain execution. For every honest party $P_j$, the simulator $\mathcal{S}$ acting as the simulated honest party $P_j$ prepares a value $\mathsf{best}_j$ which corresponds to the transcript corresponding to the most recent execution (and in particular includes messages from corrupt parties that did send the decommitments to their output secret shares). $\mathcal{S}$ then notifies $\mathcal{A}$ of the value of $\mathsf{st}, e$ that results from a trigger $\mathsf{best}_j$. Note that notifications are sent separately for each honest $P_j$. Accordingly, $\mathcal{S}$ will update the value $\mathsf{st}.\text{TT}$ as specified in Figure 10 for each $\mathsf{best}_j$ resulting from the simulated honest parties. Note that additionally it is possible to allow $\mathcal{A}$ to arbitrarily reorder the $\{\mathsf{best}_j\}$ triggers from the simulated honest parties. In particular, it is possible to consider a situation where $\mathcal{A}$ interleaves its triggers to $\mathcal{F}^*_{\text{StCon}}$ with triggers from honest parties. These will not affect the overall simulation as the messages sent by the simulated honest parties are determined before they decide to trigger $\mathcal{F}^*_{\text{StCon}}$ and is independent of the triggers supplied by $\mathcal{A}$. Thus we focus only on the harder case where $\mathcal{A}$ is essentially a rushing adversary which waits to see messages from honest triggers before deciding on how to trigger $\mathcal{F}^*_{\text{StCon}}$. Of course this does not preclude $\mathcal{A}$ to trigger $\mathcal{F}^*_{\text{StCon}}$ with an alternate partial/completed transcript *any time* during the execution of the protocol. This case is similar to the case where $\mathcal{S}$ handles a witness $w = \mathsf{exit}$ from $\mathcal{A}$ (except here $w$ might be a partial transcript). Suppose $\mathcal{A}$ on behalf of corrupt party $P_k$ triggered $\mathcal{F}^*_{\text{StCon}}$ with witness $w$ at time $t$. Then, $\mathcal{S}$ faithfully applies the trigger $(k, w, t)$ and invokes $\mathsf{Prog}$ and notifies $\mathcal{A}$ of the resulting output $\mathsf{st}, e$ (i.e., if trigger $(k, w, t)$ was valid and induced a state transition). As before, $\mathcal{S}$ simulates the response from the honest parties by running Step 3 of Figure 11 to generate new triggers if any, and applies them to $\mathcal{F}^*_{\text{StCon}}$. Then acting as $\mathcal{F}^*_{\text{StCon}}$, $\mathcal{S}$ notifies the parties of state transitions in $\mathcal{F}^*_{\text{StCon}}$. Note that the trigger $w$ is supposed to contain only values that were sent to $\mathcal{A}$ by $\mathcal{S}$ (where $\mathcal{S}$ either acted as $\mathcal{F}^*_{\text{StCon}}$, or the simulated honest parties, or during the simulation of the MPC protocol). If not and if the value $w$ results in a successful state transition, then the simulator outputs $\mathsf{fail}$ and terminates the protocol. Note that this additional restriction, namely that $w$ only contains values sent by $\mathcal{S}$ does not mean that $\mathcal{A}$ behaves honestly, as $\mathcal{A}$ could satisfy the constraint and still submit a partial transcript corresponding to an older completed execution even when the current execution is incomplete.

We briefly discuss events which can lead to a simulation failure. As mentioned before, this happens when $\mathcal{A}$ submits a witness $w$ that does not contain the values received from $\mathcal{S}$. This can happen if either $\mathcal{A}$ successfully provides (1) $\boldsymbol{h}' \neq \boldsymbol{h}^{(\mathsf{id})}$ for any value of $\mathsf{id}$ along with the corresponding signatures $\boldsymbol{\sigma}'$ on $\boldsymbol{h}'$ with signatures from all parties; or (2) $x'_k \neq x_k^{(\mathsf{id})}$ as a valid decommitment to the value $h_k$ contained in $\boldsymbol{h}^{(\mathsf{id})}$. It is easy to see that failure event (1) happens with negligible probability due to the unforgeability of the underlying digital signature scheme. In particular, in order to induce event (1) above, the adversary must be able to forge a simulated honest party's signature without knowing the secret signing key. Failure event (2) happens with negligible probability due to the binding property of the honest binding commitment scheme. In particular, the binding property of the honest binding commitment scheme implies that it is computationally infeasible to come up with a value $x'_k \neq x_k^{(\mathsf{id})}$ that is a valid decommitment to the value $h_k = \mathsf{com}(x_k^{(\mathsf{id})})$. It is fairly straightforward to create hybrids and prove that the simulation failure does not affect the indistinguishability of the real world execution and the simulated execution in the ideal world.

Finally we discuss the simulation at the end of the protocol when the simulated $\mathcal{F}^*_{\text{StCon}}$ has been in exit mode for longer than the time period $\Delta$. In this case, $\mathcal{S}$ acting as the simulated honest parties will provide $w = \mathsf{exit}$ to get their (simulated) deposits out of $\mathcal{F}^*_{\text{StCon}}$. The simulator $\mathcal{S}$ accordingly notifies $\mathcal{A}$ of these triggers. $\mathcal{S}$ then handles triggers $w = \mathsf{exit}$ from corrupt party $P_k$ controlled by $\mathcal{A}$ depending on whether the simulated $\mathcal{F}^*_{\text{StCon}}$ has an incomplete transcript or a completed transcript. Suppose $\mathsf{st}$ is the current state of the contract, and if $|\mathsf{st}.\text{TT}.X| = n$ (i.e., completed transcript), then $\mathcal{S}$ issues $(\mathsf{exit}, sid, k)$ to $\mathcal{F}^*_{\text{MSFE}}$ and collects its deposit $\mathsf{coins}(hq)$ from $\mathcal{F}^*_{\text{MSFE}}$. Now $\mathcal{S}$ possesses all the initial deposits, then it can distribute these coins back to the adversary, i.e., $\mathsf{coins}((n-1)q)$ to each corrupt party and then put the simulated $\mathcal{F}^*_{\text{StCon}}$ into inactive mode and terminate the simulation. On the other hand, if $|\mathsf{st}.\mathrm{TT}.X| \neq n$, then this means there is an off-chain execution which was aborted, and which when escalated on-chain is also aborted. In this case, we need to distribute compensations to the honest parties. In this case, $\mathcal{S}$ sends $(\mathsf{abort}, sid, ssid)$ to $\mathcal{F}^*_{\text{MSFE}}$, and for each $k \in C$ such that $\mathsf{st}.\mathrm{TT}.X[k] \neq \perp$, $\mathcal{S}$ sends $\mathsf{coins}((n-1)q)$ to $P_k$. Finally, $\mathcal{S}$ sends $(\mathsf{extra}, sid, ssid, \{(i, q_i)\}_{i \in H}, \mathsf{coins}(\sum_{i \in H} q_i))$ to $\mathcal{F}^*_{\text{MSFE}}$ where $q_i = (n - |\mathsf{st}.\mathrm{TT}.X|)(n-1)q/|\mathsf{st}.\mathrm{TT}.X|$ which ensures that the simulated $\mathcal{F}^*_{\text{StCon}}$ is not left with any money and therefore can go into inactive mode. This completes the description of the simulation. Please refer to Figures 17, 15, 16 for more details. We now state and prove helper propositions which analyzes the properties of the real/simulated execution and ensures that the simulation is indistinguishable from the real execution.

**Proposition 3.** *At the end of the execution, honest parties get back their initial deposit.*

*Proof sketch.* Note that $\mathsf{st.mode}$ = "inactive" at the end of the execution. It should be clear from Figures 10, 16 that $\mathcal{F}^*_{\text{StCon}}$ must have given out all the $\mathsf{coins}(n(n-1)q)$ that it initially received as the deposits. If $\mathcal{F}^*_{\text{StCon}}$ was in payout mode before it entered the inactive mode, again it is clear that all honest parties would have received their initial deposits back. If $\mathcal{F}^*_{\text{StCon}}$ was in abort mode before it entered the inactive mode, then in this case we have to prove that each honest party $P_j$ receives its initial deposit plus compensation out of $\mathcal{F}^*_{\text{StCon}}$. It follows from the description in Figure 10 that $\mathcal{F}^*_{\text{StCon}}$ would be in exec mode before it enters abort mode. Then $\mathcal{F}^*_{\text{StCon}}$ pushes notifications about the current transcript of $\mathsf{id}' = \mathsf{st.id}$-th execution. Note that this transcript must contain $\boldsymbol{\sigma}^{(\mathsf{id}')}$ which includes a signature from all the honest parties including $P_j$. Then it follows from the unforgeability property of the underlying signature scheme, except with negligible probability that this signature must have been generated by $P_j$ in Step 2 of the $\mathsf{id}'$-th local execution in Figure 9. Then at this point we are assured that $P_j$ possesses the decommitment corresponding to $h_j$ which is contained $\boldsymbol{h}^{(\mathsf{id}')}$ (on which all parties generated signatures). This is because the MPC step must have completed successfully before honest parties broadcast their signatures. Therefore honest $P_j$ will be able to trigger $\mathcal{F}^*_{\text{StCon}}$ with the value $\mathsf{best}_j$ which in particular includes the decommitment corresponding to honest $P_j$ and which satisfies $\mathsf{best}_j.\mathsf{id} \geq \mathsf{id}'$. This ensures that $\mathsf{st}.\mathrm{TT}.X[j] \neq \perp$. This in turn ensures that honest $P_j$ will keep triggering $\mathcal{F}^*_{\text{StCon}}$ until $\mathsf{st}.L[j] = 0$ and therefore obtain its initial deposit plus compensation.

**Proposition 4.** *Suppose Step 3 of the* $\mathsf{id}$*-th* local *execution was completed by the honest parties. Then, either the* $\mathsf{id}$*-th execution was completed (either on-chain or off-chain) and the honest parties learned the output, or all honest parties obtained compensation.*

*Proof sketch.* Note that honest parties reveal their decommitments (i.e., output secret shares) in Step 3 of the local execution and therefore, the adversary knows the output of the $\mathsf{id}$-th execution. The interesting case is when there was an abort in the $\mathsf{id}$-th local execution. In this case, honest parties would trigger $\mathcal{F}^*_{\text{StCon}}$ with the value $\mathsf{best}_j$ that includes the decommitments obtained from all honest parties (from the share broadcast step). From the unforgeability property of the underlying digital signature scheme, we have that the adversary cannot trigger $\mathcal{F}^*_{\text{StCon}}$ with a witness $w = (\mathsf{id}', \mathrm{TT}')$ where $\mathsf{id}' > \mathsf{id}$. Further when $\mathsf{id}' < \mathsf{id}$, this trigger would be overridden by the honest parties' trigger. Therefore, it follows that if the adversary wants to avoid paying compensation to the honest parties, then each corrupt party would need to reveal its decommitment and complete the execution on-chain.

**Proposition 5.** *Suppose Step 2 of the* $\mathsf{id}$*-th* local *execution was completed by the honest parties. Then, either the* $\mathsf{id}$*-th execution was completed (either on-chain or off-chain) and the honest parties learned the output, or all honest parties obtained compensation, or no party learned the output of the* $\mathsf{id}$*-th execution.*

*Proof sketch.* Clearly, if Step 3 of the $\mathsf{id}$-th local execution was completed by the honest parties, then by Proposition 4 it follows that either the $\mathsf{id}$-th execution was completed (either on-chain or off-chain) and the honest parties learned the output, or all honest parties obtained compensation. The interesting case is when Step 3 of the $\mathsf{id}$-th local execution was not completed by the honest parties which effectively means that not all corrupt parties completed Step 2 of the $\mathsf{id}$-th local execution. Note that honest parties reveal their signature on $\boldsymbol{h}^{(\mathsf{id})}$ in Step 2, and therefore the adversary possesses all signatures and therefore the values $\boldsymbol{\sigma}^{(\mathsf{id})}$ and $\boldsymbol{h}^{(\mathsf{id})}$. This means that the corrupt parties could potentially trigger $\mathcal{F}^*_{\text{StCon}}$ with a valid partial transcript of the $\mathsf{id}$-th execution. Note that since Step 2 was not completed by all parties (i.e., each $P_j$ would have $|S_j| \neq n$), it follows that honest party $P_j$ would not have updated $\mathsf{best}_j$ (and in particular $\mathsf{best}_j.\mathsf{id} = \mathsf{id} - 1$ would still hold). Now it is possible that the adversary never triggers $\mathcal{F}^*_{\text{StCon}}$ with a witness $w = (\mathsf{id}', \text{TT}')$ where $\mathsf{id}' = \mathsf{id}$. In this case, we would have that no party learned the output of the $\mathsf{id}$-th execution. Furthermore, $\mathsf{st.id} = \mathsf{id} - 1 = \mathsf{best}_j.\mathsf{id}$, and $\mathcal{F}^*_{\text{StCon}}$ would move into payout mode and give deposits back to all parties. On the other hand, the adversary could trigger $\mathcal{F}^*_{\text{StCon}}$ with a witness $w = (\mathsf{id}', \text{TT}')$ where $\mathsf{id}' = \mathsf{id}$. This is exactly the scenario handled by Step 3 of the Figure 11 where we have $\mathsf{st.id} = \mathsf{id}' = \mathsf{id} = \mathsf{best}_j.\mathsf{id} + 1$. The key thing to note is that in this case, honest parties only need their decommitments (which they received before Step 2) to continue the $\mathsf{id}$-th execution on-chain. (The values $\boldsymbol{h}$ and $\boldsymbol{\sigma}$ are taken from $\mathsf{st}.\boldsymbol{h}$ and $\mathsf{st}.\boldsymbol{\sigma}$ where each honest party receives $\mathsf{st}$ as a notification of state change in $\mathcal{F}^*_{\text{StCon}}$.) The rest of the proof is similar to the proof of Proposition 4.

□

**Proposition 6.** *Suppose Step 1 of the* $\mathsf{id}$*-th* local *execution was completed by the honest parties. Then, either the* $\mathsf{id}$*-th execution was completed (either on-chain or off-chain) and the honest parties learned the output, or all honest parties obtained compensation, or no party learned the output of the* $\mathsf{id}$*-th execution.*

*Proof sketch.* Clearly, if Step 2 of the $\mathsf{id}$-th local execution was completed by the honest parties, then by Proposition 5 it follows that either the $\mathsf{id}$-th execution was completed (either on-chain or off-chain) and the honest parties learned the output, or all honest parties obtained compensation. The interesting case is when Step 1 of the $\mathsf{id}$-th local execution was not completed by the honest parties which effectively means that at least one honest party did not receive the output secret shares. Note that an abort at this step means that honest parties do not continue this local execution (either on-chain or off-chain). In particular, honest parties do not send their signature on $\boldsymbol{h}^{(\mathsf{id})}$ to any party. Given this, except with negligible probability it is impossible for the adversary to trigger $\mathcal{F}^*_{\text{StCon}}$ with a witness $w = (\mathsf{id}', \text{TT}')$ where $\mathsf{id}' = \mathsf{id}$. Note that the joint view of the corrupt parties does not contain any information on the output of the $\mathsf{id}$-th local execution. This concludes the proof.

### A.6 An Optimization

We detail an optimization that enables us to make the script complexity independent of the functions $\{g^{(\mathsf{id})}\}$ (in particular, independent of the output length of the functions). The core idea is to use a technique analogous to hybrid encryption—we encrypt the output inside the secure computation step using a freshly generated symmetric key, and then distribute secret shares (and corresponding commitments) of this fresh key. For this technique to work and to ensure that the script complexity of our protocol is independent of the output length of the functions, we will need to work in the programmable random oracle model. This idea was one of two main ideas used in the "compact ladder" protocol in [32] to reduce the script complexity of their protocol in the $\mathcal{F}^\star_{\text{CR}}$-hybrid model.[2] We refer the reader to [32] for additional details.

# Appendix B Proof of Security for $\mathcal{F}^*_{\text{MSFE}}$ Realization

In this section, we provide full details on the simulation of the protocol that realizes $\mathcal{F}^*_{\text{MSFE}}$. Please see Figures 15, 17, and 16.

# Appendix C Proof of Theorem 2

*Proof sketch.* We briefly discuss the optimistic case when all parties are honest. In this case, there are no aborts in any of the local executions. Finally, when parties agree to exit the contract, they each send $w = \mathsf{exit}$

[2]They use a second idea which is more complicated to reduce the script complexity of their $\mathcal{F}^\star_{\text{CR}}$-hybrid protocol from $O(n^2)$ to $O(n)$. While it is possible to use that idea in our context, it does not improve our theorem, and furthermore imposes minor additional restrictions on how parties can trigger $\mathcal{F}^*_{\text{StCon}}$ (namely the triggers must be serialized in say the lexicographic order).

Preliminaries: Each simulated honest party $P_j$ also maintains a local variable $\mathsf{best}_j$.

1. (MPC step) Running the SFE simulator $\mathcal{S}_{\mathsf{sfe}}$:
   - $\mathcal{S}$ extracts inputs $\{y_s\}_{s\in C}$ from $\mathcal{A}$.
   - randomly generates $\{x_s = (z_s; \omega_s)\}_{s\in C}$ where each $z_s, \omega_s$ is generated uniformly at random, and further computes $h_s = \mathsf{com}(x_s)$ for each $s \in C$; and
   - generates dummy commitments $\{(\mathsf{st}_j, h_j)\}_{j\in H}$ by invoking the equivocal algorithm $S_1$; and
   - outputs $(x_s, \boldsymbol{h} = h_1\|\cdots\|h_n)$ to each corrupt $P_s$.
2. (Signature broadcast) Generate signatures from simulated honest parties $\sigma_j$ under $pk_j$ on the message $(\mathsf{id}, \boldsymbol{h})$. Receive from each $P_s$ a signature $\sigma_s$ on message $(\mathsf{id}, \boldsymbol{h})$ and check if the signature is valid under $pk_s$ and then broadcasts it. Let $S$ denote the set of corrupt parties whose signature on message $(\mathsf{id}, \boldsymbol{h})$ was received by $\mathcal{S}$. If $S = C$, then let $\boldsymbol{\sigma} = \sigma_1\|\cdots\|\sigma_n$ and update $\mathsf{best}_j \leftarrow (\mathsf{id}, ((j, x_j), \boldsymbol{h}^{(\mathsf{id})}, \boldsymbol{\sigma}^{(\mathsf{id})}))$ for each $j \in H$. Else, $\mathcal{S}$ suspends execution and goes to the main thread (Figure 17).
3. (Share broadcast) $\mathcal{S}$ contacts $\mathcal{F}^*_{\mathrm{MSFE}}$ with the corrupt inputs and receives the output $z$. Then it randomly samples $\{z_j\}_{j\in H}$ such that $\bigoplus_{k\in[n]} z_k = z$, and generates dummy openings to the commitments $\{x_j\}_{j\in H}$ using the equivocation algorithm $S_2$. Then acting as each honest $P_j$, the simulator $\mathcal{S}$ broadcasts $x_j$. Let $X$ include all the honest decommitments and also those that were broadcasted by $\mathcal{A}$. Each simulated honest party $P_j$ updates $\mathsf{best}_j \leftarrow (\mathsf{id}, (X, \boldsymbol{h}, \boldsymbol{\sigma}))$. If $|X| \neq [n]$, then $\mathcal{S}$ suspends the lcoal executions and goes to the main thread (Figure 17).

Figure 15: Simulating the off-chain local executions for implementing $\mathcal{F}^*_{\mathrm{MSFE}}$.

**Notation.** The variable $\mathsf{st}$ denotes the current state of the contract. The values $\boldsymbol{pk}$ and $\Delta$ are constant parameters to the contract and are always maintained as part of the state. Note that $\boldsymbol{pk}$ is initialized as in Figure 15. We represent the state variable $\mathsf{st}$ as a five tuple ($\mathsf{st.mode}$, $\mathsf{st.id}$, $\mathsf{st}.\textsc{tt}$, $\mathsf{st}.t$, $\mathsf{st}.L$). We omit $\boldsymbol{pk}$ and $\Delta$ from the state to keep the presentation simple.

**Parameters.** The initial deposits $d_1 = \cdots = d_n = (n-1)q$. The initial state is ("init", $-1$, $\perp$, $-1$, $\mathbf{1}$).

**Subroutine $\mathsf{pred}_{\mathcal{S}}$.** For $k \in C$: Let $\mathsf{pred}_{\mathcal{S}}(k, w, t; \mathsf{st}) = 1$ if

- $w$ is parsed as $(\mathsf{id}, \textsc{tt} = (X', \boldsymbol{h}', \boldsymbol{\sigma}'))$ and $\mathsf{pred}(k, w, t; \mathsf{st}) = 1$ (where $\mathsf{pred}$ is as defined in Figure 10) and $\mathsf{st}$ is the state of the simulated $\mathcal{F}^*_{\mathrm{StCon}}$; and
- the $\mathsf{id}$-th local execution simulated full transcript (cf. Figure 15) was $X, \boldsymbol{h}, \boldsymbol{\sigma}$; and $X' \subseteq X$ and $\boldsymbol{h}' = \boldsymbol{h}$ and $\boldsymbol{\sigma} = \boldsymbol{\sigma}'$. If any of the last three conditions fail, then $\mathcal{S}$ outputs $\mathsf{fail}$ and terminates the simulation.

**State transitions.** The function $\mathsf{Prog}(k, w, t; \mathsf{st})$ is defined as follows: Initialize $e \leftarrow 0$.

- If $w = (\mathsf{id}, \textsc{tt})$ and $\mathsf{pred}_{\mathcal{S}}(k, w, t; \mathsf{st}) = 1$: If $\mathsf{st.id} = \mathsf{id}$, then update $\mathsf{st}.\textsc{tt}.X \leftarrow \mathsf{st}.\textsc{tt}.X \cup \textsc{tt}.X$, else set $\mathsf{st}.\textsc{tt} \leftarrow \textsc{tt}$. Set $\mathsf{st} \leftarrow$ ("exec", $\mathsf{id}$, $\mathsf{st}.\textsc{tt}$, $t$, $\mathsf{st}.L$).
- Else if $w = \mathsf{exit}$: (this part is the same as Figure 10)
  - If (1) $\mathsf{st.mode}$ = "init" or (2) $\mathsf{st.mode}$ = "exec" and $|\mathsf{st}.\textsc{tt}.X| = n$: Set $\mathsf{st.mode} \leftarrow$ "exit" and $\mathsf{st}.t \leftarrow t$.
  - If $\mathsf{st.mode}$ = "exec" or "abort", and $t > \mathsf{st}.t + \Delta$ and $\mathsf{st}.L[j] = 1$ and $|\mathsf{st}.\textsc{tt}.X| \neq n$: Then update $\mathsf{st}.L[j] \leftarrow 0$ and $\mathsf{st.mode} \leftarrow$ "abort" and $\mathsf{st}.L[k'] \leftarrow 0$ for all $k$ such that $\mathsf{st}.\textsc{tt}.X[k'] = \perp$. Further, if $\mathsf{st}.\textsc{tt}.X[k] \neq \perp$, then set $e \leftarrow n(n-1)q/|\mathsf{st}.\textsc{tt}.X|$.
  - If $\mathsf{st.mode}$ = "exit" or "payout", and $t > \mathsf{st}.t + \Delta$ and $\mathsf{st}.L[k] = 1$: Set $e \leftarrow (n-1)q$ and update $\mathsf{st.mode} \leftarrow$ "payout" and $\mathsf{st}.L[k] \leftarrow 0$.

If at the end of a transition, it holds that $\mathsf{st}.L[k] = 0$, for all $k \in [n]$ then we update $\mathsf{st.mode} \leftarrow$ "inactive".

Figure 16: Simulating $\mathcal{F}^*_{\mathrm{StCon}}$ for $\mathcal{F}^*_{\mathrm{MSFE}}$.

$\mathcal{S}$ initializes $\boldsymbol{pk}$ with the public keys of the corrupt parties and freshly generated public/signing keys for the honest parties. $\mathcal{S}$ accepts an initialization of parameters as in Figure 16. Then for $\mathsf{id} = 1, 2, \cdots$, the simulator $\mathcal{S}$ follows the steps in Figure 15. Recall that each simulated honest party $P_j$ maintains a variable $\mathsf{best}_j$ during the local executions. $\mathcal{S}$ executes the following steps if the simulated local executions were suspended.

1. If a notification $(\mathsf{exit}, sid, j)$ was obtained from $\mathcal{F}^*_{\text{MSFE}}$ for $j \in H$, then then $\mathcal{S}$ acting as honest $P_j$ sends $w = \mathsf{exit}$ to the simulated $\mathcal{F}^*_{\text{StCon}}$.
2. If there is an abort during a local off-chain execution, then acting as each honest $P_j$, the simulator $\mathcal{S}$ parties trigger the simulated $\mathcal{F}^*_{\text{StCon}}$ with the value $\mathsf{best}_j$.
3. Each simulated honest party $P_j$ waits and respond to state changes in the simulated $\mathcal{F}^*_{\text{StCon}}$ depending on the current state $\mathsf{st}$:
   (a) If $\mathsf{st.mode}$ = "payout" or "abort" and $\mathsf{st}.L[j] = 1$, then send $w = \mathsf{exit}$ to the simulated $\mathcal{F}^*_{\text{StCon}}$.
   (b) If $\mathsf{st.id} < \mathsf{best}_j.\mathsf{id}$, or $\mathsf{st.id} = \mathsf{best}_j.\mathsf{id}$ and $\mathsf{st}.\text{TT}.X[j] = \perp$, then submit $\mathsf{best}_j$ to the simulated $\mathcal{F}^*_{\text{StCon}}$.
   (c) If $\mathsf{st.id} > \mathsf{best}_j.\mathsf{id}$, then submit $\mathsf{best}_j \leftarrow (\mathsf{st.id},\ ((j, x_j^{(\mathsf{st.id})}),\ \mathsf{st}.\text{TT}.\boldsymbol{h},\ \mathsf{st}.\text{TT}.\boldsymbol{\sigma}))$ to the simulated $\mathcal{F}^*_{\text{StCon}}$.

Finally, $\mathcal{S}$ also keeps track of whether $\mathsf{st.mode}$ = "exit" or "exec" and the current time $t$ is such that $t > \mathsf{st}.t + \Delta$. In this case, acting as the simulated honest parties, it sends $w = \mathsf{exit}$ to the simulated $\mathcal{F}^*_{\text{StCon}}$. Finally, $\mathcal{S}$ sends $(\mathsf{extra}, sid, ssid, \{(i, q_i)\}_{i \in H}, \mathsf{coins}(\sum_{i \in H} q_i))$ to $\mathcal{F}^*_{\text{MSFE}}$ where $q_i = (n - |\mathsf{st}.\text{TT}.X|)(n-1)q/|\mathsf{st}.\text{TT}.X|$ which ensures that the simulated $\mathcal{F}^*_{\text{StCon}}$ is not left with any money and therefore can go into inactive mode.

Figure 17: Main simulator for $\mathcal{F}^*_{\text{MSFE}}$.

to $\mathcal{F}^*_{\text{StCon+}}$, get back their initial deposits and put $\mathcal{F}^*_{\text{StCon+}}$ in inactive mode. Note that we calculate the complexity of a state transition (cf. Definition 3) as the sum of the size of the input states and output states. Since the input/output states contain $\boldsymbol{pk}$, we have that the complexity of each state transition is $O(n\lambda)$. We now focus on the more interesting case where some of the parties are corrupt.

We describe the simulator $\mathcal{S}$ and then prove indistinguishability of the simulation and the real world execution. Note that in the simulation, $\mathcal{S}$ would act as $\mathcal{F}^*_{\text{StCon+}}$ and wait for each corrupt party to make the initial deposits $\mathsf{coins}((n-1)q)$. It then uses $\mathsf{coins}(hq)$ from this deposit amount to submit the initial deposit to $\mathcal{F}^*_{\text{MSCD}}$. (Note that there is at least one corrupt party, and so $\mathcal{S}$ will be able to complete this step.) $\mathcal{S}$ initializes $\boldsymbol{pk}$ with public keys from corrupt parties, and supplies randomly generated public keys for honest parties (and stores the corresponding secret keys). Then for $\mathsf{id} = 1, 2, \cdots$: $\mathcal{S}$ waits to receive deposits from the parties (including notifications from $\mathcal{F}^*_{\text{MSCD}}$ of honest parties' deposits) receive the reactive function $g^{(\mathsf{id})}$ from $\mathcal{F}^*_{\text{MSCD}}$, and then picks the transcript validation function for this execution, i.e., $\mathsf{tv}^{(\mathsf{id})}$, for a reactive MPC protocol $\pi$ that implements $g^{(\mathsf{id})}$. If some party wants to add a deposit then $\mathcal{S}$ notifies $\mathcal{F}^*_{\text{MSCD}}$ and forwards the coins from $\mathcal{A}$ to $\mathcal{F}^*_{\text{MSCD}}$ if it accepts the deposit. (Here we implicitly assume that parties are notified of messages received by $\mathcal{F}^*_{\text{StCon+}}$.) Simulation of this step involves signing the updated balance vector using simulated honest parties' public keys depending on the notification from $\mathcal{F}^*_{\text{MSCD}}$. $\mathcal{S}$ then simulates the parameter agreement phase of the $\mathsf{id}$-th local execution by signing $\mathsf{tv}^{(\mathsf{id})}$ and the current balance $\boldsymbol{b}$ using signing keys of the simulated honest parties and then broadcasts these to the corrupt parties. $\mathcal{S}$ then waits to receive their signatures on $\mathsf{tv}^{(\mathsf{id})}$ from the corrupt parties. $\mathcal{S}$ parses $g^{(\mathsf{id})} = \{g_k^{(\mathsf{id})}\}_{k \in [\rho]}$ Then within each execution, for $k = 1, \ldots, \rho$: $\mathcal{S}$ invokes the MPC simulator $\mathcal{S}_{\mathsf{mpc}}$ for the MPC protocol $\pi$ used to implement $g^{(\mathsf{id})}$ and in particular $g_k^{(\mathsf{id})}$. (Note that this simulator works in the $\mathcal{F}_{\text{OT}}$-hybrid model.) $\mathcal{S}_{\mathsf{mpc}}$ extracts the $k$-th stage inputs of the corrupt parties $\{y'_s\}_{s \in C}$ where $C$ represents the set of corrupt parties. Note that for typical MPC protocols, e.g., GMW, this extraction step is possible right after the first round (when parties commit to their inputs and randomness) of that stage. Thus, once the first round of the secure protocol implementing $g_k^{(\mathsf{id})}$ is completed, $\mathcal{S}$ sends the extracted inputs to $\mathcal{F}^*_{\text{MSCD}}$ and waits to receive the output. $\mathcal{S}$ then continues to use $\mathcal{S}_{\mathsf{mpc}}$ along with the output received from $\mathcal{F}^*_{\text{MSCD}}$ to generate simulated honest messages for the execution of $g_k^{(\mathsf{id})}$ which $\mathcal{S}$ sends to $\mathcal{A}$. Note that the simulator keeps track of the value of $\mathsf{best}_j$ for each honest party $P_j$ as in Figure 12. This completes the simulation of the MPC step of the $\mathsf{id}$-th local execution as long as there is no abort or notification from the parties to exit the contract.

Now we discuss how to handle the case when parties try to exit the contract. Suppose $\mathcal{S}$ receives notification of the form $(\mathsf{exit}, sid, ssid, j)$ from $\mathcal{F}^*_{\mathrm{MSFE}}$ for $j \in H$, then acting as $\mathcal{F}^*_{\mathrm{StCon+}}$, $\mathcal{S}$ sends a notification $(\mathsf{output}, sid, ssid, j, w = \mathsf{exit}, t, \mathsf{st}, e)$ to each corrupt party where $e = 0$ and $\mathsf{st}$ represents the current state of the simulated $\mathcal{F}^*_{\mathrm{StCon+}}$. Note that in this case, $\mathcal{S}$ would also receive its deposit $\mathsf{coins}(hq)$ from $\mathcal{F}^*_{\mathrm{MSCD}}$ and it will be able to distribute $\mathsf{coins}((n-1)q)$ to all corrupt parties. On the other hand, suppose $\mathcal{S}$ acting as $\mathcal{F}^*_{\mathrm{StCon+}}$ receives a message $w = \mathsf{exit}$ from the adversary $\mathcal{A}$, then it changes the state of $\mathcal{F}^*_{\mathrm{StCon+}}$ using the function $\mathsf{Prog}(k', w, t; \mathsf{st})$ (as specified in Figure 13) where $\mathsf{st}$ represents the state of the simulated contract, and $k'$ represents the index of the corrupt party on behalf of whom $\mathcal{A}$ submitted the trigger. Let $\mathsf{st}, e$ be the output of $\mathsf{Prog}$. Now $\mathcal{S}$ simulates honest parties response to the notification of $\mathsf{st}, e$ from $\mathcal{F}^*_{\mathrm{StCon+}}$. That is, it simulates the honest parties' execution of Step 3 of Figure 14. Note that the value $\mathsf{best}_j$ for each honest $P_j$ is prepared using the MPC simulator $\mathcal{S}_{\mathsf{mpc}}$ for $g_k^{(\mathsf{id})}$ (that guarantees that the output is consistent with the output of $\mathcal{F}^*_{\mathrm{MSCD}}$). Now Step 3 instructs honest parties to send their messages to $\mathcal{F}^*_{\mathrm{StCon+}}$. Therefore, $\mathcal{S}$ acting as $\mathcal{F}^*_{\mathrm{StCon+}}$ needs to react to the messages received from the simulated honest parties. That is, $\mathcal{S}$ acting as $\mathcal{F}^*_{\mathrm{StCon+}}$ will invoke $\mathsf{Prog}$ using the inputs submitted by the simulated honest parties, and then distribute the output $\mathsf{st}, e$ from $\mathcal{F}^*_{\mathrm{StCon+}}$ to $\mathcal{A}$. This process continues until $\mathcal{F}^*_{\mathrm{StCon+}}$ is in inactive mode. Next we will discuss how to handle aborts in the off-chain execution. When $\mathcal{S}$ is simulating the off-chain executions, it sends the simulated honest parties' messages immediately, and waits for time period $\delta$ to receive messages from $\mathcal{A}$ on behalf of the corrupt party that is supposed to send the next message. If the message was not received within this time period, then this counts as an abort in the off-chain execution. For every honest party $P_j$, the simulator $\mathcal{S}$ acting as the simulated honest party $P_j$ prepares a value $\mathsf{best}_j$ which corresponds to the transcript corresponding to the most recent execution. $\mathcal{S}$ then notifies $\mathcal{A}$ of the value of $\mathsf{st}, e$ that results from a trigger $\mathsf{best}_j$. Note that notifications are sent separately for each honest $P_j$. Accordingly, $\mathcal{S}$ will update the value $\mathsf{st}.\mathrm{TT}$ as specified in Figure 13 for each $\mathsf{best}_j$ resulting from the simulated honest parties. Note that $\mathcal{A}$ could trigger $\mathcal{F}^*_{\mathrm{StCon+}}$ with an alternate partial/completed transcript *any time* during the execution of the protocol. This case is similar to the case where $\mathcal{S}$ handles a witness $w = \mathsf{exit}$ from $\mathcal{A}$ (except here $w$ might be a partial transcript). Suppose $\mathcal{A}$ on behalf of corrupt party $P_{k'}$ triggered $\mathcal{F}^*_{\mathrm{StCon+}}$ with witness $w$ at time $t$. Then, $\mathcal{S}$ faithfully applies the trigger $(k', w, t)$ and invokes $\mathsf{Prog}$ and notifies $\mathcal{A}$ of the resulting output $\mathsf{st}, e$ (i.e., if trigger $(k, w, t)$ was valid and induced a state transition). As before, $\mathcal{S}$ simulates the response from the honest parties by running Step 3 of Figure 14 to generate new triggers if any, and applies them to $\mathcal{F}^*_{\mathrm{StCon+}}$. Then acting as $\mathcal{F}^*_{\mathrm{StCon+}}$, $\mathcal{S}$ notifies the parties of state transitions in $\mathcal{F}^*_{\mathrm{StCon+}}$. Note that the trigger $w$ is supposed to contain only those honest parties' messages that were sent to $\mathcal{A}$ by $\mathcal{S}$ (where $\mathcal{S}$ either acted as $\mathcal{F}^*_{\mathrm{StCon+}}$, or the simulated honest parties, or during the simulation of the MPC protocol). If not and if the value $w$ results in a successful state transition, then the simulator outputs $\mathsf{fail}$ and terminates the protocol. Note that this additional restriction, namely that $w$ only contains values sent by $\mathcal{S}$, does not mean that $\mathcal{A}$ behaves honestly as $\mathcal{A}$ could satisfy the constraint and still submit a partial transcript corresponding to an older completed execution even when the current execution is incomplete.

We briefly discuss events which can lead to a simulation failure. As mentioned before, this happens when $\mathcal{A}$ submits a witness $w$ that modifies the simulated honest parties' messages in the transcript. This happens with negligible probability as each honest party's message is signed with the corresponding signing key, and thus by the unforgeability property of the underlying digital signature scheme, it follows that the honest messages will be left in tact. This property also implies that $\mathcal{A}$ cannot modify even corrupt parties' messages that are followed by an honest party's message (as it would have to change the signatures). It is fairly straightforward to create hybrids and prove that the simulation failure does not affect the indistinguishability of the real world execution and the simulated execution in the ideal world.

Finally we discuss the simulation at the end of the protocol when the simulated $\mathcal{F}^*_{\mathrm{StCon+}}$ has been in exit mode for longer than the time period $\Delta$. In this case, $\mathcal{S}$ acting as the simulated honest parties will provide $w = \mathsf{exit}$ to get their (simulated) deposits out of $\mathcal{F}^*_{\mathrm{StCon+}}$. The simulator $\mathcal{S}$ accordingly notifies $\mathcal{A}$ of these triggers. $\mathcal{S}$ then handles triggers $w = \mathsf{exit}$ from corrupt party $P_{k'}$ controlled by $\mathcal{A}$ then it simulates $\mathcal{F}^*_{\text{StCon+}}$ and notifies all parties including the simulated honest parties. Since no valid triggers were received within time period $\Delta$ (including messages from simulated honest parties that do not induce transitions in $\mathcal{F}^*_{\text{StCon+}}$—this case can happen only when the all off-chain and on-chain executions have been completed, i.e., $|\mathsf{st}.\text{TT}| = |\mathsf{st.tv}|$), $\mathcal{S}$ issues $(\mathsf{exit}, sid, k')$ to $\mathcal{F}^*_{\text{MSCD}}$ and collects its deposit $\mathsf{coins}(hq)$ from $\mathcal{F}^*_{\text{MSCD}}$. Now $\mathcal{S}$ possesses all the initial deposits, then it can distribute these coins back to the adversary, i.e., $\mathsf{coins}((n-1)q)$ along with the coins specified by the balance vector $\mathsf{st}.\boldsymbol{b}$ and the coins specified by the algorithm $\mathsf{cash}_{k'}$ applied to the current completed on-chain transcript $\mathsf{st}.\text{TT}$, to party corrupt party $P_{k'}$. $\mathcal{S}$ will be able to obtain these coins from $\mathcal{F}^*_{\text{MSCD}}$. Finally, $\mathcal{S}$ puts the simulated $\mathcal{F}^*_{\text{StCon+}}$ into inactive mode and terminate the simulation. On the other hand, if $|\mathsf{st}.\text{TT}| \neq |\mathsf{st.tv}|$, then this means there is an off-chain execution which was aborted, and which when escalated on-chain was also aborted. In this case, we need to distribute compensations to the honest parties. In this case, $\mathcal{S}$ sends $(\mathsf{abort}, sid, ssid)$ to $\mathcal{F}^*_{\text{MSFE}}$. Let $j_a = 1 + |\mathsf{st}.\text{TT}| \bmod n$. (We effectively penalize corrupt party $P_{j_a}$.) For each $k' \in C$ such that $k' \neq j_a$, $\mathcal{S}$ sends $\mathsf{coins}(nq)$ to $P_{k'}$. Note that in this case $\mathcal{S}$ does not receive its initial deposit $\mathsf{coins}(hq)$ from $\mathcal{F}^*_{\text{MSCD}}$ but will still be able to obtain the coins specified by the balance vector (which was not updated to include the aborted execution). Observe that the simulated $\mathcal{F}^*_{\text{StCon+}}$ is not left with any money and therefore can go into inactive mode. This completes the description of the simulation. We now state and prove helper propositions which analyzes the properties of the real/simulated execution and ensures that the simulation is indistinguishable from the real execution.

**Proposition 7.** *At the end of the protocol, honest parties do not pay a penalty, i.e., they get their initial deposit back. Furthermore, they obtain the coins specified by the balance vector that corresponds to the last completed execution (on-chain or off-chain).*

*Proof sketch.* Since we penalize exactly one party $P_{j_a}$ with $j_a = 1 + |\mathsf{st}.\text{TT}| \bmod n$, it suffices to show that $j_a \in C$. For the sake of contradiction, suppose $j_a \in H$. Clearly, $\mathsf{st.id} \leq \mathsf{best}_j.\mathsf{id}$ holds except with negligible probability since otherwise $\mathcal{A}$ will not be able to generate the signature $\sigma'_{j_a}$ on $\mathsf{tv}^{(\mathsf{st.id})}$. Also note that the value $\mathsf{best}_{j_a}$ generated in Step 3(b) is guaranteed to be a valid trigger since $\pi$ is a secure protocol for implementing $g^{(\mathsf{best}_{j_a}.\mathsf{id})}$ and $\mathsf{best}_{j_a}$ is set using the transcript of $\pi$. Then it follows that Step 3(c) or 3(d) in Figure 14 must have failed to produce a legitimate extension of $\mathsf{st}.\text{TT}$. Failure of Step 3(c) or 3(d) implies the failure of the next message function $\mathsf{nmf}$ which in turn implies by the correctness of $\pi$ that the input transcript $\mathsf{st}.\text{TT}$ must not correspond to the local execution that was aborted. This is a contradiction since $\mathsf{st}.\text{TT}$ must satisfy the predicate $\mathsf{st.tv}$ which since was signed by all honest parties and hence is guaranteed to be a faithful transcript validator for the underlying MPC protocol $\pi$. The second part of the proposition statement follows from the fact that honest parties trigger $\mathcal{F}^*_{\text{StCon+}}$ with $\mathsf{exit}$ following which $\mathcal{F}^*_{\text{StCon+}}$computes (1) a valid balance vector that corresponds to the latest completed off-chain execution which will be used if the on-chain contract aborted, or (2) a valid balance vector obtained by applying the algorith $\mathsf{cash}$ which will be used if the on-chain execution was completed. In either case, it is easy to see that the statement holds. This concludes the proof.

**Proposition 8.** *Suppose Step 2 of the* $\mathsf{id}$*-th* local *execution was completed by* **all** *parties. Then, either the* $\mathsf{id}$*-th execution was completed (either on-chain or off-chain) and the honest parties learned the output of the entire execution, or all honest parties obtained compensation. Furthermore, honest parties obtain the coins specified by the balance vector that corresponds to the last completed execution (on-chain or off-chain).*

*Proof sketch.* Clearly, if the execution was completed off-chain, then parties learned the output, and the proposition holds. We focus on the interesting case when the off-chain execution was aborted. Since Step 1 was completed, honest parties update $\mathsf{best}_j \leftarrow (\mathsf{transcript},\ \mathsf{id},\ \perp,\ \mathsf{tv}^{(\mathsf{id})},\ \boldsymbol{\sigma}^{(\mathsf{id})})$. Observe that by the logic in Figures 14, 12 it holds that updates to $\mathsf{best}_j$ only increase the value of $\mathsf{best}_j.\mathsf{id}$ or keep it equal to its previous value. Therefore, when the honest parties escalate the aborted off-chain execution to $\mathcal{F}^*_{\text{StCon+}}$ (cf. Step 2 in Figure 14), they would trigger $\mathcal{F}^*_{\text{StCon+}}$ with a value $\mathsf{best}_j$ satisfying $\mathsf{best}_j.\mathsf{id} = \mathsf{id}$. Furthermore, whenever (honest) parties receive a notification from $\mathcal{F}^*_{\text{StCon+}}$ that it is their turn to send the next message (Step 3(d) in Figure 14) they compute the next message using the next message function $\mathsf{nmf}$ and send the newly generated message to $\mathcal{F}^*_{\text{StCon+}}$. As argued in the proof of Proposition 7, this newly generated message is guaranteed except with negligible probability to be accepted by $\mathcal{F}^*_{\text{StCon+}}$. Therefore, an honest party $P_j$ is never stuck in an execution, and therefore will not be penalized. Furthermore, if the execution is not completed on-chain, then we have by the logic in Figure 13 that one party (which must be corrupt) will be penalized and lose his deposit of $\mathsf{coins}((n-1)q)$ which will be distributed evenly among the remaining parties. The last part of the proposition statement can be proved using an argument identical to the one in the proof of Proposition 7. This completes the proof of the proposition.

**Proposition 9.** *Suppose Step 2 of the* $\mathsf{id}$*-th* local *execution was completed by* **honest** *parties. Then, either the* $\mathsf{id}$*-th execution was completed (either on-chain or off-chain) and the honest parties learned the output of the entire execution, or all honest parties obtained compensation. Furthermore, they obtain the coins specified by the balance vector that corresponds to the last completed execution (on-chain or off-chain).*

*Proof sketch.* Suppose Step 1 was completed by all parties, then we can apply Proposition 8 and we are done. We focus on the remaining case where at least one corrupt party did not broadcast its signature on $\mathsf{tv}^{(\mathsf{id})}$. Note that in this case, the corrupt parties possess the signatures from all honest parties, and therefore can trigger $\mathcal{F}^*_{\mathrm{StCon}+}$ with a witness that *begins* the $\mathsf{id}$-th execution *on-chain* (i.e., the $\mathsf{id}$-th off-chain reactive computation was not even started). This is precisely the case handled by Step 3(c) in Figure 14. Note that $\mathsf{st.id} = \mathsf{id}$ will equal $\mathsf{best}_j.\mathsf{id} + 1$ for an honest $P_j$. In this case, we instruct $P_j$ to choose an appropriate fresh input $y_j$ and randomness $\omega_j$ and continue the execution on-chain. After this step, $\mathsf{best}_j$ is updated and will satisfy $\mathsf{st.id} = \mathsf{best}_j.\mathsf{id}$. Then by an argument identical to the one made in Proposition 8, we have that an honest party $P_j$ is never stuck in an execution and therefore will not be penalized. As before, if the execution is not completed on-chain, then by the logic in Figure 13, we have that one party (which must be corrupt) will be penalized and lose his deposit of $\mathsf{coins}((n-1)q)$ which will be distributed evenly among the remaining parties. The last part of the proposition statement can be proved using an argument identical to the one in the proof of Proposition 7. This completes the proof. □

# Appendix D Honest Binding Commitments

**Definition 4** (Honest binding commitments [20])**.** *A (non-interactive)* commitment scheme *for message space* $\{\mathcal{M}_\lambda\}$ *is a pair of* PPT *algorithms* $S, R$ *such that for all* $\lambda \in \mathbb{N}$*, all messages* $m \in \mathcal{M}_\lambda$*, and all random coins* $\omega$ *it holds that* $R(m, S(1^\lambda, m; \omega), \omega) = 1$*. A commitment scheme for message space* $\{\mathcal{M}_\lambda\}$ *is* honest-binding *if:*

**Binding (for an honest sender)** *For all* PPT *algorithms* $\mathcal{A}$ *(that maintain state throughout their execution), the following is negligible in* $\lambda$*:*

$$\Pr\left[\begin{array}{l} m \leftarrow \mathcal{A}(1^\lambda); \\ \omega \leftarrow \{0,1\}^*; \mathsf{com} \leftarrow S(1^\lambda, m; \omega); \\ (m', \omega') \leftarrow \mathcal{A}(\mathsf{com}, \omega) : \\ \qquad R(m', \mathsf{com}, \omega') = 1 \bigwedge m' \neq m \end{array}\right]$$

**Equivocation** *There is an algorithm* $\tilde{S} = (\tilde{S}_1, \tilde{S}_2)$ *such that for all* PPT $\mathcal{A}$ *(that maintain state throughout their execution) the following is negligible:*

$$\left| \Pr\left[\begin{array}{l} m \leftarrow \mathcal{A}(1^\lambda); \\ \omega \leftarrow \{0,1\}^*; \mathsf{com} \leftarrow S(1^\lambda, m; \omega) : \\ \qquad \mathcal{A}(1^\lambda, \mathsf{com}, \omega) = 1 \end{array}\right] - \Pr\left[\begin{array}{l} (\mathsf{com}, \mathsf{st}) \leftarrow \tilde{S}_1(1^\lambda); \\ m \leftarrow \mathcal{A}(1^\lambda); \omega \leftarrow \tilde{S}_2(\mathsf{st}, m) : \\ \qquad \mathcal{A}(1^\lambda, \mathsf{com}, \omega) = 1 \end{array}\right] \right|$$

Equivocation implies the standard hiding property. Also, observe that binding holds for commitments generated by $(\tilde{S}_1, \tilde{S}_2)$. As observed in [6], we can construct highly efficient heuristically secure honest binding commitment schemes in the *programmable random oracle* model. In the following let $\mathsf{Hash}$ be a programmable hash function, and let $\omega \in \{0,1\}^\lambda$. We describe the algorithms $S, R$ (algorithms $\tilde{S}_1, \tilde{S}_2$ are obtained by standard oracle programming techniques).

$\underline{S(1^k, m; \omega)}$
return $\mathsf{com} := \mathsf{Hash}(m \| \omega)$;

$\underline{R(m, \mathsf{com}, \omega)}$
If $\mathsf{com} \stackrel{?}{=} \mathsf{Hash}(m \| \omega)$
    return 1;
else return 0;

# Appendix E Code

Here, we provide the implementation code for the stateful contracts that we discussed in Sections 6 and 7. Please see Figures 18 to 22.

```
contract SmartAmortize {
  address[] public players;
  mapping (address => uint) playermap;
  int bestRound = -1;
  mapping(int => bytes32[]) commits;
  mapping(bytes32 => bytes32) openings;
  event EventTriggered(uint T1, uint T2);
  uint T1;
  uint T2;
  uint T3;
  modifier after_ (uint T) { if (T > 0 && block.number >= T) _; else throw;}
  modifier before(uint T) { if (T > 0 && block.number <  T) _; else throw; }
  modifier onlyplayers { if (playermap[msg.sender] > 0) _; else throw; }
  modifier beforeTrigger { if (T1 == 0) _; else throw; }

  function SmartAmortize(address[] _players) {
    for (uint i = 0; i < _players.length; i++) {
      players.push(_players[i]);
      playermap[_players[i]] = (i+1);
    }
  }

  function trigger() onlyplayers beforeTrigger {
    T1 = block.number;
    T2 = block.number + 10;
    T3 = block.number + 20;
    EventTriggered(T1, T2);
  }

  function submitClaim(Signature[] sigs, uint r, bytes32[] _commits)
  after_(T1) before(T2) {
    // Advance to a new claim with a larger number
    var _h = sha3(r, _commits);
    for (uint j = 0; j < players.length; j++) {
      verifySignature(players[j], _h, sigs[j]);
    }
    commits[r] = _commits;
    if (r > bestRound) bestRound = r;
  }

  function openCommitment(bytes32 opening) {
    openings[sha3(opening)] = opening;
  }

  function finalize() after_(T3) {
    uint DEPOSIT = COMPENSATION * (players.length - 1);

    var anyCorrupt = false;
    if (bestRound >= 0)
      for (var i = 0; i < players.length; i++) {
        // Did any player fail to open?
        if (openings[commits[bestRound][i]] == 0)
          anyCorrupt = true;
      }

    for (i = 0; i < players.length; i++) {
      if (anyCorrupt && openings[commits[bestRound][i]] != 0 ) {
        assert(players[i].send(DEPOSIT + COMPENSATION));
      } else {
        assert(players[i].send(DEPOSIT));
      }
    }
  }
```

Figure 18: Implementation of the SmartAmortize contract in Solidity.

```
contract PrepareDeposits {
    modifier after_ (uint T) {if (T > 0 && block.number >= T) _;else throw;}
    modifier before(uint T) {if (T == 0 || block.number < T) _; else throw;}

    uint T1;
    uint threshold;
    address recipient;

    bool complete;
    mapping (address => uint) deposits;

    function PrepareDeposits(uint _T1, uint _threshold, address _recipient){
        T1 = _T1;
        threshold = _threshold;
        recipient = _recipient;
    }

    function finalize() after_(T1) {
        if (complete) recipient.send(this.balance);
    }

    function withdraw() after_(T1) {
        if (!complete) {
            msg.sender.send(deposits[msg.sender]);
            deposits[msg.sender] = 0;
        }
    }

    function deposit() before(T1) payable {
        deposits[msg.sender] += msg.value;
        if (this.balance >= threshold) complete = true;
    }
}
```

Figure 19: Implementation of the Deposits contract in Solidity.

```
contract SmartAmortize {
  ...

  int256 secondBestRound = -1;
  bytes32 INITIAL_STATE = {hardcoded};
  ...

  function submitClaim(Signature[] sigs, uint r, bytes32[] _commits)
  after_(T1) before(T2) {
    // Advance to a new claim with a larger number
    var _h = sha3(r, _commits);

    for (uint j = 0; j < players.length; j++) {
      verifySignature(players[j], _h, sigs[j]);
    }
    commits[r] = _commits;

    if (r > bestRound) {
      secondBestRound = bestRound;
      bestRound = r;
    } else if (r > secondBestRound) {
      secondBestRound = r;
    }
  }
  ...
  function finalize() after_(T3) {
    uint DEPOSIT = COMPENSATION * (players.length - 1);
    uint[] memory PAYMENT;

    var anyCorrupt = false;
    if (bestRound >= 0) {
      for (var i = 0; i < players.length; i++) {
        // Did any player fail to open?
        if (openings[commits[bestRound][i]] == 0)
          anyCorrupt = true;
      }
    }

    bytes32 state = 0;
    if (!anyCorrupt) {
      for (i = 0; i < players.length; i++) {
        state ^= openings[commits[bestRound][i]];
      }
    } else if (secondBestRound >= 0) {
      for (i = 0; i < players.length; i++) {
        state ^= openings[commits[bestRound][i]];
      }
    } else {
      state = INITIAL_STATE;
    }
    PAYMENT = applicationSpecificPayment(state);

    for (i = 0; i < players.length; i++) {
      if (openings[commits[bestRound][i]] != 0) {
        if (anyCorrupt) {
          assert(players[i].send(DEPOSIT + COMPENSATION + PAYMENT[i]));
        } else {
          assert(players[i].send(DEPOSIT + PAYMENT[i]));
        }
      }
    }
  }
```

Figure 20: Extension of the SmartAmortize contract to payments that depend on the (most recent) state.

```
contract SmartDuplex {
  address[2] players;
  mapping (address => uint) playermap;
  int bestRound = -1;
  uint[2] deposits;
  int net;
  uint[2] withdrawals;
  uint[2] withdrawn;
  event EventTriggered(uint T1, uint T2);
  uint T1;
  uint T2;
  modifier after_ (uint T) { if (T > 0 && block.number >= T) _; else throw;}
  modifier before(uint T) { if (T > 0 && block.number <  T) _; else throw; }
  modifier onlyplayers { if (playermap[msg.sender] > 0) _; else throw; }
  modifier beforeTrigger { if (T1 == 0) _; else throw; }

  function SmartDuplex(address[2] _players) {
    for (uint i = 0; i < 2; i++) {
      players[i] = _players[i];
      playermap[_players[i]] = (i+1);
    }
  }

  function deposit() onlyplayers beforeTrigger payable {
    deposits[playermap[msg.sender]-1] += msg.value;
  }

  function trigger() onlyplayers afterInit beforeTrigger {
    T1 = block.number;
    T2 = block.number + 10;
    EventTriggered(T1, T2);
  }

  function update(Signature[2] sigs, int r, int _net,
                  uint[2] _withdrawals) before(T2) {
    if (r <= bestRound) return;
    bestRound = r;
    var _h = sha3(r, _net, _withdrawals);
    for (uint i = 0; i < 2; i++) {
      verifySignature(players[i], _h, sigs[i]);
      withdrawals[i] = _withdrawals[i];
    }
    net = _net;
  }

  function withdraw() onlyplayers {
    uint i = playermap[msg.sender]-1;
    uint toWithdraw = 0;

    // Before finalizing, can withdraw up to withdrawals[i]
    if (T2 == 0 || block.number < T2) {
      toWithdraw = withdrawals[i] - withdrawn[i];
    }
    // After finalizing, can withdraw up to deposit+/-net
    else {
      // positive net: Alice gets money
      int net2 = (i == 0) ? net : -net;
      var finalBalance = uint(int(deposits[i]) + net2);
      toWithdraw = finalBalance - withdrawn[i];
    }

    withdrawn[i] = toWithdraw;
    assert(msg.sender.send(toWithdraw));
  }
}
```

Figure 21: Implementation of the SmartDuplex contract in Solidity.

```
contract equalityDLOGs {
    // Modulus for public keys
    uint constant pp =
     0xFFFFFFFFFFFFFFFFFFFFFFFFFFFFFFFFFFFFFFFFFFFFFFFFFFFFFFFEFFFFFC2F;

    function verifyHalf(uint[2] G, uint[2] X, uint[3] K, uint s, uint c)
     internal constant returns (bool) {
        if (!Secp256k1.isPubKey(X)) return false;
        if (!Secp256k1.isPubKey(K)) return false;

        uint[3] memory sG = Secp256k1._mul(s, G);
        uint[3] memory cX = Secp256k1._mul(c, X);

        // Add both points together
        uint[3] memory KcX = Secp256k1._add(K,cX);

        // Convert to Affine Co-ordinates
        ECCMath.toZ1(KcX, pp);
        ECCMath.toZ1(sG, pp);

        // Verify. Do they match?
        if(KcX[0] == sG[0] &&
           KcX[1] == sG[1]) {
            return true;
        } else {
            return false;
        }
    }

    function verifyFull(uint[2] memory G, uint[2] memory H,
     uint[2] memory X, uint[2] memory Y,
     uint[3] KX, uint[3] KY, uint s) constant returns (uint) {

        //The base points (generators) are G[0]=Gx,G[1]=Gy,H[0]=Hx,H[1]=Hy

        // Get c = H(KX, KY);
        bytes32 b_c = sha256(KY[0], KX[0]);
        uint c = uint(b_c);

        if (!verifyHalf(G, X, KX, s, c)) return 0;
        if (!verifyHalf(H, Y, KY, s, c)) return 0;
        return 1;
    }
}
```

Figure 22: Solidity code for proof of knowledge of equality of discrete logarithms.